\documentclass[11pt]{article}
\usepackage{url,subfigure,fullpage,ifthen,calc,xspace,graphicx,wrapfig,mdwlist}
\usepackage{latexsym,amssymb,pst-node,pstricks}

\newcommand{\mylabel}[1]{\textnormal{\textsf{[#1]}}

\label{#1}}
\renewcommand{\mylabel}[1]{\label{#1}}

\let\latexcite=\cite
\def\cite{\nolinebreak\latexcite}
\let\latexref=\ref
\def\ref{\nolinebreak\latexref}

\psset{unit=0.1in}
\newcommand{\np}{\mathsf{NP}}

\newcommand{\poly}{\mathrm{poly}}

\newtheorem{theorem}{Theorem}[section]
\newtheorem{fact}[theorem]{Fact}
\newtheorem{proposition}[theorem]{Proposition}
\newtheorem{lemma}[theorem]{Lemma}

\newtheorem{corollary}[theorem]{Corollary}

\newenvironment{proof}{
\vspace*{-\parskip}\noindent\textit{Proof.}}{$\qed$

\medskip
}

\newenvironment{proof*}{\noindent\textit{Proof.}}{

\vspace{1pc}
}

\newcommand{\tup}[1]{\langle #1 \rangle}

\newcommand{\qed}{\hfill\Box}
\renewcommand{\epsilon}{\varepsilon}

\newcommand{\modulo}{~\mathsf{mod}~}
\newcommand{\mod}{\mathsf{mod}~}

\newcounter{romanlistcounter}
  {\setcounter{romanlistcounter}{0}%
   \begin{list}{\textit{(\roman{romanlistcounter})}}{%
        \usecounter{romanlistcounter}%
      \setlength{\itemsep}{0pc}%
      \setlength{\itemindent}{1pc}%
      \setlength{\topsep}{0pc}%
      \setlength{\mylabelwidth}{3em}}}
  {\end{list}}

\newcounter{alphalistcounter}
  {\setcounter{alphalistcounter}{0}%
   \begin{list}{\textit{(\alph{alphalistcounter})}}{%
        \usecounter{alphalistcounter}%
      \setlength{\itemsep}{0pc}%
      \setlength{\itemindent}{1pc}%
      \setlength{\topsep}{0pc}%
      \setlength{\mylabelwidth}{3em}}}
  {\end{list}}

\def\compactify{\itemsep=0pt \topsep=0pt \partopsep=0pt \parsep=0pt}
\let\latexusecounter=\usecounter
\newenvironment{CompactEnumerate}
  {\def\usecounter{\compactify\latexusecounter}
   \begin{enumerate}}
  {\end{enumerate}\let\usecounter=\latexusecounter}

\newenvironment{oldtheorem}[1][Theorem]{
  \noindent
  \textbf{#1} \em
  }{

  ~

  }

\newif\ifabstract
\abstractfalse
\newif\iffull
\fulltrue

\newif\ifredraw
\redrawfalse

\newrgbcolor{darkgreen}{0 0.5 0}
\newcmykcolor{orange}{0 0.20 0.84 0}
\newcommand{\fillcolor}{white}

\newcommand{\comment}[1]{}

\iffull
\renewenvironment{itemize*}{\begin{itemize}\vspace*{0ex}}{\end{itemize}\vspace*{-1ex}}
\renewenvironment{enumerate*}{\begin{enumerate}}{\end{enumerate}}
\renewenvironment{description*}{\begin{description}}{\end{description}}
\else
\def\compactify{\topsep=0ex \itemsep=0ex  \partopsep=0pt \parsep=0pt}
\let\latexusecounter=\usecounter

\let\realbibitem=\bibitem
\def\bibitem{\par \vspace{-1.2ex}\realbibitem}

\def\captionfont{\rm\em\footnotesize}
\def\captionlabelfont{\bf\footnotesize}
{\makeatletter
 \global\let\plainfont@makecaption=\@makecaption
 \long\gdef\@makecaption#1#2{%
   \plainfont@makecaption{\captionlabelfont #1}{\captionfont #2}}}

\fi

\psset{fillstyle=solid}
\psset{linewidth=0.2pt}

\newcommand{\floor}[4][white]{
  \begin{pspicture}(0,0)(4,4)
    \pspolygon[fillcolor=#1](0,0)(0,1)(0,0)(0,#2)(1,#2)(1,#3)(2,#3)(2,#4)(3,#4)(3,0)(3,1)(3,0)
  \end{pspicture}
}

\newcommand{\gridline}[1][nogrid]{
  \begin{pspicture}(0,0)(4,4)
    \pspolygon[fillcolor=\fillcolor,linecolor=\fillcolor,linewidth=0pt](0,0)(0,1)(3,1)(3,0)
    \psline[linecolor=black](0,0)(0,1)
    \psline[linecolor=black](3,0)(3,1)

    \ifthenelse{\equal{grid}{#1}}{
      \psline(0,0)(3,0)
      \psline(0,1)(3,1)
      \psline(1,0)(1,1)
      \psline(2,0)(2,1)
      }{}
  \end{pspicture}
  }

\newcommand{\notchline}[1][nogrid]{
 \begin{pspicture}(0,0)(4,4)

   \pspolygon[fillcolor=\fillcolor,linecolor=\fillcolor,linewidth=0pt](0,0)(0,1)(5,1)(5,0)
   \psline[linecolor=black](0,0)(0,1)
   \psline[fillstyle=none,linecolor=black](3,-1)(3,0)(5,0)(5,1)(3,1)(3,2)

      \ifthenelse{\equal{grid}{#1}}{
        \psline(0,0)(5,0)
        \psline(0,1)(5,1)
        \psline(1,0)(1,1)
        \psline(2,0)(2,1)
        \psline(3,0)(3,1)
        \psline(4,0)(4,1)
      }{}
  \end{pspicture}
}

\newcounter{currentcolumn}

\newcommand{\backgroundline}[3][nogrid]{
 \begin{pspicture}(0,0)(4,4)
    \pspolygon[fillcolor=#3,linecolor=#3,linewidth=0pt](0,0)(0,1)(#2,1)(#2,0)
      \psline[linecolor=black](0,0)(0,1)
      \psline[linecolor=black](#2,0)(#2,1)

    \ifthenelse{\equal{grid}{#1}}{
      \setcounter{currentcolumn}{0}
      \psline(0,0)(#2,0)
      \psline(0,1)(#2,1)
      
      \whiledo{\value{currentcolumn} < #2} {
        \psline(\value{currentcolumn},0)(\value{currentcolumn},1)
        \stepcounter{currentcolumn}        
      }}{}
  \end{pspicture}
}

\newcounter{notchheight}
\newcounter{currentheight}

\newcommand{\background}[4][nogrid]{
  \setcounter{currentheight}{0}
  \whiledo{\value{currentheight} < #3}{
    \piece{\backgroundline[#1]{#2}{#4}}0{\value{currentheight}}
    \stepcounter{currentheight}
    }
}

\newcommand{\column}[3][nogrid]{
  \setcounter{notchheight}{#2}
  \addtocounter{notchheight}{-1}
  \setcounter{currentheight}{0}
  \whiledo{\value{currentheight} < #3}{
    \ifthenelse{\value{currentheight} < \value{notchheight}}
    {
      \piece{\gridline[#1]}0{\value{currentheight}}
      \stepcounter{currentheight}
      }
    {
      \piece{\notchline[#1]}0{\value{currentheight}}
      \stepcounter{currentheight}
      \addtocounter{notchheight}{6}
      }
    }
  \piece{\floor000}00
  }

\newcommand{\leftgunleft}[1][lightgray]{
  \begin{pspicture}(0,0)(4,4)
    \pspolygon[fillcolor=#1](0,1)(0,2)(3,2)(3,0)(2,0)(2,1)
  \end{pspicture}
  }

\newcommand{\leftgunright}[1][lightgray]{
  \begin{pspicture}(0,0)(4,4)
    \pspolygon[fillcolor=#1,fillstyle=solid](3,1)(3,0)(0,0)(0,2)(1,2)(1,1)
  \end{pspicture}
  }

\newcommand{\leftgunup}[1][lightgray]{
  \begin{pspicture}(0,0)(4,4)
    \pspolygon[fillcolor=#1](1,3)(2,3)(2,0)(0,0)(0,1)(1,1)
  \end{pspicture}
 }

\newcommand{\leftgundown}[1][lightgray]{
  \begin{pspicture}(0,0)(4,4)
    \pspolygon[fillcolor=#1](1,0)(0,0)(0,3)(2,3)(2,2)(1,2)
  \end{pspicture}
  }

\newcommand{\rightgunleft}[1][lightgray]{
 \begin{pspicture}(0,0)(4,4)
    \pspolygon[fillcolor=#1](0,1)(0,0)(3,0)(3,2)(2,2)(2,1)
  \end{pspicture}
  }

\newcommand{\rightgunright}[1][lightgray]{
  \begin{pspicture}(0,0)(4,4)
    \pspolygon[fillcolor=#1](3,1)(3,2)(0,2)(0,0)(1,0)(1,1)
  \end{pspicture}
  }

\newcommand{\rightgundown}[1][lightgray]{
 \begin{pspicture}(0,0)(4,4)
    \pspolygon[fillcolor=#1](1,3)(0,3)(0,0)(2,0)(2,1)(1,1)    
  \end{pspicture}
  }

\newcommand{\rightgunup}[1][lightgray]{
  \begin{pspicture}(0,0)(4,4)
    \pspolygon[fillcolor=#1](1,0)(2,0)(2,3)(0,3)(0,2)(1,2)
  \end{pspicture}
  }

\newcommand{\leftsnakehorizontal}[1][lightgray]{
  \begin{pspicture}(0,0)(4,4)
    \pspolygon[fillcolor=#1](0,1)(0,2)(2,2)(2,1)(3,1)(3,0)(1,0)(1,1)
  \end{pspicture}
}

\newcommand{\leftsnakevertical}[1][lightgray]{%
 \begin{pspicture}(0,0)(4,4)%
  \pspolygon[fillcolor=#1](1,0)(0,0)(0,2)(1,2)(1,3)(2,3)(2,1)(1,1)%
  \end{pspicture}%
}

\newcommand{\rightsnakehorizontal}[1][lightgray]{
 \begin{pspicture}(0,0)(4,4)
    \pspolygon[fillcolor=#1](3,1)(3,2)(1,2)(1,1)(0,1)(0,0)(2,0)(2,1)
  \end{pspicture}
}

\newcommand{\rightsnakevertical}[1][lightgray]{
 \begin{pspicture}(0,0)(4,4)
    \pspolygon[fillcolor=#1](1,0)(2,0)(2,2)(1,2)(1,3)(0,3)(0,1)(1,1)
  \end{pspicture}
}

\renewcommand{\square}[1][lightgray]{
 \begin{pspicture}(0,0)(4,4)
    \pspolygon[fillcolor=#1](0,0)(0,2)(2,2)(2,0)
  \end{pspicture}
  }

\newcommand{\ivertical}[1][lightgray]{
 \begin{pspicture}(0,0)(4,4)
    \pspolygon[fillcolor=#1](0,0)(1,0)(1,4)(0,4)
  \end{pspicture}
}

\newcommand{\ihorizontal}[1][lightgray]{
 \begin{pspicture}(0,0)(4,4)
    \pspolygon[fillcolor=#1](0,0)(4,0)(4,1)(0,1)
  \end{pspicture}
}

\newcommand{\teeup}[1][lightgray]{
  \begin{pspicture}(0,0)(4,4)
    \pspolygon[fillcolor=#1](0,0)(3,0)(3,1)(2,1)(2,2)(1,2)(1,1)(0,1)
  \end{pspicture}
}

\newcommand{\teedown}[1][lightgray]{
  \begin{pspicture}(0,0)(4,4)
    \pspolygon[fillcolor=#1](0,2)(3,2)(3,1)(2,1)(2,0)(1,0)(1,1)(0,1)
  \end{pspicture}
}

\newcommand{\teeleft}[1][lightgray]{
 \begin{pspicture}(0,0)(4,4)
    \pspolygon[fillcolor=#1](2,0)(2,3)(1,3)(1,2)(0,2)(0,1)(1,1)(1,0)
  \end{pspicture}
}

\newcommand{\teeright}[1][lightgray]{
 \begin{pspicture}(0,0)(4,4)
    \pspolygon[fillcolor=#1](0,0)(0,3)(1,3)(1,2)(2,2)(2,1)(1,1)(1,0)
  \end{pspicture}
}

\def\LGl{\leftgunleft}
\def\LGr{\leftgunright}
\def\LGu{\leftgunup}
\def\LGd{\leftgundown}
\def\RGl{\rightgunleft}
\def\RGr{\rightgunright}
\def\RGu{\rightgundown}
\def\RGd{\rightgunup}
\def\Il{\ihorizontal}
\def\Ir{\ihorizontal}
\def\Iu{\ivertical}
\def\Id{\ivertical}
\def\Sq{\square}
\def\LSl{\leftsnakehorizontal}
\def\LSr{\leftsnakehorizontal}
\def\LSu{\leftsnakevertical}
\def\LSd{\leftsnakevertical}
\def\RSl{\rightsnakehorizontal}
\def\RSr{\rightsnakehorizontal}
\def\RSu{\rightsnakevertical}
\def\RSd{\rightsnakevertical}
\def\Tl{\teeleft}
\def\Tr{\teeright}
\def\Tu{\teeup}
\def\Td{\teedown}

\newcommand{\piece}[3]{
  \rput(#2,#3){#1}
}

\newcounter{height}
\newcounter{width}

\newenvironment{block}[2]{
  \setcounter{width}{#1}
  \addtocounter{width}{-2}
  \setcounter{height}{#2}
  \addtocounter{height}{-3}
  \begin{pspicture}(0,-1)(\value{width},\value{height})}{\end{pspicture}}

\newcommand{\emptycol}[2][\fillcolor]{
\begin{pspicture}(0,0)(4,4)
    \pspolygon[fillcolor=#1](0,0)(1,0)(1,#2)(0,#2)
  \end{pspicture}
}

\newcommand{\annotation}[1]{
  \begin{pspicture}(0,0)(4,4)
    #1
  \end{pspicture}
}

\def\tLG{\mathsf{LG}}
\def\tLS{\mathsf{LS}}
\def\tRS{\mathsf{RS}}
\def\tRG{\mathsf{RG}}
\def\tSq{\mathsf{Sq}}
\def\tT{\mathsf{T}}  
\def\tI{\mathsf{I}}  

\newcommand{\partition}{\textsc{3-Partition}\xspace}
\newcommand{\tetris}[1]{\textsc{Tetris}[#1]\xspace}

\newcommand{\crm}[1][max]{\textnormal{\textsf{#1-cleared-rows}}}
\newcommand{\tm}[1][max]{\textnormal{\textsf{#1-tetrises}}}
\newcommand{\hm}[1][min]{\textnormal{\textsf{#1-height-filled}}}
\newcommand{\ppm}[1][max]{\textnormal{\textsf{#1-placed-pieces}}}

\def\unprepped{unprepped\xspace}
\def\Unprepped{Unprepped\xspace}
\def\iprepped{($\tI$-UP)\xspace}
\newcommand{\lgprepped}[1]{($\tLG$-UP-$#1$)\xspace}
\newcommand{\ipreppedlg}[1]{($\tI$-UP-$\tLG$-$#1$)\xspace}
\newcommand{\lgpreppedlg}[2]{($\tLG$-UP-$\{#1,#2\}$)\xspace}
\def\underflat{underflat\xspace}
\def\Underflat{Underflat\xspace}
\newcommand{\lgunderflat}[1]{($\tLG$-UF-$#1$)\xspace}
\def\unapproachable{unapproachable\xspace}

\newcommand{\lgunapproachable}[1]{($\tLG$-UA-$#1$)\xspace}
\newcommand{\lglgunapproachable}[2]{($\tLG$-UA-$\{#1,#2\}$)\xspace}
\newcommand{\lgoverflatA}{($\tLG$-OF-1)\xspace}
\newcommand{\lgoverflatB}{($\tLG$-OF-2)\xspace}
\newcommand{\lgoverflatC}{($\tLG$-OF-3)\xspace}
\newcommand{\lgoverflatD}[1]{($\tLG$-OF-4-$#1$)\xspace}
\def\overflat{overflat\xspace}
\def\Overflat{Overflat\xspace}
\def\thappy{trigger-happy\xspace}
\def\Thappy{Trigger-Happy\xspace}
\newcommand{\lgthappy}[1]{($\tLG$-TH-$#1$)\xspace}
\newcommand{\lglgthappy}[2]{($\tLG$-TH-$\{#1,#2\}$)\xspace}
\def\splat{short-plateau\xspace}
\def\Splat{Short-Plateau\xspace}
\def\tplat{tall-plateau\xspace}
\def\Tplat{Tall-Plateau\xspace}
\newcommand{\lgtplat}[1]{($\tLG$-TP-$#1$)\xspace}

\def\balcony{balcony\xspace}
\def\OXX{011-floored notch\xspace}
\def\XXO{110-floored notch\xspace}
\def\XXX{111-floored notch\xspace}
\def\bOXX{balconied \OXX}
\def\bXXO{balconied \XXO}
\def\bXXX{balconied \XXX}
\def\brect{balconied unfilled sub-notch rectangle\xspace}
\def\bfive{balconied \bl{5} in non-notch rows\xspace}
\def\btwo{balconied \bl{2}}
\def\bone{balconied \bl{1} in a non-notch row\xspace}
\def\bmod{balconied flat-bottomed 2/3-ceiling discrepancy\xspace}
\def\gap{notch-spanning 2/3-ceiling discrepancy\xspace}
\def\bgap{balconied \gap{}\xspace}
\def\spurn{spurned notch\xspace}

\def\Spurnes{Spurned notches\xspace}
\def\veranda{veranda\xspace}
\newcommand{\bl}[1]{$#1$-\veranda{}\xspace}

\newcommand{\bver}[1][(\alpha\not\equiv_4 0)]{balconied \bl{#1} in
  non-notch rows\xspace} 
\def\bone{\bver[1]} 
\def\bfive{\bver[5]}

\def\reasonable{reasonable\xspace}
\def\Reasonable{Reasonable\xspace}

\def\pbd{T}
\def\resarea{R}

\begin{document}

\title{Tetris is Hard, Even to Approximate}

\author{Erik D.\ Demaine\thanks{Laboratory for Computer Science;
    Massachusetts Institute of Technology; 200 Technology Square;
    Cambridge, MA 02139, USA.  Email:
    \{\texttt{edemaine,srhohen,dln}\}\texttt{@theory.lcs.mit.edu}.}
  \and Susan Hohenberger$^*$ \and David Liben-Nowell$^*$}

\maketitle
\begin{abstract}
In the popular computer game of \emph{Tetris}, the player is given a
sequence of tetromino pieces and must pack them into a rectangular
gameboard initially occupied by a given configuration of filled
squares; any completely filled row of the gameboard is cleared and all
pieces above it drop by one row.  We prove that in the offline version
of Tetris, it is $\np$-complete to maximize the number of cleared
rows, maximize the number of tetrises (quadruples of rows
simultaneously filled and cleared), minimize the maximum height of an
occupied square, or maximize the number of pieces placed before the
game ends.  We furthermore show the extreme inapproximability of the
first and last of these objectives to within a factor of
$p^{1-\varepsilon}$, when given a sequence of $p$ pieces, and the
inapproximability of the third objective to within a factor of $2 -
\varepsilon$, for any $\varepsilon >0$.  Our results hold under
several variations on the rules of Tetris, including different models
of rotation, limitations on player agility, and restricted piece sets.
\end{abstract}

\newpage

\section{Introduction}
\mylabel{sec:intro}

Tetris \cite{tetris} is a popular computer game invented by
mathematician Alexey Pazhitnov in the mid-1980s.  Tetris is one of the
most widespread computer games ever created.  By 1988, just a few
years after its invention, it was already the best-selling game in the
United States and England.  Over 50 million copies have been sold
worldwide.  (Incidentally, Sheff \cite{sheff} gives a fascinating
account of the tangled legal debate over the profits, ownership, and
licensing of Tetris.)


In this paper, we embark on the study of the computational 
complexity of playing Tetris.  We consider the \emph{offline} 
version of Tetris, in which the sequence of pieces that will be 
dropped is specified in advance.  Our main result is a proof that 
optimally playing offline Tetris is $\np$-complete, and 
furthermore is highly inapproximable.

\paragraph{The game of Tetris.} 
Concretely, the game of Tetris is as follows.  (We give precise
definitions in Section \ref{sec:rules}, and discuss some variants 
on these definitions in Section \ref{sec:variants}.)  We are 
given an initial \emph{gameboard}, which is a rectangular grid 
with some gridsquares filled and some empty.  (In typical Tetris 
implementations, the gameboard is 20-by-10, and ``easy'' levels 
have an initially empty gameboard, while ``hard'' levels have 
non-empty initial gameboards, usually with the gridsquares below 
a certain row filled independently at random.)

\begin{figure}[t]
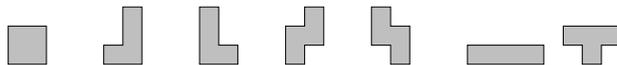

  \begin{center}
    \begin{tabular}{c}
      \Sq \LGu \RGu
      \LSd \RSd \Ir  \Td
    \end{tabular}
    \caption{The tetrominoes $\tSq$ (``square''), $\tLG$ (``left
      gun''), $\tRG$ (``right gun''), $\tLS$ (``left snake''), $\tRS$
      (``right snake''), $\tI$ (``I''), and $\tT$ (``T'').}
    \mylabel{fig:tetris-pieces}
  \end{center}
\end{figure}

A sequence of \emph{tetrominoes}---see Figure
\ref{fig:tetris-pieces}---is generated, typically probabilistically;
the next piece appears in the middle of the top row of the gameboard.
The piece falls, and as it falls the player can rotate the piece and
slide it horizontally.  It stops falling when it lands on a filled
gridsquare, though the player has a final opportunity to slide or
rotate it before it stops moving permanently.  If, when the piece
comes to rest, all gridsquares in an entire row $h$ of the game board
are filled, row $h$ is \emph{cleared}.  All rows above $h$ fall one
row lower; the top row of the gameboard is replaced by an entirely
unfilled row.  As soon as a piece is fixed in place, the next piece
appears at the top of the gameboard.  To assist the player, typically
a one-piece \emph{lookahead} is provided: when the $i$th piece begins
falling, the identity of the $(i+1)$st piece is revealed.

A player \emph{loses} when a new piece is blocked by filled
gridsquares from entirely entering the gameboard.  Normally, the
player can never win a Tetris game, since pieces continue to be
generated until the player loses.  Thus the player's objective is to
maximize his or her score, which increases as pieces are placed and as
rows are cleared.

\paragraph{Our results.}

We introduce the natural full-information (offline) version of Tetris:
we have a \emph{deterministic, finite} piece sequence, and the player
knows the identity and order of all pieces that will be presented.
(\emph{Games} magazine has, incidentally, posed several Tetris
puzzles based on the offline version of the game \cite{gamesmag}.)  We
study the offline version because its hardness captures much of the
difficulty of playing Tetris; intuitively, it is only easier to play
Tetris with complete knowledge of the future, so the difficulty of
playing the offline version suggests the difficulty of playing the
online version.  It also naturally generalizes the one-piece lookahead
of implemented versions of Tetris.

It is natural to generalize the Tetris gameboard to $m$-by-$n$, since
a relatively simple dynamic program solves the case of a constant-size
gameboard in time polynomial in the number of pieces.  Furthermore, in
an attempt to consider the inherent difficulty of the game---and not
any accidental difficulty due to the limited reaction time of the
player---we begin by allowing the player an arbitrary number of shifts
and rotations before the current piece drops in height.  (We will
restrict these moves to realistic levels later.)

In this paper, we prove that it is $\np$-complete to optimize any of
several natural objective functions for Tetris:
\begin{itemize*}
\item maximizing the number of rows cleared while playing the given
  piece sequence;
\item maximizing the number of pieces placed before a loss occurs;
\item maximizing the number of \emph{tetrises}---the simultaneous
  clearing of four rows;
\item minimizing the height of the highest filled gridsquare over the
  course of the sequence.
\end{itemize*}
We also prove the extreme inapproximability of the first two (and the
most natural) of these objective functions: given an initial gameboard
and a sequence of $p$ pieces, for any constant $\varepsilon > 0$, it
is $\np$-hard to approximate to within a factor of $p^{1-\varepsilon}$
the maximum number of pieces that can be placed without a loss, or the
maximum number of rows that can be cleared.  We also show that it is
$\np$-hard to approximate the minimum height of a filled gridsquare to
within a factor of $2 - \varepsilon$.

To prove these results, we first show that the cleared-row
maximization problem is $\np$-hard, and then give various extensions
of our reduction for the remaining objectives.  Our initial proof of
hardness proceeds by a reduction from \partition, in which we are
given a set $S$ of $3s$ integers and a bound $\pbd$, and asked to
partition $S$ into $s$ sets of three numbers each, so that the sum of
the numbers in each set is exactly $\pbd$.  Intuitively, we define an
initial gameboard that forces pieces to be placed into $s$ piles, and
give a sequence of pieces so that all of the pieces associated with
each integer must be placed into the same pile.  The player can clear
all rows of the gameboard if and only if all $s$ of these piles have
the same height.

A key difficulty in our reduction is that there are only a constant
number of piece types, so any interesting component of a desired
$\np$-hard problem instance must be encoded by a sequence of multiple
pieces.  The bulk of our proof of soundness is devoted to showing
that, despite the decoupled nature of a sequence of Tetris pieces, the
only way to possibly clear the entire gameboard is to place in a
single pile all pieces associated with each integer.

Our reduction is robust to a wide variety of modifications to the
rules of the game.  In particular, we show that our results also hold
in the following settings:
\begin{itemize*}
\item with restricted player agility---allowing only two
  rotation/translation moves before each piece drops in height;
\item with restricted piece sets---using $\{\tLG,\tLS,\tI,\tSq\}$ or
  $\{\tRG,\tRS,\tI,\tSq\}$, plus at least one other piece;
\item without any losses---i.e., with an infinitely tall gameboard;
\item under a wide variety of different rotation models---including
  the somewhat non-intuitive model that we have observed in real
  Tetris implementations.
\end{itemize*}

\paragraph{Related work:  Tetris.}  This paper is, to the best of our
knowledge, the first consideration of the complexity of playing
Tetris.  Kostreva and Hartman \cite{kostreva} consider Tetris from a
control-theoretic perspective, using dynamic programming to choose the
``optimal'' move---defined using a heuristic measure of the quality of
a configuration.  Other previous work has concentrated on the
possibility of a \emph{forced eventual loss} (or a \emph{perpetual
  loss-avoiding strategy}) in the online, infinite version of the
game.  In other words, under what circumstances can the player be
forced to lose, and how quickly?  Under what circumstances can the
player make the game last for infinitely many moves?

Brzustowski \cite{brzustowski} has shown a number of results on forced
eventual losses, both positive and negative.  He has given a strategy
for perpetually avoiding a loss in any (sufficiently large) even-width
gameboard using any one-piece pieceset, or any two of the pieces
$\{\tI, \tSq, \tRG, \tLG\}$.  (The strategies for the piecesets
$\{\tI, \tLG\}$, $\{\tI, \tRG\}$, and $\{\tRG,\tLG\}$ rely on the
one-piece lookahead.)  He has also given such a perpetual
loss-avoiding strategy for any (sufficiently large) odd-width
gameboard for the piecesets $\{\tI\}$, $\{\tRG\}$, $\{\tLG\}$, and
$\{\tT\}$.

On the negative side, Brzustowski has shown that perpetual
loss-avoidance is impossible for the piecesets $\{\tRS\}$, $\{\tLS\}$,
and $\{\tSq\}$ in odd-width boards.  More fundamentally, he has proven
that in any size board, if the machine can adversarially choose the
next piece (following the lookahead piece) in reaction to the player's
moves, then the machine can force an eventual loss using any pieceset
that contains $\{\tLS,\tRS\}$.  Burgiel \cite{burgiel} has
strengthened this last result for gameboards of width $2n$ for odd
$n$, showing that an alternating sequence of $\tLS$'s and $\tRS$'s
will eventually cause a loss, regardless of the way in which the
player places them.  This implies that, if pieces are chosen
independently at random with a non-zero probability mass assigned to
each of $\tLS$ and $\tRS$, then there is a forced eventual loss with
probability one for any such gameboards.

\paragraph{Related work:  other games and puzzles.}
A number of other popular one-player computer games have been shown to
be $\np$-hard, most notably Minesweeper---or, more precisely, the
Minesweeper ``consistency'' problem \cite{minesweeper}.  See the
survey of the first author \cite{demaine:survey} for a summary of
other games and puzzles that have been studied from the perspective of
computational complexity. These results form the emerging area of
\emph{algorithmic combinatorial game theory}, in which many new
results have been established in the past few years.

\section{Rules of Tetris}
\mylabel{sec:rules}

Here we rigorously define the game of Tetris, formalizing the
intuition of the previous section.  While tedious, we feel that such
rigor is necessary so that the many subtle nuances of Tetris become
transparent (following immediately from the rules).  For concreteness,
we have chosen to give very specific rules, but in fact the remainder
of this paper is robust to a variety of modifications to these rules;
in Section \ref{sec:variants}, we discuss some variations on these
rules for which our results still apply.

In particular, here we consider a particular set of rotation
rules---what we call \emph{instantaneous rotation}---that, in our
opinion, is the most intuitive and natural model for Tetris rotation.
However, we have observed that many Tetris implementations use a
different set of rotation rules; in Section \ref{sec:variants}, we
show that our results continue to hold under this observed rotation
model.

\paragraph{The gameboard.}

The \emph{gameboard} is a grid of $m$ rows and $n$ columns, indexed
from bottom-to-top and left-to-right.  The $\tup{i,j}$th
\emph{gridsquare} is either \emph{unfilled} (\emph{open},
\emph{unoccupied}) or \emph{filled} (\emph{occupied}).  In a legal
gameboard, no row is completely filled, and there are no completely
empty rows that lie below any filled gridsquare.  When determining the
legality of certain moves, we consider all gridsquares outside the
gameboard as always-occupied sentinels.

\paragraph{Game pieces.}

\begin{figure}[tp]
  \begin{center}
    \ifredraw
\begin{tabular}{ccccccc}
\begin{block}{4}{2}
  \piece{\Sq}00
  \piece{\annotation{\pscircle[fillstyle=solid,fillcolor=black](0,0){0.2}}}{1}{1}
\end{block}
\begin{block}{4}{2}
  \piece{\LGu}00
  \piece{\annotation{\pscircle[fillstyle=solid,fillcolor=black](0,0){0.2}}}{1.5}{1.5}
\end{block}
\begin{block}{4}{2}
  \piece{\RGu}00
  \piece{\annotation{\pscircle[fillstyle=solid,fillcolor=black](0,0){0.2}}}{0.5}{1.5}
\end{block}
\ifabstract \\\\ \fi
\begin{block}{4}{2}
  \piece{\LSu}00
  \piece{\annotation{\pscircle[fillstyle=solid,fillcolor=black](0,0){0.2}}}{1.5}{1.5}
\end{block}
\begin{block}{4}{2}
  \piece{\RSu}00
  \piece{\annotation{\pscircle[fillstyle=solid,fillcolor=black](0,0){0.2}}}{1.5}{1.5}
\end{block}
\begin{block}{4}{2}
  \piece{\Ir}00
  \piece{\annotation{\pscircle[fillstyle=solid,fillcolor=black](0,0){0.2}}}{1.5}{0.5}
\end{block}
\begin{block}{4}{2}
  \piece{\Td}00
  \piece{\annotation{\pscircle[fillstyle=solid,fillcolor=black](0,0){0.2}}}{1.5}{1.5}
\end{block}\\\\
\end{tabular}
\else
\includegraphics{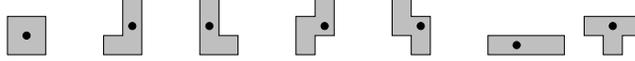}
\fi


  \caption{Piece \emph{centers}.}
  \label{fig:pieces-centers}
  \end{center}
\end{figure}

The seven Tetris pieces, shown in Figure \ref{fig:tetris-pieces}, are
exactly those connected rectilinear polygons that can be created by
assembling four $1$-by-$1$ gridsquares.  The \emph{center} of each
piece is shown in Figure \ref{fig:pieces-centers}.  A \emph{piece
  state} $P$ is a 4-tuple, consisting of:
\begin{enumerate*}
\item a \emph{piece type}---$\tSq, \tLG, \tRG, \tLS, \tRS, \tI,$ or
  $\tT$;
\item an \emph{orientation}---$0^\circ,90^\circ, 180^\circ,$ or
  $270^\circ$---the number of degrees clockwise from the piece's
  \emph{base orientation} (shown in Figure \ref{fig:tetris-pieces});
\item a \emph{position} of the piece's center on the gameboard, chosen
  from $\{1, \ldots, m\} \times \{1, \ldots n\}$.  The \emph{position}
  of a $\tSq$ is the location of the upper-left gridsquare of the
  $\tSq$, since its center falls on the boundary of four gridsquares
  rather than in the interior of one;
\item the value \emph{fixed} or \emph{unfixed}, indicating whether the
  piece can continue to move.
\end{enumerate*}
In an \emph{initial piece state}, the piece is in its base
orientation, and the initial position places the highest gridsquares
of the piece into row $m$, and the center into column $\lfloor n/2
\rfloor$, and the piece is unfixed.

\paragraph{Rotating pieces.}  A \emph{rotation model} is a
computable function $R: \tup{P,\theta,B} \mapsto P'$, where $P$ and
$P'$ are piece states, $\theta \in \{-90^\circ,90^\circ\}$ is the
rotation angle, and $B$ is a gameboard.  We impose the following
conditions on $R$:
\begin{enumerate*}
\item If $P = \tup{t,o,\tup{i,j},f}$ and the rotation is \emph{legal},
  then $P' = \tup{t,(o+\theta) \modulo 360^\circ,\tup{i',j'},f}$ for
  some $i'$ and $j'$.  If the rotation is \emph{illegal}, then $P' =
  P$.
  
\item In determining the legality of a rotation, $R$ only examines an
  $O(1)$-sized \emph{neighborhood} of the piece $P$---i.e., only
  gridsquares within a constant distance of the original position are
  relevant---and depends only on the neighborhood, and not its
  location in the gameboard.
  
\item If all the gridsquares in the neighborhood of $P$ are unfilled,
  then the rotation is legal.
  
\item If the rotation is legal, then $P'$ does not occupy any
  gridsquare already filled in $B$.  \label{cant-overlap}
\end{enumerate*}
We will impose additional constraints on $R$ in Section
\ref{sec:soundness} to restrict us to \emph{reasonable} rotation
models.  Intuitively, a reasonable rotation model simply allows for
the turning of a piece on the board without any unnatural powers of
translation, such as ``jumping'' to a distant point in the gameboard.
Our proof assumes an arbitrary reasonable rotation model; in Section
\ref{sec:variants}, we discuss a number of important reasonable
rotation models.

For now, we will consider the \emph{instantaneous rotation model}: fix
the piece center (shown in Figure \ref{fig:pieces-centers}), and
rotate the piece around that point.  The position after rotation is
unchanged---i.e., $\tup{i',j'} = \tup{i,j}$.  A rotation is illegal
only if it violates Condition \ref{cant-overlap}.

\paragraph{Playing the game.}

No moves are legal for a piece $P =
\tup{t,o,\tup{i,j},\mathit{fixed}}$.  The following moves are legal
for a piece $P = \tup{t,o,\tup{i,j},\mathit{unfixed}}$, with current
gameboard $B$:
\begin{enumerate*}
\item A \emph{clockwise rotation}.  The new piece state is
  $R(P,90^\circ,B)$.
  
\item A \emph{counterclockwise rotation}.  The new state is
  $R(P,-90^\circ,B)$.
  
\item A \emph{slide to the left}.  If the gridsquares to the left of
  the piece are open in $B$, we can translate $P$ to the left by one
  column.  The new piece state is
  $\tup{t,o,\tup{i,j-1},\mathit{unfixed}}$.
  
\item A \emph{slide to the right}, similarly.  The new piece state is
  $\tup{t,o,\tup{i,j+1},\mathit{unfixed}}$.
  
\item A \emph{drop} by one row, if all of the gridsquares beneath the
  piece are open in $B$.  The new piece state is
  $\tup{t,o,\tup{i-1,j},\mathit{unfixed}}$.
  
\item A \emph{fix}, if at least one gridsquare below the piece is
  filled in $B$.  The new piece state is
  $\tup{t,o,\tup{i,j},\mathit{fixed}}$.
\end{enumerate*}
A \emph{trajectory} $\sigma$ of a piece $P$ is a sequence of legal
moves starting from an initial state and ending with a fix move.  The
result of a trajectory for a piece $P$ on gameboard $B$ is a new
gameboard $B'$, defined as follows:
\begin{enumerate*}
\item The new gameboard $B'$ is initially $B$ with the gridsquares of
  $P$ filled.
  
\item If the piece is fixed so that, for some row $r$, every
  gridsquare in row $r$ of $B'$ is full, then row $r$ is
  \emph{cleared}.  For each $r' \ge r$, replace row $r'$ of $B'$ by
  row $r'+1$ of $B'$.  Row $m$ of $B'$ is an empty row.  Multiple rows
  may be cleared by the fixing of a single piece.
  
\item If the next piece's initial state is blocked in $B'$, the game
  ends and the player \emph{loses}.
\end{enumerate*}
For a \emph{game} $\tup{B_0,P_1,\ldots,P_p}$, a \emph{trajectory
  sequence} $\Sigma$ is a sequence $B_0, \sigma_1, B_1, \ldots,
\sigma_p, B_p$ so that, for each $i$, the trajectory $\sigma_i$ for
piece $P_i$ on gameboard $B_{i-1}$ results in gameboard $B_i$.
However, if there is a losing move $\sigma_q$ for some $q\le p$ then
the sequence $\Sigma$ terminates at $B_q$ instead of $B_p$.

\paragraph{The Tetris problem.} 
We will consider a variety of different objectives for Tetris (e.g.,
maximizing the number of cleared rows, maximizing the number of pieces
placed without a loss, etc.) For the decision version of a particular
objective $\Phi$, the Tetris problem $\tetris{\Phi}$ is formally as
follows:
\begin{description*}
\item[Given:] A Tetris game $\mathcal{G} = \tup{B,P_1, P_2, \ldots,
    P_p}$.
  
\item[Output:] Does there exist a trajectory sequence $\Sigma$ so that
  $\Phi(\mathcal{G},\Sigma)$ holds?
\end{description*}
We say that an objective function $\Phi$ is \emph{acyclic} when, for
all games $\mathcal{G}$, if there is a trajectory sequence $\Sigma$ so
that $\Phi(\mathcal{G},\Sigma)$ holds, then there is a trajectory
sequence $\Sigma'$ so that $\Phi(\mathcal{G},\Sigma')$ holds and there
are no repeated piece states in $\Sigma'$.  Most interesting Tetris
objective functions are acyclic; in fact, many depend only on the
final placement of each piece.

An objective function $\Phi$ is \emph{checkable} when, given a game
$\mathcal{G}$ and a trajectory sequence $\Sigma$, we can compute the
truth value of $\Phi(\mathcal{G},\Sigma)$ in time
$\poly(|\mathcal{G}|,|\Sigma|)$.

\begin{theorem}  \mylabel{thm:tetris-in-np}
  For any checkable acyclic objective $\Phi$ we have $\tetris{\Phi}
  \in \np$.
\end{theorem}
\begin{proof}
  We are given a Tetris game $\tup{B,P_1,\ldots,P_p}$.  Here is an
  $\np$ algorithm for $\tetris{\Phi}$:
  
  Guess an acyclic trajectory sequence $\Sigma$, and confirm that
  $\Sigma$ is a legal, acyclic trajectory in time $\poly(|\Sigma|)$.
  Confirming that all rotations in $\Sigma$ are legal depends on the
  computability of the rotation function, and the fact that legality
  can only depend on the constant-sized neighborhood of the piece.
  
  Since $\Sigma$ is acyclic, each of its $p$ trajectories can only
  contain at most $4 \cdot |B|+1$ states---unfixed once in each
  position and each orientation, and one final fixed state.  Thus
  $|\Sigma| = \poly(\mathcal{G})$.  Since $\Phi$ is checkable, we can
  then in time $\poly(|\mathcal{G}|,|\Sigma|) = \poly(|\mathcal{G}|)$
  verify that $\Phi(\mathcal{G},\Sigma)$ holds, and since $\Phi$ is
  acyclic, guessing an acyclic trajectory sequence $\Sigma$ suffices.
\end{proof}
The $\Phi(\mathcal{G},\Sigma)$ that we will initially concern
ourselves with is $\crm[k](\mathcal{G},\Sigma)$: in the game
$\mathcal{G}$, does $\Sigma$ clear at least $k$ rows without incurring
a loss?  In Section \ref{sec:variants}, we will consider a variety of
other objective functions.

\begin{lemma} \mylabel{lemma:max-clear-easy}
  The objective $\crm[k]$ is checkable and acyclic.
\end{lemma}
\begin{proof}
  The objective is acyclic because it only depends on the fixed piece
  state at the end of each trajectory, so the path in the trajectory
  is irrelevant; it is checkable since it results from a simple scan
  of the status of the gameboard after each trajectory in $\Sigma$.
\end{proof}

\section{The Reduction}
\mylabel{sec:reduction}

In this section, we define a mapping from instances of \partition
\cite[p.\ 224]{garey:1979} to instances of \tetris{\crm[k]}.  Recall
the \partition problem:
\begin{description*}
\item[Given:] A sequence $a_1, \ldots, a_{3s}$ of non-negative
  integers and a non-negative integer $\pbd$, so that $\pbd/4 < a_i <
  \pbd/2$ for all $1 \le i \le 3s$ and so that $\sum_{i=1}^{3s} a_i =
  s\pbd$.
  
\item[Output:] Can $\{a_1, \ldots, a_{3s}\}$ be partitioned into $s$
  disjoint subsets $A_1, \ldots, A_s$ so that, for all $1 \le j \le
  s$, we have $\sum_{a_i \in A_j} a_i= \pbd$?
\end{description*}
\begin{figure}[tb]
  \begin{center}
    \input{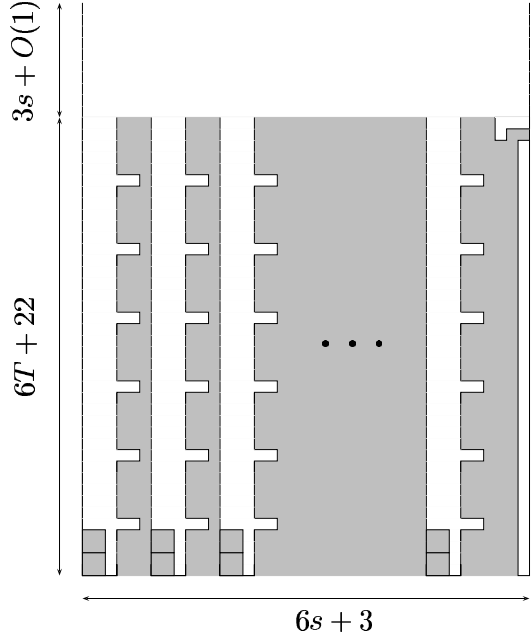}
    \caption{The initial gameboard for a Tetris game mapped from an instance
      of \partition.}  \mylabel{fig:reduction-initial}
  \end{center}
\end{figure}
We limit our attention to \partition instances that obey the following
properties, for technical reasons that will later become apparent:
\begin{enumerate*}
\item For any set $S \subseteq \{a_1, \ldots, a_{3s}\}$, if $\sum_{a_i
    \in S} a_i = \pbd$ then $|S| = 3$.
\item $\pbd$ is even.
\item If $\sum_{a_i \in A_j} a_i \not= \pbd$ then $\left|\pbd - \sum_{a_i
      \in A_j} a_i \right| \ge 3s$.
\end{enumerate*}
\noindent
We can map an arbitrary \partition instance into one obeying these
properties by multiplying each $a_i$ and $\pbd$ by $4s$.  This mapping
does not affect whether or not the instance has a valid 3-partition.
Property (2) is obviously guaranteed; for property (3), note that,
before the multiplication, if $\sum_{a_i \in A_j} a_i \not= \pbd$ then
$\smash{|\pbd - \sum_{a_i \in A_j} a_i|} \ge 1$ since all values are
integral, and multiplying by $4s$ multiplies differences by $4s$ as
well.  For Property (1), note that we still have $\pbd/4 < a_i <
\pbd/2$, so it is still the case that if $\sum_{a_i \in S} a_i = \pbd$
then $|S| = 3$.

We choose to reduce from this problem because it is $\np$-hard to
solve \partition even if the inputs $a_i$ and $\pbd$ are provided in
unary:
\begin{theorem}[Garey and Johnson \cite{garey:1975}]
  \mylabel{thm:partition-hard} \partition is $\np$-complete in the
  strong sense.  $\qed$
\end{theorem}
Given a \partition instance $\mathcal{P} = \tup{a_1, \ldots, a_{3s},
  \pbd}$ obeying the three conditions above, we will produce a Tetris
game $\mathcal{G(P)}$ whose gameboard can be completely cleared
precisely if $\mathcal{P}$ is a ``yes'' instance of \partition.

The initial gameboard is shown in Figure \ref{fig:reduction-initial}.
Intuitively, there are $s$ \emph{buckets} corresponding to the sets
$A_1, \ldots, A_s$ for the \partition problem.  The piece sequence
will consist of a number of tetrominoes corresponding to each $a_i$,
chosen carefully so that all pieces corresponding to $a_i$ must be
placed into the same bucket.  There is a legal 3-partition for $\{a_1,
\ldots, a_{3s}\}$ exactly when the piles of pieces in each bucket have
the same height.  The last three columns of the gameboard form a
\emph{lock} which prevents any rows from being cleared until the end
of the piece sequence; if all buckets are filled exactly to the same
height, then the entire board can be cleared using the last portion of
our piece sequence.

Formally, our game $\mathcal{G}$ consists of the following:
\begin{description}
\item[Initial board:] Our gameboard will have $6\pbd + 22 + (3s+O(1))$
  rows and $6s+3$ columns.  Intuitively, the factor of six in the
  height is because each $a_i$ will be represented by $a_i+1$ blocks
  of six rows and six columns each; since the set $A_j =
  \{a_i,a_j,a_k\}$ sums to $\pbd$, this is $6(\pbd+3) = 6\pbd + 18$.
  In addition to these $6\pbd + 18$ rows, there are four rows at the
  bottom ensuring that the initial blocks are placed correctly.
%
 
  The top $3s+O(1)$ rows---the $O(1)$ is exactly the size of the
  neighborhood in the rotation model---are initially empty, and are
  there solely as a staging area in which to rotate and translate
  pieces before they fall into the bottom $6\pbd+22$ rows.  We will
  not mention them again in the construction (and, below, the
  \emph{highest} row is the ($6\pbd+22$)nd).  Our choice of $3s+O(1)$
  as the number of staging rows will be discussed in Section
  \ref{sec:variants}.
  
  The remainder of the initial board can be thought of in $s+1$
  logical pieces, the first $s$ of which are six columns wide and the
  last of which is three columns wide.  The first $s$ logical pieces
  are \emph{buckets}, arranged in the following six-column pattern:
  \begin{itemize}
  \item the first and second columns are empty except that the four
    lowest rows are full;
  \item the third column is completely empty;
  \item the fourth and fifth columns are full in each row $h
    \not\equiv 5 ~(\mod 6)$ and empty in each $h \equiv 5 ~(\mod 6)$;
  \item the sixth column is completely full;
  \end{itemize}
  We call a row $r \equiv 5 ~(\mod 6)$ a \emph{notch row}, and refer
  to the unfilled rows of columns 4 and 5 of row $r$ as a
  \emph{notch}.
  
  The last logical piece is a three-column \emph{lock}, and consists
  of the following:
  \begin{itemize}
  \item the first column is full except that the highest and
    second-highest rows are empty;
  \item the second column is full except that the topmost row is
    empty;
  \item the third column is empty except that the second-highest row
    is full.
  \end{itemize}

\item[Pieces:] The sequence of pieces for our game consists of a
  sequence of pieces for each $a_i$, followed by a number of
  additional pieces after all the $a_i$'s.  For each integer $a_1,
  \ldots, a_{3s}$, we have the following pieces:
  \begin{itemize}
  \item the \emph{initiator}, which consists of the sequence
    $\tup{\tI, \tLG, \tSq}$;
  \item the \emph{filler}, which consists of the sequence $\tup{\tLG,
      \tLS, \tLG, \tLG, \tSq}$ repeated $a_i$ times;
  \item the \emph{terminator}, which consists of the sequence
    $\tup{\tSq, \tSq}$.
  \end{itemize}
  These pieces are given for $a_1$, $a_2$, etc., in exactly this
  order.  After the pieces corresponding to $a_{3s}$, we have the
  following pieces:
  \begin{itemize}
  \item $s$ successive $\tI$'s;
  \item one $\tRG$;
  \item $3\pbd/2+5$ successive $\tI$'s.  (Since we enforced that
    $\pbd$ is even, this is an integral number of pieces.)
  \end{itemize}
\end{description}

\begin{theorem}  \mylabel{lemma:reduction-polynomial}
  The game $\mathcal{G(P)}$ is polynomial in the size of
  $\mathcal{P}$.
\end{theorem}
\begin{proof}
  The gameboard has size $6\pbd+22+3s+O(1)$ by $6s+3$, and the total
  number of pieces is
  \[ \sum_{i=1}^{3s} \left[3 + 5a_i + 2\right] + s + 1 +
  \left(\frac{3\pbd}{2} + 5\right) ~~=~~ 16s + 5s\pbd +
  \frac{3\pbd}{2} + 6.
  \]
  The $a_i$'s and $\pbd$ are represented in unary, so the size of the
  game is polynomial.
\end{proof}

\section{Completeness}
\mylabel{sec:completeness}

\begin{figure}[t!]
  \begin{center}
    \ifredraw
\psset{unit=0.04in}
\setlength{\tabcolsep}{-2pt}
\vspace*{-1.5ex}
\[
\underbrace{
  \begin{tabular}{ccccccccccccccccccccccc}
    \begin{block}{4}{30}
      \column{5}{30}
    \piece{\floor[lightgray]440}00
  \end{block}
  &
  \begin{block}{4}{30}
    \column{5}{30}
    \piece{\floor[lightgray]440}00
    \piece{\Iu[green]}20
  \end{block}
  &
  \begin{block}{4}{30}
    \column{5}{30}
    \piece{\floor[lightgray]444}00
    \piece{\LGr[green]}24
  \end{block}
  &
  \begin{block}{4}{30}
    \column{5}{30}
    \piece{\floor[lightgray]444}00
    \piece{\LGr[lightgray]}24
    \piece{\Sq[green]}04
  \end{block}
\end{tabular}
}_{\mbox{initiator}}
~~~
\underbrace{
  \begin{tabular}{ccccccccccccccccccccccc}
    \begin{block}{4}{30}
    \column{5}{30}
    \piece{\floor[lightgray]666}00
    \piece{\LGr[lightgray]}24
    \piece{\LGr[green]}06
  \end{block}
  &
  \begin{block}{4}{30}
    \column{5}{30}
    \piece{\floor[lightgray]877}00
    \piece{\LGr[lightgray]}24
    \piece{\LSl[green]}07
  \end{block}
  &
  \begin{block}{4}{30}
    \column{5}{30}
    \piece{\floor[lightgray]998}00
    \piece{\LGr[lightgray]}24
    \piece{\LGl[green]}08
  \end{block}
&
    \begin{block}{4}{30}
    \column{5}{30}
    \piece{\floor[lightgray]{10}{10}{10}}00
    \piece{\LGr[lightgray]}24
    \piece{\LGr[green]}2{10}
  \end{block}
   &
   \begin{block}{4}{30}
     \column{5}{30}
     \piece{\floor[lightgray]{10}{10}{10}}00
     \piece{\LGr[lightgray]}24
     \piece{\LGr[lightgray]}2{10}
     \piece{\Sq[green]}0{10}
   \end{block}
   &
   \begin{block}{4}{30}
     \column{5}{30}
     \piece{\floor[lightgray]{12}{12}{10}}00
     \piece{\LGr[lightgray]}24
     \piece{\LGr[lightgray]}2{10}
     \piece{\LGr[green]}0{12}
   \end{block}
&
  \begin{block}{4}{30}
    \column{5}{30}
     \piece{\floor[lightgray]{14}{13}{13}}00
     \piece{\LGr[lightgray]}24
     \piece{\LGr[lightgray]}2{10}
     \piece{\LSl[green]}0{13}
  \end{block}
  &
  \begin{block}{4}{30}
    \column{5}{30}
     \piece{\floor[lightgray]{15}{15}{14}}00
     \piece{\LGr[lightgray]}24
     \piece{\LGr[lightgray]}2{10}
    \piece{\LGl[green]}0{14}
  \end{block}
  &
  \begin{block}{4}{30}
    \column{5}{30}
     \piece{\floor[lightgray]{16}{16}{16}}00
     \piece{\LGr[lightgray]}24
     \piece{\LGr[lightgray]}2{10}
    \piece{\annotation{\vdots}}{1.5}{17}
  \end{block}
  &
  \begin{block}{4}{30}
    \column{5}{30}
     \piece{\floor[lightgray]{16}{16}{16}}00
     \piece{\LGr[lightgray]}24
     \piece{\LGr[lightgray]}2{10}
     \piece{\annotation{\vdots}}{1.5}{17}
     \piece{\LGl[lightgray]}0{20}
  \end{block}
&
\begin{block}{4}{30}
    \column{5}{30}
     \piece{\floor[lightgray]{16}{16}{16}}00
     \piece{\LGr[lightgray]}24
     \piece{\LGr[lightgray]}2{10}
     \piece{\annotation{\vdots}}{1.5}{17}
     \piece{\LGl[lightgray]}0{20}
     \piece{\LGr[green]}2{22}
  \end{block}
&
\begin{block}{4}{30}
    \column{5}{30}
     \piece{\floor[lightgray]{16}{16}{16}}00
     \piece{\LGr[lightgray]}24
     \piece{\LGr[lightgray]}2{10}
     \piece{\annotation{\vdots}}{1.5}{17}
     \piece{\LGl[lightgray]}0{20}
     \piece{\LGr[lightgray]}2{22}
     \piece{\Sq[green]}0{22}
  \end{block}
\end{tabular}
}_{\mbox{filler, filler, \ldots, filler}}
~~~
\underbrace{
\begin{tabular}{ccccccccccccccccccccccc}
\begin{block}{4}{30}
    \column{5}{30}
     \piece{\floor[lightgray]{16}{16}{16}}00
     \piece{\LGr[lightgray]}24
     \piece{\LGr[lightgray]}2{10}
     \piece{\annotation{\vdots}}{1.5}{17}
     \piece{\LGl[lightgray]}0{20}
     \piece{\floor[lightgray]222}0{22}
     \piece{\LGr[lightgray]}2{22}
     \piece{\Sq[green]}0{24}
  \end{block}
&
\begin{block}{4}{30}
    \column{5}{30}
     \piece{\floor[lightgray]{16}{16}{16}}00
     \piece{\LGr[lightgray]}24
     \piece{\LGr[lightgray]}2{10}
     \piece{\annotation{\vdots}}{1.5}{17}
     \piece{\LGl[lightgray]}0{20}
     \piece{\floor[lightgray]442}0{22}
     \piece{\LGr[lightgray]}2{22}
     \piece{\Sq[green]}0{26}
  \end{block}
&
\begin{block}{4}{30}
    \column{5}{30}
     \piece{\floor[lightgray]{16}{16}{16}}00
     \piece{\LGr[lightgray]}24
     \piece{\LGr[lightgray]}2{10}
     \piece{\annotation{\vdots}}{1.5}{17}
     \piece{\LGl[lightgray]}0{20}
     \piece{\floor[lightgray]662}0{22}
     \piece{\LGr[lightgray]}2{22}
  \end{block}
\end{tabular}
}_{\mbox{terminator}}
\]
\else
\includegraphics{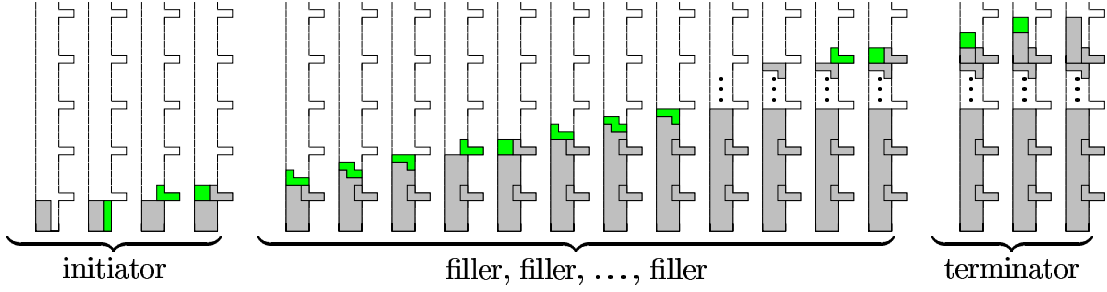}
\fi

    \caption{A valid sequence of moves within a bucket.}
    \mylabel{fig:valid-moves}
  \end{center}
\end{figure}

\begin{figure}[t!]
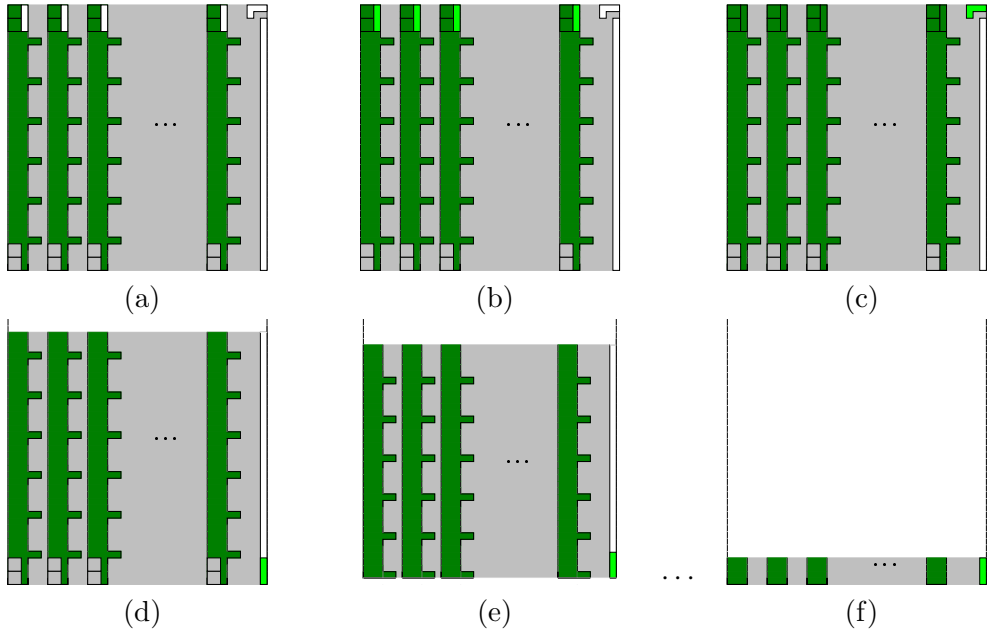

\begin{center}
  \psset{unit=0.035in}
  \begin{tabular}{ccccc}
    \input{figures/valid/filled1.tex}
    & ~~ & 
    \input{figures/valid/filled2.tex}
    & ~~ &
    \input{figures/valid/filled3.tex}
    \\
    (a) & & (b) & & (c) 
    \\
    \input{figures/valid/filled4.tex}
    &&
    \input{figures/valid/filled5.tex}
    & $\cdots$ &
    \input{figures/valid/filled6.tex}
    \\
    (d) & & (e) & & (f)
 \end{tabular}
 \caption{Finishing a valid move sequence.}
 \mylabel{fig:valid-finish}
\end{center}
\end{figure}

Here we show the easier direction of the correctness of our reduction:
for a ``yes'' instance of \partition, we can clear the entire
gameboard.

\begin{theorem}[Completeness]  \mylabel{thm:completeness}
  For any ``yes'' instance $\mathcal{P}$ of \partition, there is a
  trajectory sequence $\Sigma$ that clears the entire gameboard of
  $\mathcal{G(P)}$ without triggering a loss.
\end{theorem}
\begin{proof}
  Since $\mathcal{P}$ is a ``yes'' instance, there is a partitioning
  of the $a_i$'s into sets $A_1, \ldots, A_s$ so that $\sum_{a_i \in
    A_j} a_i = \pbd$.  We have ensured that $|A_j| = 3$ for all $j$.
  Place all pieces associated with set $A_j = \{x, y, z\}$ into the
  $j$th bucket of the gameboard, as illustrated in Figure
  \ref{fig:valid-moves}.
  
  Figure \ref{fig:valid-finish}(a) shows the configuration after all
  of the pieces associated with $a_1, \ldots, a_{3s}$ have been
  placed, as follows.  After all pieces associated with the number $x$
  have been placed into bucket $j$, the first $4+6x+2$ rows of bucket
  $j$ are full, and the left-hand two columns of bucket $j$ are filled
  four rows above that.  The pieces associated with the number $y$
  fill the next $4+6y+2$ rows, and those with $z$ the next $4+6z+2$
  rows, leaving again two columns with four additional rows filled.
  So the total number of rows filled after the ``numbers'' is $18 +
  6(x+y+z) = 18 + 6\pbd$.  Doing this for each bucket $j$ yields a
  configuration in which all buckets are filled up to row $18+6\pbd$,
  with columns one and two filled up to row $22+6\pbd$.
  
  We next get $s$ successive $\tI$'s in the sequence.  We produce the
  configuration in Figure \ref{fig:valid-finish}(b) by dropping one of
  the $\tI$'s into the third column of each bucket, to fill rows
  $19+6\pbd$ through $22+6\pbd$.  Now the configuration has the first
  $22+6\pbd$ rows filled in all of the buckets, and the lock is
  untouched.
  
  Next we get an $\tRG$.  Drop it into the slot in the lock, yielding
  the configuration of Figure \ref{fig:valid-finish}(c); the first two
  rows are then cleared.  In the resulting configuration, the first
  $20+6\pbd$ rows of the first $6s+2$ columns are full, and the last
  column is completely empty.
  
  Figure \ref{fig:valid-finish}(d--f) shows the final stage of the
  sequence, as the $3\pbd/2 + 5$ successive $\tI$'s arrive.  Drop each
  into the last column of the lock.  Each of the $\tI$'s clears four
  rows; in total, this clears $4\cdot(3\pbd/2 + 5) = 6\pbd+20$ rows,
  clearing the entire the gameboard.  The first, second, and last of
  these $\tI$'s are illustrated in Figure \ref{fig:valid-finish}(d--f).
\end{proof}

\section{Soundness}
\mylabel{sec:soundness}

Call \emph{valid} any trajectory sequence that clears $6\pbd+22$ rows
in $\mathcal{G(P)}$.  We will refer to a move or trajectory as
\emph{valid} if it can appear in a valid trajectory sequence.  In this
section, we show that the existence of a valid strategy for the Tetris
game $\mathcal{G(P)}$ implies that $\mathcal{P}$ is a ``yes'' instance
of \partition.

We will often omit reference to $\mathcal{G(P)}$, and refer to its 
parts simply as the \emph{gameboard} and the \emph{piece 
sequence}.

\subsection{Basic Counting}

The soundness of our reduction is fundamentally based upon the
observation that, in order to win $\mathcal{G(P)}$, we must fruitfully
use every gridsquare in the piece sequence.  There are exactly as many
unfilled gridsquares in the bottom $6\pbd+22$ rows of the gameboard as
there are gridsquares in the piece sequence; thus, in order to clear
the gameboard, we can never place any piece so that it extends beyond
the $(6\pbd+22)$nd row.

\begin{fact}  \mylabel{fact:unfilled-count}
  The gameboard initially has $64s + 20s\pbd + 6\pbd + 24$ unfilled
  gridsquares in the bottom $6\pbd+22$ rows.  Of these, there are $64s
  + 20s\pbd$ unfilled gridsquares in the buckets, and $6\pbd+24$
  unfilled gridsquares in the lock.
\end{fact}
\begin{proof}
  In each bucket, there are exactly $6\pbd+18$, $6\pbd+18$,
  $6\pbd+22$, and $0$ unfilled gridsquares in the first, second,
  third, and sixth columns, respectively.  The fourth and fifth
  columns are full in each row $h \not\equiv 5 ~(\mod 6)$ and empty in
  each $h \equiv 5 ~(\mod 6)$, so the first $6\pbd+18$ rows are
  exactly one-sixth unfilled; the four highest rows are filled since
  $6\pbd+23 \equiv 5 ~(\mod 6)$.  Thus there are exactly $\pbd+3$
  unfilled gridsquares in the fourth and fifth columns, and, in total,
  $20\pbd+64$ unfilled gridsquares per bucket.  Therefore, buckets
  account for exactly $64s + 20s\pbd$ unfilled gridsquares.
  
  In the lock, the first column has exactly two unfilled gridsquares,
  the second column has exactly one unfilled gridsquare, and the third
  column has exactly $6\pbd+21$ unfilled gridsquares.  Thus there are
  $2 + 1 + 6\pbd+21 = 6\pbd+24$ total in the lock.
  
  Overall, then we have $64s + 20s\pbd + 6\pbd + 24$ unfilled gridsquares.
\end{proof}

\begin{fact}  \mylabel{fact:filled-count}
  The total number of gridsquares in the piece sequence is exactly
  $64s + 20s\pbd + 6\pbd + 24$.  Of these, there are $64s + 20s\pbd$
  gridsquares in sequence of pieces up to (but not including) the
  $\tRG$, and $6\pbd+24$ gridsquares in the pieces starting at (and
  including) the $\tRG$.
\end{fact}
\begin{proof}
  As we calculated in the proof of Lemma
  \ref{lemma:reduction-polynomial}, there are $16s + 5s\pbd +
  \frac{3\pbd}{2} + 6$ total pieces in the sequence, of which the last
  $16s + 5s\pbd$ precede the $\tRG$.  Each covers exactly four
  gridsquares, so the total number of gridsquares in the piece
  sequence is $64s + 20s\pbd + 6\pbd + 24$, of which $64s + 20s\pbd$
  precede the $\tRG$.
\end{proof}

\begin{corollary} \mylabel{cor:area}
  Any move that places a filled gridsquare above row $6\pbd+22$ is
  invalid.
\end{corollary}
\begin{proof}
  The total number of filled gridsquares in the input sequence and
  initial gameboard combined is exactly the number of gridsquares in
  $6\pbd+22$ rows.  Thus to clear $6\pbd+22$ rows, every filled
  gridsquare must be in a cleared row.  In particular, we must clear
  the bottom $6\pbd+22$ rows (since each of these rows has at least
  one filled gridsquare initially), and cannot place a filled
  gridsquare above the ($6\pbd+22$)nd row.
\end{proof}

\subsection{Invalid Moves and Configurations}

In this section, we describe general types of configurations from
which the game can never be won.  In Section \ref{sec:play-right}, we
will use these results to show that the only possible valid strategy
is that of Section \ref{sec:completeness}.

\subsubsection{Security of the Lock}

\begin{lemma}  \mylabel{lemma:unlock-rg}
  In any valid strategy, none of the pieces $\{\tLG,\tI,\tLS,\tSq\}$
  can be the first piece placed in a lock column.
\end{lemma}
\begin{proof}
  It is easy to verify that none of $\{\tLG,\tI,\tLS,\tSq\}$ can be
  placed in such a way to fill even one of the unfilled gridsquares in
  the lock without also filling a gridsquare beyond the $(6\pbd+22)$nd
  row.  Therefore, by Corollary \ref{cor:area}, such a move is
  invalid.
\end{proof}

There are two important corollaries of this simple lemma:

\begin{corollary} \mylabel{lemma:no-solving} 
  In any valid strategy, no rows are cleared before the $\tRG$.
\end{corollary}
\begin{proof}
  No rows can be cleared until at least one piece enters the lock
  columns; by Lemma \ref{lemma:unlock-rg}, the first piece to do so
  must be the $\tRG$.
\end{proof}

\begin{corollary}  \mylabel{cor:all-into-bucket}
  In any valid strategy:
  \begin{CompactEnumerate}
  \item all gridsquares of all pieces preceding the $\tRG$ must all be
    placed into buckets, filling all empty bucket gridsquares.
  \item all gridsquares of all pieces starting with (and including)
    the $\tRG$ must be placed into the lock columns, filling all empty
    lock gridsquares.
  \end{CompactEnumerate}
\end{corollary}
\begin{proof}
  Immediate from Lemma \ref{lemma:unlock-rg}, Corollary
  \ref{cor:area}, and Facts \ref{fact:unfilled-count} and
  \ref{fact:filled-count}.
\end{proof}

In the remainder of this section, we will only consider move sequences
that obey these corollaries.

\subsubsection{Definition of Unfillable Buckets}
\mylabel{sec:unwinnables}

Call a bucket \emph{unfillable} if it cannot be filled completely
using arbitrarily many pieces from the set $\{\tLG, \tLS, \tSq,
\tI\}$. 

\begin{lemma} \mylabel{lemma:unfillable}
  In any valid strategy, no configuration with an unfillable bucket
  arises.
\end{lemma}
\begin{proof}
  By Corollary \ref{cor:all-into-bucket}, all gridsquares of pieces
  preceding $\tRG$ must go into buckets, and Facts
  \ref{fact:unfilled-count} and \ref{fact:filled-count} imply that
  there are exactly the same number of unfilled bucket gridsquares as
  pre-$\tRG$ gridsquares in the sequence.  If we do not completely
  fill each bucket, then at least one of these gridsquares will not go
  into a bucket, violating Corollary \ref{cor:all-into-bucket}.
  (Since no rows are cleared, an unfillable configuration can never be
  made fillable again, and therefore makes the trajectory sequence
  invalid.)
\end{proof}

Here we outline a collection of buckets which, if they arise in play,
prevent $\mathcal{G(P)}$ from being won.  See Figures
\ref{fig:unapproachable} and \ref{fig:unfillable-configs}.

\paragraph{Unapproachable Buckets.}

\begin{figure}[tbp]
  \begin{center}
    \begin{block}{5}{14}
      \column{6}{14}
      \piece{\floor[lightgray]344}00
    \end{block}
  \end{center}
  \caption{An \unapproachable bucket.}
  \label{fig:unapproachable}
\end{figure}

In Figure \ref{fig:unapproachable}, we show an
\emph{\unapproachable{}} bucket.  This bucket, which can arise during
play, is unfillable; in fact, even without a notion of gravity and
piece movements, it is impossible even to \emph{tile} such a bucket
completely using $\{\tI,\tLG,\tLS,\tSq\}$.

\begin{lemma}
  An \unapproachable bucket is unfillable.
\end{lemma}
\begin{proof}
  Suppose not, and consider a filling of a bucket that was
  \unapproachable.  Let row $n$ be the first notch row above the
  \unapproachable part of the bucket.
  
  The notch in row $n$ must have been filled by either a $\tLG$ or an
  $\tI$, either of which also fills the gridsquare in the third column
  of row $n$:
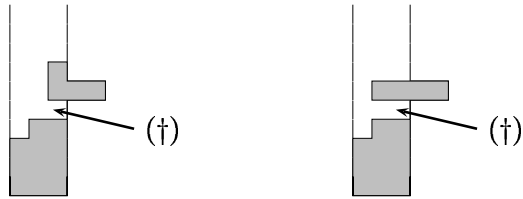
  \begin{center}
    \ifredraw
\begin{tabular}{ccc}
\begin{block}{5}{10}
  \column{6}{10}
  \piece{\floor[lightgray]344}00
  \piece{\LGr[lightgray]}25
  \piece{\annotation{($\dag$)}}{8}{3}
  \piece{\annotation{\psline[linewidth=1pt]{->}(6.5,3.5)(2.25,4.5)}}00
\end{block}
&  ~~~~~~~~~~~~~~~~~~ &
\begin{block}{5}{10}
  \column{6}{10}
  \piece{\floor[lightgray]344}00
  \piece{\Ir[lightgray]}15
  \piece{\annotation{($\dag$)}}{8}{3}
  \piece{\annotation{\psline[linewidth=1pt]{->}(6.5,3.5)(2.25,4.5)}}00
\end{block}

\end{tabular}
\else
\includegraphics{figures/unapproachable/marked.epsi.clean}
\fi


  \end{center}
  However, the gridsquare denoted by ($\dag$) can only be filled by a $\tLG$ or $\tLS$ (and an
  $\tLS$ only in the first case), and either one
  creates an region that cannot be filled by any Tetris piece:
  \begin{center}
    \ifredraw
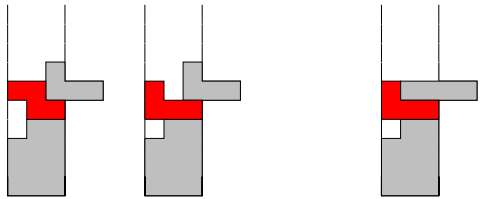
\begin{tabular}{llllllllll}
\begin{block}{5}{10}
  \column{6}{10}
  \piece{\floor[lightgray]344}00
  \piece{\LGr[lightgray]}25
  \piece{\LSl[red]}04
\end{block}
&
\begin{block}{5}{10}
  \column{6}{10}
  \piece{\floor[lightgray]344}00
  \piece{\LGr[lightgray]}25
  \piece{\LGr[red]}04
\end{block}
&
~~~~~~~
&
\begin{block}{5}{10}
  \column{6}{10}
  \piece{\floor[lightgray]344}00
  \piece{\Ir[lightgray]}15
  \piece{\LGr[red]}04
\end{block}
\end{tabular}
\else
\includegraphics{figures/unapproachable/hosed.epsi.clean}
\fi


  \end{center}
  Thus there was no such filling of the bucket, and \unapproachable
  buckets are unfillable.
\end{proof}

\begin{figure}[tb]
\begin{center}
  \ifredraw
\begin{tabular}{ccccccc}
  \begin{block}{8}{10}
    \column{5}{10}
    \piece{\floor[lightgray]404}00
    \piece{\Sq}14
  \end{block}
  &
  \begin{block}{8}{10}
    \column{3}{10}
    \piece{\floor[lightgray]330}00
  \end{block}
  &
  \begin{block}{8}{10}
    \column{6}{10}
    \piece{\annotation{\pspolygon[fillcolor=lightgray](1,0)(3,0)(3,1)(1,1)}}09
    \piece{\floor[lightgray]455}00
  \end{block}
  &
  \begin{block}{8}{10}
    \column{6}{10}
    \piece{\annotation{\pspolygon[fillcolor=lightgray](1,0)(3,0)(3,1)(1,1)}}09
    \piece{\floor[lightgray]554}00
  \end{block}
  &
  \begin{block}{8}{10}
    \column{6}{10}
    \piece{\annotation{\pspolygon[fillcolor=lightgray](1,0)(3,0)(3,1)(1,1)}}09
    \piece{\floor[lightgray]555}00
  \end{block}
  \\\\
  (a) & (b) & (c) & (d) & (e)
  \\\\
  \begin{block}{8}{10}
    \column{6}{10}
    \piece{\annotation{\pspolygon[fillcolor=lightgray](1,0)(3,0)(3,1)(1,1)}}09
    \piece{\floor[lightgray]010}00
    \piece{\annotation{\pspolygon[fillcolor=lightgray](1,0)(2,0)(2,1)(1,1)}}04
    \piece{\annotation{\psline{<-}(1.5,1)(1.5,2)}}00
    \piece{\annotation{\rput(1.5,2.5){$\alpha$}}}00
    \piece{\annotation{\psline{->}(1.5,3)(1.5,4)}}00
  \end{block}
&
  \begin{block}{8}{10}
    \column{6}{10}
    \piece{\annotation{\pspolygon[fillcolor=lightgray](1,0)(3,0)(3,1)(1,1)}}09
    \piece{\floor[lightgray]040}00
    \piece{\annotation{\pspolygon[fillcolor=lightgray](1,0)(2,0)(2,1)(1,1)}}06
    \piece{\annotation{\psline{<->}(1.5,4)(1.5,6)}}00
  \end{block}
  &
  \begin{block}{8}{10}
    \column{6}{10}
    \piece{\annotation{\pspolygon[fillcolor=lightgray](1,0)(3,0)(3,1)(1,1)}}09
    \piece{\floor[lightgray]023}00
    \piece{\annotation{\pspolygon[linestyle=dotted,fillstyle=none](1,0)(3,0)(3,-3)(1,-3)}}05
  \end{block}
  &
  \begin{block}{8}{10}
    \column{6}{10}
    \piece{\annotation{\pspolygon[fillcolor=lightgray](1,0)(3,0)(3,1)(1,1)}}09
    \piece{\annotation{\pspolygon[fillcolor=lightgray](1,0)(3,0)(3,-2)(2,-2)(2,-1)(1,-1)}}08
    \piece{\floor[lightgray]011}00
    \piece{\annotation{\psline{<->}(2.5,1)(2.5,6)}}00
    \piece{\annotation{\psline{<->}(1.5,1)(1.5,7)}}00
  \end{block}
  &
  \begin{block}{8}{10}
    \column{5}{10}
    \piece{\annotation{\pspolygon[fillcolor=lightgray](1,0)(3,0)(3,1)(1,1)}}09
    \piece{\annotation{\pspolygon[fillcolor=lightgray](1,0)(3,0)(3,-2)(2,-2)(2,-1)(1,-1)}}08
    \piece{\floor[lightgray]010}00
    \piece{\annotation{\psline{<->}(2.5,0)(2.5,6)}}00
    \piece{\annotation{\psline{<->}(1.5,1)(1.5,7)}}00
  \end{block}
  \\\\
  (f) & (g)  & (h) & (i) & (j)
\end{tabular}
\else
\includegraphics{figures/configs/bad.epsi.clean}
\fi


\end{center}
\caption{Some unfillable buckets: (a) a hole, (b) a \spurn, (c)
  a \bOXX, (d) a \bXXO, (e) a \bXXX, (f) a \bver, (g) a \btwo, (h) a
  \brect, (i) a \bmod, (j) a \bgap.
}
\mylabel{fig:unfillable-configs}
\end{figure}

\paragraph{Holes.}

A \emph{hole} is an unfilled gridsquare at height $h$ in some bucket
so that there is a contiguous series of filled gridsquares separating
that gridsquare from the empty rows above the buckets.  (No piece can
ever fill the hole.)

\paragraph{\Spurnes.}  

A \emph{\spurn{}} is a bucket in which, for some notch row $r$, (1)
the two gridsquares in the notch are not filled, and (2) the
gridsquare in the second column of row $r$ is filled.  (The filled
gridsquare in the second column prevents any piece from entering the
notch.)

\paragraph{Balconied buckets.}

A \emph{\balcony{}} is a bucket in which, for some row $r$, two of the
three gridsquares in row $r$ are filled.  Intuitively, the \balcony
prevents any piece other than $\tI$ from ``getting past'' row $r$,
which means that any unfilled gridsquares below $r$ must be filled
entirely using $\tI$'s.  Thus any bucket with a balcony lying over an
area that cannot be filled entirely by $\tI$'s is unfillable.

Call a \emph{\bl{\alpha}{}} an $\alpha$-long maximal run of
consecutive unfilled gridsquares in a column.

We claim that none of the following buckets can be filled using only
$\tI$'s if there is a balcony in row $r$:
\begin{enumerate}
\item a \emph{\bXXX{}}, \emph{\OXX{}}, or \emph{\XXO{}}.  
  
  For some row $h < r$ where $h$ contains an unfilled notch, the three
  gridsquares in row $h-1$ are all filled (for \XXX), or the second
  and third are filled (for \OXX), or the first and second are filled
  (\XXO).
  
  The first two are unfillable because an $\tI$ cannot rotate into the
  notch; the last is because filling the notch creates a hole, and
  finishing row $h-1$ creates a \XXX (or a hole, if filling the third
  column of row $h-1$ simultaneously fills the third column of row
  $h$).

\item a \emph{\bver{}} or \emph{\btwo{}}.
  
  For some row $h < r$, we have (1) $h$ is the highest unfilled
  gridsquare in a \bl{\alpha} which spans no unfilled notch rows
  (i.e., only non-notch rows and notch rows in which the notch is
  already filled), and $\alpha \not\equiv 0 ~(\mod 4)$, or (2) there
  is a \bl{2} in the $h$th and $(h+1)$st rows.
  
  No \bl{\alpha} for $\alpha \not\equiv 0 ~(\mod 4)$ can be filled by
  any number of vertical $\tI$'s, and a horizontal $\tI$ will only fit
  in a notch row.  For the \bver, there are no relevant notch rows;
  for a \bl{2} that spans a notch row, filling that notch can only
  leave a \bver[1].

\item a \emph{\brect{}}.
  
  There is an unfilled notch at height $h < r$, and, for column $i \in
  \{2,3\}$, and some $1 \le j \le 3$, we have the following in column
  $i$: the gridsquares at heights $h, h-1, \ldots, h-j$ are unfilled,
  and the gridsquare at height $h-j-1$ is filled.
  
  The notch cannot be filled without creating an \bver[\alpha], for
  $1\le \alpha \le 3$, and any vertically-placed $\tI$ in the second
  or third columns creates a \spurn or a hole.
  
\item a \emph{\bmod{}} or a \emph{\gap{}}.
  
  In a \bmod, for some $\alpha \ge 1$, some $1 \le \beta \le 3$, and
  some row $h < r$, we have the following: in both columns two and
  three, the gridsquares between rows $h$ and $h+\alpha$ inclusive are
  empty, and the gridsquares in rows $h-1$ are full.  In column two
  (respectively, column three), rows $h+\alpha+1, \ldots,
  h+\alpha+\beta$ are also empty, and the gridsquare in row
  $h+\alpha+\beta+1$ is full.  In column three (respectively, column
  two), the gridsquare in row $h+\alpha+1$ is full.  
  
  A \bgap is just like the above, with the following exceptions: (1)
  we eliminate the requirement that the second and third gridsquares
  of row $h-1$ be full, and (2) we add the requirement that there be
  an unfilled notch in rows $h, \ldots, h+\alpha$.

  \medskip
  
  For a \bmod, since $\beta \not\equiv 0 ~(\mod 4)$, the empty
  gridsquares in columns two and three cannot both be filled by
  vertically-oriented $\tI$'s, and any horizontally-oriented $\tI$'s
  placed into notches in these rows fill one gridsquare in both
  columns, and thus do not resolve this discrepancy.  For a \bgap, we
  must fill the notch with a horizontal $\tI$; once we have done so,
  the result is a \bmod.

\end{enumerate}
These arguments are formalized in Appendix \ref{app:unfillable}.

\subsection{\Reasonable Rotation Models}  

Which moves are legal, and therefore which instances have valid
trajectory sequences, depends on certain properties of the rotation
model.  So far our only mention of the details of our rotation model
has been in the intuition of Section \ref{sec:unwinnables}; here we
formalize the dependence on the details of the model.  Call a rotation
model \emph{\reasonable{}} if it satisfies the following four
conditions:
\begin{enumerate}
\item A piece cannot ``jump'' from one bucket to another, or into a
  disconnected region.
\item An $\tLG$ cannot enter a 
notch if the gridsquare in the second
  column of the notch row is filled.
\item For any \balcony at height $h$ with the $i$th column empty in
  the \balcony, the only gridsquares at height $h' \le h$ that can be
  filled by any piece other than an $\tI$ are those in the $i$th column
  in rows $h$ and $h-1$.
\item An $\tI$ cannot enter a notch if the gridsquares in the second
  and third columns of the row immediately beneath the notch row are
  filled.
\end{enumerate}
These four conditions are the only ones that we rely on, so this 
is the only place where we depend on the details of the rotation 
model. Thus, to show that our reduction holds for any particular 
rotation model, it suffices to prove that the model is 
\reasonable.

\begin{lemma}  \label{lemma:bad-configs}
  In any \reasonable rotation model, a configuration with a bucket
  containing any of the following is unfillable: (1) a hole, (2) a
  \spurn, (3) a \bXXO, \OXX, or \XXO, (4) a \brect, (5) a \bver, (6) a
  \btwo, (7) a \bmod, and (8) a \bgap.  $\qed$
\end{lemma}
\begin{lemma}  \label{lemma:instantaneous-reasonable}
  The instantaneous rotation model of Section \ref{sec:rules} is
  \reasonable.  $\qed$
\end{lemma}
We prove Lemmas \ref{lemma:bad-configs} and
\ref{lemma:instantaneous-reasonable} in Appendices
\ref{app:unfillable} and \ref{app:rotmodels}, respectively. The proof
of Lemma \ref{lemma:bad-configs} formalizes the intuition of the
previous section, explaining why we cannot completely fill each of
these buckets.  The proof of Lemma
\ref{lemma:instantaneous-reasonable} simply checks the conditions of
reasonability.

\subsection{The Only Way to Play ...}
\mylabel{sec:play-right}

Armed with the results from the previous section, we will show that
the only valid moves are those in which we play correctly according to
the sequence described in Section \ref{sec:completeness}.  We assume
(by Corollary \ref{lemma:unlock-rg}) that no pieces are placed into
the lock columns until the $\tRG$.

\begin{figure}[tbp]
  \begin{center}
    \ifredraw
\begin{tabular}{llllll}
  \begin{block}{7}{8}
    \column{5}{8}
    \piece{\floor[lightgray]044}00
  \end{block}
  &
  \begin{block}{7}{8}
    \column{5}{8}
    \piece{\floor[lightgray]440}00
  \end{block}
  &
  \begin{block}{7}{8}
    \column{5}{8}
    \piece{\floor[lightgray]666}00
    \piece{\LGr[lightgray]}24
  \end{block}
  &
  \begin{block}{7}{14}
    \column{5}{14}
    \piece{\floor[lightgray]099}00
    \piece{\LGr[lightgray]}24
  \end{block}
  &
  \begin{block}{5}{14}
    \column{5}{14}
    \piece{\floor[lightgray]998}00
    \piece{\LGr[lightgray]}24
  \end{block}  
\\\\
(a) & (b) & (c) & (d) & (e) 
\end{tabular}
\else
\includegraphics{figures/configs/good.epsi.clean}
\fi


    \caption{The possibly valid buckets:  (a) and (b) unprepped, (c)
      \overflat, (d) \tplat, (e) \thappy.}
    \mylabel{fig:good-configs}
  \end{center}
\end{figure}

In Figure \ref{fig:good-configs}, we give a collection of buckets
which can arise during play. We call \emph{unprepped} a bucket, 
as in (a) or (b), which is filled up to the base of a notch 
except the top four rows of exactly one of the first and third 
columns are unfilled.  A bucket is \emph{\overflat{}}, as in (c), 
if it is exactly filled up the row above the top of a notch.  A 
bucket is a \emph{\tplat{}} as in (d), if it has only the first 
column unfilled in the nine rows starting one row beneath a 
notch.  Finally, a bucket is \emph{\thappy{}} if it is unfilled 
in exactly the row beneath a notch row plus the third column of 
row below that.

Call a configuration \emph{unprepped} if each of its buckets is
unprepped.  Call a configuration \emph{one-$x$} (respectively,
\emph{one-$x$-one-$y$}) if exactly one of its buckets is of type $x$
(respectively, one each of types $x$ and $y$) and the rest are
unprepped.

\begin{lemma}[Initiator Soundness]  \mylabel{lemma:initiator-correctness}
  In an unprepped configuration, the only possibly valid strategy for
  $\tup{\tI,\tLG,\tSq}$ is to place all three pieces in some bucket to
  produce an overflat bucket, yielding a one-\overflat configuration.
  $\qed$
\end{lemma}

\begin{lemma}[Filler Soundness] \mylabel{lemma:filler-correctness}
For the sequence $\tup{\tLG,\tLS,\tLG,\tLG,\tSq}$:
  \begin{enumerate}
  \item In a one-\overflat configuration, the only possibly valid
    strategy is either (1) to place all pieces in the \overflat bucket
    to produce an \overflat bucket, yielding a one-\overflat
    configuration, or (2) to place $\tup{\tLG,\tLS}$ into the
    \overflat bucket and $\tup{\tLG,\tLG,\tSq}$ in an unprepped
    bucket, yielding a one-\tplat-one-\thappy configuration.
    
  \item In a one-\tplat-one-\thappy configuration, there is no valid
    strategy.  $\qed$
\end{enumerate}
\end{lemma}

\begin{lemma}[Terminator Soundness] \mylabel{lemma:terminator-correctness}
  For the sequence $\tup{\tSq,\tSq}$,
  \begin{enumerate}
  \item In a one-\overflat configuration, the only possibly valid
    strategy is to place both pieces in the \overflat bucket to
    produce an unprepped bucket, yielding an unprepped configuration.
    
  \item In a one-\tplat-one-\thappy configuration, there is no valid
    strategy. $\qed$
\end{enumerate}
\end{lemma}
The proofs of these propositions, found in Appendix
\ref{sec:soundness-proofs}, all follow the same (tedious) outline: we
exhaustively enumerate all possible moves that can be made in the
initial configuration using the given pieces, and show that, except
for the ``correct'' moves leading to the stated final
configuration(s), each move yields an unfillable configuration, in the
sense of Lemma \ref{lemma:bad-configs}.

\begin{lemma}  \mylabel{lemma:play-right}
  For any $r \ge 0$, in an unprepped configuration, the only possibly
  valid strategy for the sequence
  \[
  \tI, \tLG, \tSq, r \times \tup{\tLG, \tLS, \tLG, \tLG, \tSq}, \tSq,
  \tSq
  \]
  is to place all of the pieces into a single bucket, yielding an
  unprepped configuration.
\end{lemma}
\begin{proof}
  By Lemma \ref{lemma:initiator-correctness}, the only valid moves for
  the initial $\tup{\tI, \tLG, \tSq}$ place them all in some bucket,
  making it \overflat.  By induction on $r$ and by Lemma
  \ref{lemma:filler-correctness}, each successive $\tup{\tLG, \tLS,
    \tLG, \tLG, \tSq}$ must be placed into the same bucket, yielding
  either a one-\overflat or one-\tplat-one-\thappy configuration.
  Furthermore, if a one-\tplat-one-\thappy arises, then the next
  pieces (another $\tup{\tLG, \tLS, \tLG, \tLG, \tSq}$ or a
  $\tup{\tSq,\tSq}$) cannot be validly placed. By Lemma
  \ref{lemma:terminator-correctness}, the final pieces
  $\tup{\tSq,\tSq}$ must go into the same bucket, making it---and the
  configuration---unprepped.
\end{proof}

\begin{proposition}  \mylabel{prop:is-correct}
  Consider an unprepped configuration with exactly $4s$ total unfilled
  gridsquares in all buckets.  Then there is a winning strategy for
  $s$ successive $\tI$'s only if there are exactly four unfilled
  gridsquares per bucket, and this strategy is to place one in each
  bucket to fill it up to the $(6\pbd+22)$nd row.
\end{proposition}
\begin{proof}
  If there are fewer than four unfilled gridsquares in any bucket,
  then the initial unprepped configuration must have a filled
  gridsquare above the $(6\pbd+22)$nd row.  Thus there is a valid
  strategy only if there are exactly four unfilled gridsquares in each
  bucket.  
  
  In this configuration, any placement of the $\tI$'s other than one
  per bucket would fill a gridsquare beyond the $(6\pbd+22)$nd row.
\end{proof}

\begin{theorem}[Soundness]  \mylabel{thm:soundness}
  If there is a valid strategy for $\mathcal{G(P)}$, then
  $\mathcal{P}$ is a ``yes'' instance of \partition.
\end{theorem}
\begin{proof}
  If there is a valid strategy for $\mathcal{G(P)}$, then by
  Corollary \ref{cor:all-into-bucket}, there is a way of placing all
  pieces preceding the $\tRG$ into buckets to exactly fill all the
  empty bucket gridsquares.  By Lemma \ref{lemma:play-right}, for each
  number $a_i$, we must place all of the pieces associated with $a_i$
  into a single bucket, yielding an unprepped bucket.  Thus, after the
  pieces associated with $a_1, \ldots, a_{3s}$ have been placed, the
  result is an unprepped configuration with a total of $4s$ unfilled
  gridsquares in the buckets.  By Proposition \ref{prop:is-correct},
  we place exactly one $\tI$ in each of these buckets to fill it up to
  the ($6\pbd+22$)nd row.  Thus after the numbers, each bucket must have
  been completely filled up to the ($6\pbd+18$)th row, with two of the
  three columns filled up to the ($6\pbd+22$)nd row.
  
  Define $A_j$ to be the set of $a_i$'s so that all the pieces
  associated with $a_i$ are placed into bucket $j$.  The number of
  unfilled gridsquares in bucket j, by Fact \ref{fact:unfilled-count},
  is $20\pbd + 64$.  Thus the number of gridsquares filled by the pieces
  associated with $a_i \in A_j$ must be exactly $20\pbd +60$, and the
  last four gridsquares in the bucket must have be filled by the
  $\tI$.  The total number of gridsquares associated with the sequence
  corresponding to $a_i$ is $4 \cdot [3 + 5a_i + 2] = 20a_i + 20$, by
  Fact \ref{fact:filled-count}.  Thus the total number of gridsquares
  associated with numbers $a_i \in A_j$ is $\sum_{a_i \in A_j} (20a_i
  + 20) + 4$.  Thus $20\pbd + 60 = \sum_{a_i \in A_j} (20a_i + 20)$.
  Recall from Section \ref{sec:reduction} that we chose the $a_i$'s and
  $\pbd$ so that the following facts hold:
  \begin{enumerate}
  \item if $\sum_{a_i \in A_j} a_i = \pbd$ then $|A_j| = 3$;
  \item if $\sum_{a_i \in A_j} a_i \not= \pbd$ then $\left|\pbd - \sum_{a_i
        \in A_j} a_i \right| \ge 3s$.
  \end{enumerate}
  
  Then $20\pbd + 60 = \sum_{a_i \in A_j} (20a_i + 20)$ implies that
  $\sum_{a_i \in A_j} a_i = \pbd$ and thus that $|A_j| = 3$.
  
  This holds for all $j$, so the sets $A_1, \ldots, A_s$ are a valid
  3-partition.  Thus $\mathcal{P}$ is indeed a ``yes'' instance of
  \partition.
\end{proof}

\begin{theorem}
  $\tetris{\crm}$ is $\np$-complete.
\end{theorem}
\begin{proof}
  Immediate from Lemmas \ref{lemma:max-clear-easy} and
  \ref{lemma:instantaneous-reasonable} and Theorems
  \ref{thm:tetris-in-np}, \ref{thm:partition-hard},
  \ref{lemma:reduction-polynomial}, \ref{thm:completeness}, and
  \ref{thm:soundness}.
\end{proof}

\section{$\np$-Completeness for Other Objectives and Inapproximability}
\mylabel{sec:other-objectives}

In our original definition, we considered the maximization of the
number of rows cleared over the course of play.  This is a fundamental
component of a player's score, but in fact the score may be more
closely aligned with the number of \emph{tetrises}---that is, the
number of times during play that four rows are cleared simultaneously,
by the vertical placement of an $\tI$---that occur during play.
(Maximizing tetrises is a typical goal in real play.)

Another type of metric---considered by Brzustowski \cite{brzustowski}
and Burgiel \cite{burgiel}, for example---is that of \emph{survival}.
How many pieces can be placed before a loss must occur?

Define $\tm[k](\mathcal{G},\Sigma)$, $\hm[h](\mathcal{G},\Sigma)$, and
$\ppm[p](\mathcal{G},\Sigma)$, respectively, as the following
objectives: in the game $\mathcal{G}$, does $\Sigma$, respectively,
contain at least $k$ tetrises, never fill a gridsquare above height
$h$, and place at least $p$ pieces before losing?

In this section, we describe reductions extending that of Section
\ref{sec:reduction} to establish the hardness of optimizing each of
these objectives.  In Section \ref{sec:inapprox}, we give results on
the hardness of approximating the number of rows cleared and the
number of pieces survived.  Note that all of these objectives are
checkable and acyclic, and therefore the corresponding problems are in
$\np$.

\subsection{Maximizing Tetrises}

\begin{figure}[tp]
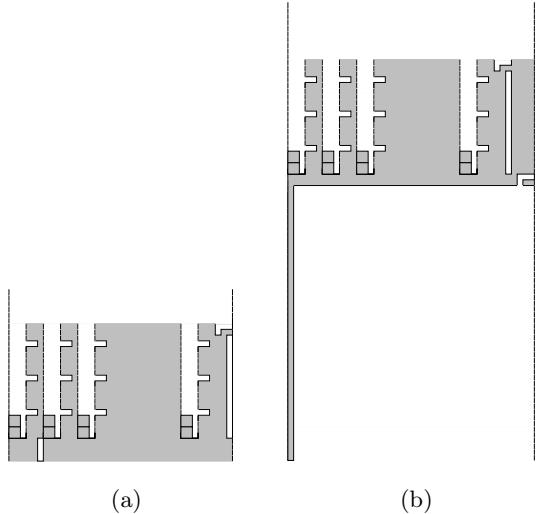

  \begin{center}
    \begin{tabular}{cc}
      \subfigure[]{\input{figures/initial/tetrises}
        \label{fig:initial-tetrises}} & 
      \subfigure[]{\input{figures/initial/survive}
        \label{fig:initial-survive}}
    \end{tabular}
    \caption{
      The initial gameboards showing hardness for (a) maximizing
      tetrises and (b) maximizing survival time.  The top of these
      gameboards is an exact reproduction of our previous reduction.}
  \end{center}
\end{figure}

We use a reduction very similar to that of Section
\ref{sec:reduction}, as shown in Figure \ref{fig:initial-tetrises}.
The new game is as follows:
\begin{itemize}
\item The top $6\pbd + 22 + 3s+O(1)$ rows of the gameboard are exactly
  the same as in our previous reduction.  We add four rows below
  these, entirely full except in the sixth column.
  
\item The piece sequence is exactly the same as in the previous
  reduction, with a single $\tI$ appended.
\end{itemize}
Our gameboard, shown in Figure \ref{fig:initial-tetrises}, is that of
Section \ref{sec:reduction}, augmented with four new bottom rows that
are full in all but the sixth column.  We append a single $\tI$ to our
previous piece sequence.

For a ``yes'' instance of \partition---namely, one in which we can
clear the top $6\pbd+22$ rows using the original part of the piece
sequence---$(6\pbd+20)/4 + 1$ tetrises are achievable.  (The last occurs
when the appended $\tI$ is placed into the new bottom rows.)

For a ``no'' instance, we cannot clear the top $6\pbd+22$ rows using 
the original pieces, and since the sixth column is full in all of 
the original rows, we cannot clear the bottom four rows with the 
last $\tI$ in the sequence.  Thus we clear at most $6\pbd+22$ rows.  
This implies that there were at most $(6\pbd+20)/4 < (6\pbd+20)/4 + 1$ 
tetrises.

Therefore we can achieve $(6\pbd+24)/4$ tetrises just in the case that
the top $6\pbd+22$ rows can be cleared by the first part of the sequence,
which occurs exactly when the $\partition$ instance is a ``yes''
instance.  Therefore it is $\np$-hard to maximize the number of
tetrises achieved.

\begin{theorem}
  $\tetris{\tm}$ is $\np$-complete.  $\qed$
\end{theorem}

\subsection{Maximizing Lifetime}

Our original reduction yields some initial intuition on the hardness
of maximizing lifetime.  In the ``yes'' case of \partition, there is a
trajectory sequence that fills no gridsquares above the $(6\pbd+22)$nd
row, while in the ``no'' case we must fill some gridsquare in the
$(6\pbd+23)$rd row:

\begin{theorem}  \label{thm:min-max-height-hard}
  $\tetris{\hm}$ is $\np$-complete.  $\qed$
\end{theorem}

However, this does not immediately imply the hardness of maximizing
the number of pieces that the player can place without losing, because
Theorem \ref{thm:min-max-height-hard} only applies for certain
heights---and, in particular, does not apply for height $m$ (the top
row of the gameboard), because our trajectory sequence from Section
\ref{sec:completeness} requires some operating space above the
$(6\pbd+22)$nd row for rotations and translations.

To show the hardness of maximizing survival time, we need to do some
more work.  We augment our previous reduction as shown in Figure
\ref{fig:initial-survive}.  Intuitively, we have created a large area
at the bottom of the gameboard that can admit a large number of
$\tSq$'s, but we place a lock so that $\tSq$'s can reach this area
only if the gameboard of the original reduction is cleared.
Crucially, the gameboard has odd width, so after a large number of
$\tSq$'s a loss must occur.

Our new gameboard consists of the following layers, for a value $r$ to
be determined below:
\begin{itemize}
\item The top $6\pbd + 22 + 3s+O(1)$ rows are exactly the same as in
  our previous reduction, with the addition of four completely-filled
  columns on the right-hand side of the gameboard.
\item The two next-highest rows form a second \emph{lock}, preventing
  access to the rows beneath.  This lock requires a $\tRG$ to be
  unlocked, just as in the lock columns at the top of the previous
  reduction.  The unfilled squares in the lock are in the four new
  columns.
\item The bottom $r$ rows form a \emph{reservoir}, and are empty in
  all columns but the first.
\end{itemize}
The gameboard has $6\pbd+3s+r+O(1)$ rows and $6s+7$ columns.  Let $A=
O(\pbd s^2)$ be the total area in and above the lock rows, and let
$\resarea = r(6s+6)$ be the total initially unfilled area in the
reservoir.

Our piece sequence is augmented as follows: first we have all pieces
of our original reduction, then a single $\tRG$, and finally $\resarea/4$
successive $\tSq$'s.

For the moment, choose $r = \poly(\pbd,s)$ so that $\resarea \ge 2A +1$.

In the ``yes'' case of \partition, the first part of the sequence can
be used to entirely clear the $6\pbd+22$ rows of the original
gameboard.  The $\tRG$ clears the second lock, and the $\resarea/4$
successive $\tSq$'s can then be packed into the reservoir to clear all
of the reservoir rows.

In the ``no'' case of \partition, the first part of the sequence
cannot entirely clear the top $6\pbd+22$ rows of the gameboard.  Since
all rows above the second lock are filled, this means that the $\tRG$
cannot unlock the reservoir, and crucially the $\tRG$ is the last
chance to do so---no number of $\tSq$'s can ever subsequently clear
the lock rows.  We claim that within $A/2+1$ $\tSq$'s (which cover
$2A+4$ gridsquares), a loss will occur.  Since there are an odd number
of columns, only rows that initially contain an odd number of filled
gridsquares can be cleared by a sequence of $\tSq$'s; thus each row
can be cleared at most once in the $\tSq$ sequence.  In order to
survive $2A+4$ gridsquares from a $\tSq$ sequence, at least one row
must be cleared more than once.  Therefore after $A/2 + 1 \le
\resarea/4$ successive $\tSq$'s, a loss must occur.

\begin{theorem}  \label{thm:survive-max-hard}
  $\tetris{\ppm}$ is $\np$-complete.  $\qed$
\end{theorem}

\subsection{Hardness of Approximation} 
\mylabel{sec:inapprox}

By modifying the reduction of Theorem \ref{thm:survive-max-hard}, we
can prove extreme inapproximability for either maximizing the number
of rows cleared or maximizing the number of pieces placed without a
loss, and a weaker inapproximability result for minimizing the maximum
height of a filled gridsquare.
\begin{theorem}
  Given a sequence of $p$ pieces, approximating $\tetris{\ppm}$ to
  within a factor of $p^{1 - \varepsilon}$ for any constant
  $\varepsilon > 0$ is $\np$-hard.
\end{theorem}
\begin{proof}
  Our construction is as in Figure \ref{fig:initial-survive}, but with
  a larger reservoir: choose $r$ so that the $r$-row reservoir's
  unfilled area $R$ is larger than $(2A)^{1/\varepsilon}$, where $A$
  is the total area of the gameboard excluding the reservoir rows.  As
  before, we append to the original piece sequence one $\tRG$ followed
  by exactly enough $\tSq$'s to completely fill the reservoir.  As in
  Theorem \ref{thm:survive-max-hard}, in the ``yes'' case of
  \partition, we can place all of the pieces in the given sequence
  (which in total cover an area of at least $\resarea$), while in the
  ``no'' case we can place pieces covering at most $2A$ area before a
  loss must occur.  Thus it is $\np$-hard to distinguish the case in
  which we can survive all $p$ pieces of the original sequence from
  the case in which we can survive at most $2A/4 < (R^\varepsilon)/4 <
  p^\varepsilon$ pieces.
\end{proof}
\begin{theorem}
  Given a sequence of $p$ pieces, approximating
  $\tetris{\crm}$ to within a factor of $p^{1 - \varepsilon}$ for any
  constant $\varepsilon > 0$ is $\np$-hard.
\end{theorem}
\begin{proof}
  Our construction is again as in Figure \ref{fig:initial-survive},
  with $r>a^{2/\varepsilon}$ rows in the reservoir, where there are
  $a$ total rows at or above the second lock.  As above, in the
  ``yes'' case of \partition, we can completely fill and clear the
  gameboard, and in the ``no'' case we can clear at most $a$ rows.
  Thus it is $\np$-hard to distinguish the case in which at least $r$
  rows can be cleared from the case in which at most $a <
  r^{\varepsilon/2}$ rows can be cleared.
  
  Note that the number of columns $c$ in our gameboard is fixed and
  independent of $r$, and that the number of pieces in the sequence is
  constrained by $r < p < (r+a)c$.  We also require that $r$ be large
  enough that $p < (r+a)c < r^{2/(2-\varepsilon)}$.  (Note that $r$,
  and thus our game, is still polynomial in the size of the \partition
  instance.)  Thus in the ``yes'' case we clear at least $r >
  p^{1-\varepsilon/2}$ rows, and in the ``no'' case we clear at most
  $a < r^{\varepsilon/2} < p^{\varepsilon/2}$.  Thus it is $\np$-hard
  to approximate the number of cleared rows to within a factor of
  $(p^{1-\varepsilon/2})/(p^{\varepsilon/2}) = p^{1-\varepsilon}$.
\end{proof}

\begin{theorem}
  Given a sequence of $p$ pieces, approximating $\tetris{\hm}$ to
  within a factor of $2 - \varepsilon$ for any constant $\varepsilon >
  0$ is $\np$-hard.
\end{theorem}
\begin{proof}
  Once again, our construction follows Figure
  \ref{fig:initial-survive}.  Let $F = O(Ts)$ be the total number of
  filled gridsquares at or above the rows of the lower lock, and let
  $P = O(Ts)$ be the total number of gridsquares in the piece sequence
  up to and including the second $\tRG$.  Choose $r = (F+P)/\delta$,
  where $\delta = \varepsilon/(3-\varepsilon)$.
  
  As before, in the ``yes'' instance of $\partition$, we can place the
  pieces of the given sequence so that the highest filled gridsquare
  is in the $(6\pbd+22)$nd row of the original gameboard, which is
  height $6\pbd+24 + r \le r + P + F \le r(1+\delta)$ in our
  gameboard.
  
  In the ``no'' case, all of the $\tSq$'s appended to the piece
  sequence will have to be placed at or above the second lock, since
  we can never break into the reservoir.  Note that, if rows are not
  cleared, we can never pack the appended $\tSq$'s into fewer than $r$
  rows.  Thus the height of the highest filled gridsquare is at least
  $2r-\kappa$, where $\kappa$ is the number of rows cleared during the
  sequence.  As before, we can only clear rows that have an odd number
  of filled gridsquares in them before the $\tSq$'s in the sequence.
  Since there are only $F+P$ gridsquares in total in this part of the
  sequence, obviously $\kappa \le F+P = r\delta$.  Thus there is a
  filled gridsquare at height at least $r(2 - \delta)$.
  
  Therefore it is $\np$-hard to approximate the minimum height of the
  maximum filled gridsquare to within a factor of $r(2 - \delta)/r(1 +
  \delta) = (2-\delta)/(1+\delta) = 2 - \varepsilon$.
\end{proof}

\section{Varying the Rules of Tetris}
\mylabel{sec:variants}

The completeness of our reduction does not depend on the full
set of allowable moves in Tetris, and the soundness does not depend on
all of its limitations.  Thus our results continue hold in some
modified settings.  

\subsection{Limitations on Player Agility}
We have phrased the rules of Tetris so that the player can, in
principle, make infinitely many translations or rotations before
moving the piece down to the next-highest row.  When actually playing
Tetris, there is a fixed amount of time (varying with the difficulty
level) in which to make manipulations at height $h$; one cannot slide
pieces arbitrarily far to the left or right before the piece falls.

Our reduction requires only that the player be able to make two
translations before the piece falls by another row (or is fixed), to
slide a $\tLG$ into a notch.  This is why we have chosen to have
$3s+O(1)$ empty rows at the top of the game board---this gives us
enough room to do any desired translation and rotation before the
piece reaches the top of a bucket while still only making two moves at
any given height.  (In the ``no'' case for loss-avoidance, this may
cause the game to end more quickly, but the ``yes'' case remains
feasible.)

Thus the problem remains $\np$-hard even when move sequences are
restricted to at most two moves between drops, for any of the
objectives.

\subsection{Piece Set}

Our reduction uses only the pieces $\{\tLG,\tLS,\tI,\tSq,\tRG\}$, so
Tetris remains $\np$-complete when the pieceset is thus restricted.
By taking the mirror image of our reduction, the hardness also holds
for the pieceset $\{\tRG,\tRS,\tI,\tSq,\tLG\}$.  In fact, the use of
the $\tRG$ in the lock was not required; we simply need some piece
that does not appear elsewhere in the piece sequence.  Thus Tetris
remains $\np$-hard for any piece set consisting of
$\{\tLG,\tLS,\tI,\tSq\}$ or $\{\tRG,\tRS,\tI,\tSq\}$, plus at least
one other piece.  (The reduction works exactly as before if the key
for the lock is $\tT$, pointing downward; for a snake, the bottom
three gridsquares of a vertically-oriented snake serves as the key,
and we observe that no other piece---except $\tT$, which is not in our
sequence---can be placed without filling at least \emph{two}
gridsquares above the $(6T+22)$nd row.)

\subsection{Losses}

We defined a loss as the fixing of a piece so that it does not fit 
entirely within the gameboard; i.e., the piece fills some 
gridsquare in the would-be $(m+1)$st row of the $m$-row 
gameboard.  Instead, we might define losses as occurring only 
\emph{after} rows have been cleared---that is, a piece can be 
fixed so that it extends into the would-be $(m+1)$st row, so long 
as this is not the case once all filled rows are cleared.  Since 
the completeness sequence (of Lemma \ref{thm:completeness}) never 
fills gridsquares anywhere near the top of the gameboard, our 
results hold for this definition as well.

In fact, for our original reduction, we do not depend on the 
definition of losses at all---the completeness trajectory 
sequence does not near the top of the gameboard, and the 
soundness proof does not rely on losses.  Obviously the objective 
of Theorem \ref{thm:survive-max-hard} is nonsensical without a 
definition of losses, but all other results still hold.
 
\subsection{Rotation Rules}

In Section \ref{sec:rules}, we specified concrete rules for the 
rotation of pieces around a particular fixed point in each 
piece.  In fact, our reduction applies for a wider variety of 
rotation models---any \reasonable rotation model, as defined in 
Section \ref{sec:soundness}.  In particular, there are two other 
\reasonable rotation rules of interest: the \emph{continuous} 
model and the \emph{Tetris} model that we have observed in 
practice.

In the \emph{continuous} (or \emph{Euclidean}) rotation model, the
rotation of a piece is around its center, as before, but we
furthermore require that all gridsquares that the piece \emph{passes
  through} must be unoccupied.

\begin{lemma}  \label{lemma:continuous-reasonable}
  The continuous rotation model is \reasonable.  $\qed$
\end{lemma}

\begin{figure}[t!]
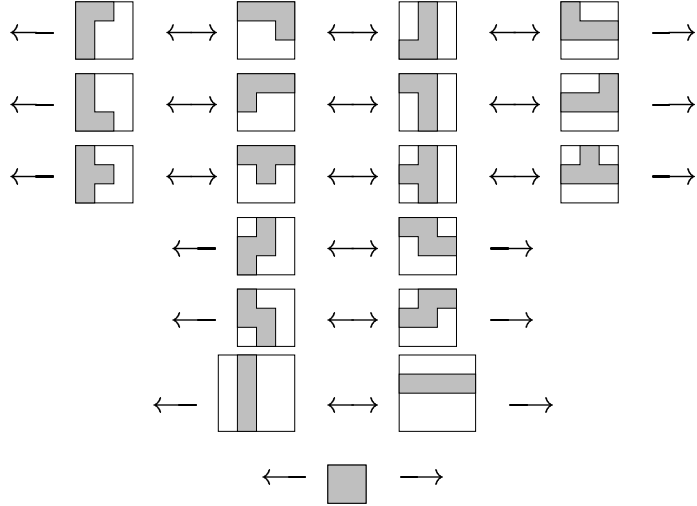

  \begin{center}
    \ifredraw
\begin{tabular}{cc}
\ifredraw
\begin{tabular}[t]{cccccccccc}
  $\longleftarrow$
  \begin{block}33
    \piece{\annotation{\pspolygon(0,0)(3,0)(3,3)(0,3)}}00
    \piece{\LGd}00
  \end{block}
  &
  $\longleftrightarrow$
  \begin{block}33
    \piece{\annotation{\pspolygon(0,0)(3,0)(3,3)(0,3)}}00
    \piece{\LGl}01
  \end{block}
  &
  $\longleftrightarrow$
  \begin{block}33
    \piece{\annotation{\pspolygon(0,0)(3,0)(3,3)(0,3)}}00
    \piece{\LGu}00
  \end{block}
  &
  $\longleftrightarrow$
  \begin{block}33
    \piece{\annotation{\pspolygon(0,0)(3,0)(3,3)(0,3)}}00
    \piece{\LGr}01
  \end{block}
  &
  $\longrightarrow$
  \\\\
\end{tabular}

\else
\includegraphics{figures/rotations/tetris/lg.epsi.clean}
\fi
 \\
\begin{tabular}[t]{cccccccccc}
$\longleftarrow$
      \begin{block}33
        \piece{\annotation{\pspolygon(0,0)(3,0)(3,3)(0,3)}}00
        \piece{\RGu}00
      \end{block}
&
$\longleftrightarrow$
      \begin{block}33
        \piece{\annotation{\pspolygon(0,0)(3,0)(3,3)(0,3)}}00
        \piece{\RGr}01
      \end{block}
&
$\longleftrightarrow$
      \begin{block}33
        \piece{\annotation{\pspolygon(0,0)(3,0)(3,3)(0,3)}}00
        \piece{\RGd}00
      \end{block}
&
$\longleftrightarrow$
      \begin{block}33
        \piece{\annotation{\pspolygon(0,0)(3,0)(3,3)(0,3)}}00
        \piece{\RGl}01
      \end{block}
&
$\longrightarrow$
\\\\
    \end{tabular}

 \\
\begin{tabular}[t]{cccccccccc}
$\longleftarrow$
\begin{block}33
  \piece{\annotation{\pspolygon(0,0)(3,0)(3,3)(0,3)}}00
  \piece{\Tr}00
\end{block}
&
$\longleftrightarrow$
\begin{block}33
  \piece{\annotation{\pspolygon(0,0)(3,0)(3,3)(0,3)}}00
  \piece{\Td}01
\end{block}
&
$\longleftrightarrow$
\begin{block}33
  \piece{\annotation{\pspolygon(0,0)(3,0)(3,3)(0,3)}}00
  \piece{\Tl}00
\end{block}
&
$\longleftrightarrow$
\begin{block}33
  \piece{\annotation{\pspolygon(0,0)(3,0)(3,3)(0,3)}}00
  \piece{\Tu}01
\end{block}
&
$\longrightarrow$
\\\\
\end{tabular}

 \\
\begin{tabular}[t]{cccccccccc}
$\longleftarrow$
\begin{block}33
  \piece{\annotation{\pspolygon(0,0)(3,0)(3,3)(0,3)}}00
  \piece{\LSd}00
\end{block}
&
$\longleftrightarrow$
\begin{block}33
  \piece{\annotation{\pspolygon(0,0)(3,0)(3,3)(0,3)}}00
  \piece{\LSl}01
\end{block}
&
$\longrightarrow$
\\\\
    \end{tabular}

 \\
\begin{tabular}[t]{cccccccccc}
$\longleftarrow$
\begin{block}33
  \piece{\annotation{\pspolygon(0,0)(3,0)(3,3)(0,3)}}00
  \piece{\RSd}00
\end{block}
&
$\longleftrightarrow$
\begin{block}33
  \piece{\annotation{\pspolygon(0,0)(3,0)(3,3)(0,3)}}00
  \piece{\RSl}01
\end{block}
&
$\longrightarrow$
\\\\
    \end{tabular}

 \\
    \begin{tabular}[t]{cccccccccc}
$\longleftarrow$
      \begin{block}44
        \piece{\annotation{\pspolygon(0,0)(4,0)(4,4)(0,4)}}00
        \piece{\Id}10
      \end{block}
&
$\longleftrightarrow$
      \begin{block}44
        \piece{\annotation{\pspolygon(0,0)(4,0)(4,4)(0,4)}}00
        \piece{\Il}02
      \end{block}
&
$\longrightarrow$
\\\\
    \end{tabular}

 \\
\begin{tabular}[t]{cccccccccc}
$\longleftarrow$
\begin{block}22
  \piece{\annotation{\pspolygon(0,0)(2,0)(2,2)(0,2)}}00
  \piece{\Sq}00
\end{block}
&
$\longrightarrow$
\\\\
\end{tabular}


\end{tabular}
\else
\includegraphics{figures/rotations/tetris/all.epsi.clean}
\fi


    \caption{The Tetris model of rotation.
      The pictured $k$-by-$k$ bounding
      box is in the same position in each configuration; each piece
      can be rotated clockwise to yield the configuration on its right
      (wrapping to the leftmost column) or counterclockwise to yield
      the configuration on its left.}
    \label{fig:rotation-tetris}
  \end{center}
\end{figure}

The \emph{Tetris rotation model}, which we have observed to be the one
used in a number of actual Tetris implementations, is illustrated in
Figure \ref{fig:rotation-tetris}.  Intuitively, this model works as
follows: for each piece type, choose the smallest $k$ so that the
piece fits within a $k$-by-$k$ bounding box ($k=2$ for $\tSq$, $k=4$
for $\tI$, and $k=3$ otherwise).  In a particular orientation, choose
the smallest $k_1$ and $k_2$ so that the piece fits in $k_1$-by-$k_2$
bounding box.  Place the piece so that the $k_1$-by-$k_2$ bounding box
is exactly centered in the $k$-by-$k$ box.  This does not in general
yield a position aligned on the grid, so shift the $k_1$-by-$k_2$
bounding box to the left and up, as necessary.  (Incidentally, it took
us some time to realize that the ``real'' rotation in Tetris did not
follow the instantaneous model, which is intuitively the most natural
one.)


\begin{lemma}  \label{lemma:tetris-reasonable}
  The Tetris rotation model is \reasonable.  $\qed$
\end{lemma}

The proofs of Lemmas \ref{lemma:continuous-reasonable} and
\ref{lemma:tetris-reasonable}, found in Appendix \ref{app:rotmodels},
are straightforward checks of the four conditions of \reasonable
rotation models.

\subsection{The Final Result}

\begin{theorem}
  It remains $\np$-hard to optimize (or approximate) the maximum
  height of a filled gridsquare, the number of rows cleared,
  tetrises attained, or pieces placed without a loss when any of the
  following hold:
  \begin{CompactEnumerate}
  \item the player is restricted to two rotation/translation moves
    before each piece drops in height.
  \item pieces are restricted to $\{\tLG,\tLS,\tI,\tSq\}$ or
    $\{\tRG,\tRS,\tI,\tSq\}$ plus at least one other piece.
  \item losses are not triggered until \emph{after} filled rows are
    cleared.
  \item rotations follow any reasonable rotation model.
  \end{CompactEnumerate}
  When losses never occur, it remains hard to optimize (or
  approximate) the number of cleared rows, number of tetrises, or
  maximum height of a filled gridsquare.  $\qed$
\end{theorem}
\section{Conclusion and Future Work}
\mylabel{sec:conclusion}

An essential part of our reduction is a complicated initial gameboard
from which the player must start.  A major open question is whether
Tetris can be played efficiently with an empty initial configuration:
\begin{itemize}
 \item What is the complexity of Tetris when the initial gameboard is
  empty?
\end{itemize}
While our results hold under a variety of modifications to the rules
of Tetris, some rule changes break our reduction.  In particular, our
completeness result relies on the translation of pieces as they fall.
At more difficult levels of the game, it may be very hard to make two
translations before the piece drops another row in height.

\begin{itemize}
\item Is translation a crucial part of the complexity of Tetris?
  Suppose the model for moves (following Brzustowski
  \cite{brzustowski}) is the following: the piece can be translated
  and rotated as many times as the player pleases, and then falls into
  place.  (That is, no translation or rotation is allowed after the
  piece takes its first downward step.)  Is the game still hard?
\end{itemize}
Another class of open questions considers versions of the Tetris game
with gameboards and piecesets of restricted size:
\begin{itemize}
\item What is the complexity of Tetris for a gameboard with a constant
  number of rows?  A constant number of columns?  Is Tetris
  fixed-parameter-tractable with respect to the number of rows or
  number of columns?  (We have polynomial-time algorithms for the
  special cases in which the total number of gridsquares is
  logarithmic in the number of pieces in the sequence, or for the case
  of a gameboard with two columns.)
  
\item We have reduced the pieceset down to five of the seven pieces.
  For what piecesets is Tetris polynomial-time solvable?  (For
  example, with the pieceset $\{\tI\}$ the problem seems polynomially
  solvable, though non-trivial because of the initial
  partially filled gameboard.)
\end{itemize}
Finally, in this paper we have concentrated our efforts on the
offline, adversarial version of Tetris.  In a real Tetris game, the
initial gameboard and piece sequence are generated probabilistically,
and the pieces are presented in an online fashion:
\begin{itemize}
\item What can we say about the difficulty of playing online Tetris if
  pieces are generated independently at random according to the
  uniform distribution, and the initial gameboard is randomly
  generated?
\end{itemize}



\paragraph{Acknowledgments.}  The first author thanks
Amos Fiat and Ming-wei Wang for helpful initial discussions.  
Thanks also to Josh Tauber for pointing out the puzzles in 
\emph{Games Magazine} \cite{gamesmag}.

\begin{small}
  \bibliography{tetris} 
\end{small}

\appendix
\section{Rotation Models}
\label{app:rotmodels}

Recall that a rotation model is \emph{\reasonable{}} if it satisfies
the following four conditions:
\begin{enumerate}
\item A piece cannot ``jump'' from one bucket to another, or into a
  disconnected region. \label{item:jump}
\item An $\tLG$ cannot enter a notch if the gridsquare in the second
  column of the notch row is filled.
\item For any \balcony at height $h$ with the $i$th column empty in
  the \balcony, the only gridsquares at height $h' \le h$ that can be
  filled by any piece other than an $\tI$ are those in the $i$th
  column in rows $h$ and $h-1$. 
\item An $\tI$ cannot enter a notch if the gridsquares in the second
  and third columns of the row immediately beneath the notch row are
  filled.
\end{enumerate}

As a preliminary step, we will give some sufficient conditions on
rotation models for two of these conditions.

\begin{lemma}  \mylabel{lemma:stationary-unfillable-holes}
  Suppose, for every rotation of every piece, there is some gridsquare
  that is filled both before and after the rotation.  Then clause (1)
  of the definition of \reasonable is satisfied.
\end{lemma}
\begin{proof}
  Partition the unfilled gridsquares of the bucket into connected
  \emph{regions} (where two gridsquares are connected if and only if
  they share an edge), counting the four gridsquares containing the
  current piece as unfilled.  Since some gridsquare remains filled by
  the piece before and after a legal rotation, the four gridsquares
  filled before and after the rotation must be in the same region.
  Furthermore, neither a horizontal translation nor drop can shift the
  piece from one region to another.
  
  By definition, a hole is a different region from that at the top of
  the bucket, where the piece enters.  A piece can never move from one
  region to another, so therefore it can never fill a hole.
  Similarly, a piece can never jump from one bucket to another, since
  doing so would force it to pass through the wall dividing the
  buckets.
\end{proof}

\begin{lemma}  \mylabel{lemma:heavy-line-unpassable-balcony}
  Suppose, for every rotation of every piece except $\tI$, there is a
  row $h$ that has at least two gridsquares filled by the piece before
  the rotation, and, after the rotation, one of the following holds:
  \begin{CompactEnumerate}
  \item[(a)] some row $h' \ge h-1$ still has at least two gridsquares
    filled by the piece.
  \item[(b)] row $h-2$ has at least two gridsquares filled by the piece,
    and at least one gridsquare in row $h-1$ that was unfilled before
    the rotation is now filled.
  \end{CompactEnumerate}
  Then clause (3) of the definition of \reasonable is satisfied.
\end{lemma}
\begin{proof}
  Suppose, as per clause (3), that there is a \balcony at height $b$,
  with the $i$th column empty in row $b$.
  
  For every piece other than $\tI$, in any orientation, there is a row
  that has at least two gridsquares filled by that piece.  Call such a row
  \emph{heavy}.  In any configuration in which a heavy row
  is at height $b+1$ or above, it is easy to confirm that clause (3)
  is satisfied. 
  
  Suppose that there is a way to move the piece so that the following
  condition holds:
  \begin{quote}
    ($\ast$) all of its heavy rows appear in row $b-1$ or lower.
  \end{quote}
  Consider the first time that ($\ast$) occurs, and consider the
  previous move $\sigma$, i.e., the one that made ($\ast$) true.
  
  It is obvious that $\sigma$ cannot be a horizontal translation or a
  fix, since these moves do not affect the vertical placement of the
  heavy rows.  Furthermore, the move $\sigma$ cannot have been a drop,
  since the filled gridsquares in the topmost heavy row would somehow have
  passed through the \balcony.  Thus $\sigma$ must have been a
  rotation.
  
  If $\sigma$ is a rotation satisfying condition (a), then there is a heavy
  row that cannot have passed below the \balcony---it is either still above
  the \balcony, or at best is in the same row as the \balcony,
  implying that the rotation was not legal (since two of the three
  gridsquares in row $b$ were already occupied).
  
  Otherwise, the rotation $\sigma$ must satisfy the condition (b).
  The only way to have made ($\ast$) true would be for a heavy row
  to have been in row $b+1$ before the rotation and $b-1$ after.  But,
  by condition (b), to do this the piece must occupy two distinct
  gridsquares in row $b$: one before and a different one after the rotation.  Since row $b$
  is a \balcony, and two of its three gridsquares are already
  occupied, this is impossible.

  Thus there is no such move $\sigma$, and clause (3) is satisfied.
\end{proof}

\subsection{Instantaneous Rotation}

\begin{oldtheorem}[Lemma \ref{lemma:instantaneous-reasonable}]
  The instantaneous rotation model is \reasonable.
\end{oldtheorem}
\begin{proof}
  Recall that in the instantaneous rotation model, rotation occurs
  around the \emph{center} of each piece:
  \begin{center}
    \ifredraw
\begin{tabular}{ccccccc}
\begin{block}{4}{2}
  \piece{\Sq}00
  \piece{\annotation{\pscircle[fillstyle=solid,fillcolor=black](0,0){0.2}}}{1}{1}
\end{block}
&
\begin{block}{4}{2}
  \piece{\LGl}00
  \piece{\annotation{\pscircle[fillstyle=solid,fillcolor=black](0,0){0.2}}}{1.5}{1.5}
\end{block}
&
\begin{block}{4}{2}
  \piece{\RGr}00
  \piece{\annotation{\pscircle[fillstyle=solid,fillcolor=black](0,0){0.2}}}{1.5}{1.5}
\end{block}
&
\begin{block}{4}{2}
  \piece{\LSl}00
  \piece{\annotation{\pscircle[fillstyle=solid,fillcolor=black](0,0){0.2}}}{1.5}{1.5}
\end{block}
&
\begin{block}{4}{2}
  \piece{\RSr}00
  \piece{\annotation{\pscircle[fillstyle=solid,fillcolor=black](0,0){0.2}}}{1.5}{1.5}
\end{block}
&
\begin{block}{4}{2}
  \piece{\Ir}00
  \piece{\annotation{\pscircle[fillstyle=solid,fillcolor=black](0,0){0.2}}}{1.5}{0.5}
\end{block}
&
\begin{block}{4}{2}
  \piece{\Tu}00
  \piece{\annotation{\pscircle[fillstyle=solid,fillcolor=black](0,0){0.2}}}{1.5}{0.5}
\end{block}\\\\
\end{tabular}
\else
\includegraphics{figures/rotations/inst/centers.epsi.clean}
\fi


  \end{center}

  \begin{enumerate}
  \item \textbf{No piece can fill a hole.}  Immediate from Lemma
    \ref{lemma:stationary-unfillable-holes}---by definition, the
    gridsquare containing the center of the piece is filled before and
    after the rotation.
    
  \item \textbf{No $\tLG$ can fully fill a \spurn.}  Suppose that the
    $\tLG$ managed to enter the notch, and consider the first time
    that it did so.  It is in the first of the following
    configurations when this occurs.  
    
    In the move before its first entry into the notch, it must be in
    one of the remaining configurations (respectively, arising from a
    clockwise rotation, counterclockwise rotation, left slide, right
    slide, and drop).  Note that it is impossible for the $\tLG$ to
    have entered the notch via a fix, which does not change the
    piece's position or orientation.
    
    \begin{center}
      \ifredraw
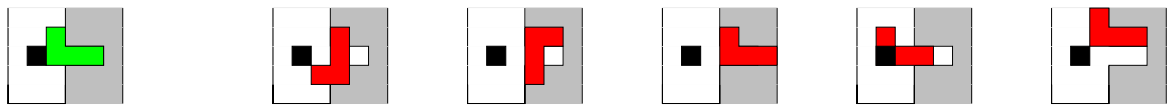
\begin{tabular}{lllllll}
\begin{block}85
  \background{6}{5}{lightgray}
  \column35
  \piece{\annotation{\pspolygon[fillstyle=solid,fillcolor=black](1,3)(1,2)(2,2)(2,3)}}00
  \piece{\LGr[green]}22
\end{block}
& ~~~~ & 
\begin{block}85
  \background{6}{5}{lightgray}
  \column35
  \piece{\annotation{\pspolygon[fillstyle=solid,fillcolor=black](1,3)(1,2)(2,2)(2,3)}}00
  \piece{\LGu[red]}21
\end{block}
&
\begin{block}85
  \background{6}{5}{lightgray}
  \column35
  \piece{\annotation{\pspolygon[fillstyle=solid,fillcolor=black](1,3)(1,2)(2,2)(2,3)}}00
  \piece{\LGd[red]}31
\end{block}
&
\begin{block}85
  \background{6}{5}{lightgray}
  \column35
  \piece{\annotation{\pspolygon[fillstyle=solid,fillcolor=black](1,3)(1,2)(2,2)(2,3)}}00
  \piece{\LGr[red]}32
\end{block}
&
\begin{block}85
  \background{6}{5}{lightgray}
  \column35
  \piece{\LGr[red]}12
  \piece{\annotation{\pspolygon[fillstyle=solid,fillcolor=black](1,3)(1,2)(2,2)(2,3)}}00
\end{block}
&
\begin{block}85
  \background{6}{5}{lightgray}
  \column35
  \piece{\annotation{\pspolygon[fillstyle=solid,fillcolor=black](1,3)(1,2)(2,2)(2,3)}}00
  \piece{\LGr[red]}23
\end{block}
\end{tabular}
\else
\includegraphics{figures/rotations/inst/spurn.epsi.clean}
\fi


    \end{center}
    In each of these configurations, the $\tLG$ intersects a filled
    gridsquare---in the fourth, in the second column of the notch row;
    for the others, one or more of the initially-filled gridsquares in
    the notch columns of the bucket.

  \item \textbf{Only an $\tI$ can fully pass a \balcony.}  Immediate
    from Lemma \ref{lemma:heavy-line-unpassable-balcony}---one can
    easily verify that, in fact, the first stated condition applies
    for each rotation.
    
  \item \textbf{An $\tI$ cannot enter a \OXX or \XXX.}  Suppose that
    the $\tI$ managed to enter the notch, and consider the first time
    that it did so.  It is in the first configuration of one of the
    following rows when this occurs.  (Note that the rotation center
    could be in either the second or third gridsquare of the $\tI$,
    reading from left to right.)
    
    In the move before the $\tI$'s first entry into the notch, it must
    be in one of the remaining configurations---respectively, arising
    from a clockwise rotation, counterclockwise rotation, left slide,
    right slide, and drop.  Note that it is impossible for the $\tI$
    to have entered the notch via a fix, which does not change the
    piece's position or orientation.
   \begin{center}
      \ifredraw
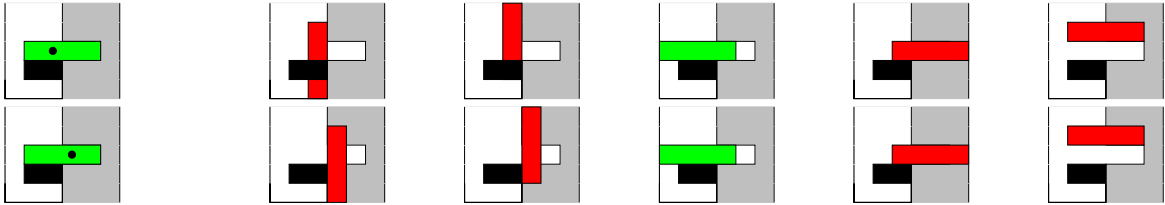
\begin{tabular}{lllllllll}
\begin{block}85
  \background{6}{5}{lightgray}
  \column35
  \piece{\annotation{\pspolygon[fillstyle=solid,fillcolor=black](1,2)(1,1)(3,1)(3,2)}}00
  \piece{\Ir[green]}12
  \piece{\annotation{\pscircle[fillstyle=solid,fillcolor=black](0,0){0.2}}}{2.5}{2.5}
\end{block}
& ~~~~ & 
\begin{block}85
  \background{6}{5}{lightgray}
  \column35
  \piece{\Iu[red]}20
  \piece{\annotation{\pspolygon[fillstyle=solid,fillcolor=black](1,2)(1,1)(3,1)(3,2)}}00
\end{block}
&
\begin{block}85
  \background{6}{5}{lightgray}
  \column35
  \piece{\Iu[red]}21
  \piece{\annotation{\pspolygon[fillstyle=solid,fillcolor=black](1,2)(1,1)(3,1)(3,2)}}00
\end{block}
&
\begin{block}85
  \background{6}{5}{lightgray}
  \column35
  \piece{\annotation{\pspolygon[fillstyle=solid,fillcolor=black](1,2)(1,1)(3,1)(3,2)}}00
  \piece{\Ir[green]}02
\end{block}
&
\begin{block}85
  \background{6}{5}{lightgray}
  \column35
  \piece{\annotation{\pspolygon[fillstyle=solid,fillcolor=black](1,2)(1,1)(3,1)(3,2)}}00
  \piece{\Ir[red]}22
\end{block}
&
\begin{block}85
  \background{6}{5}{lightgray}
  \column35
  \piece{\annotation{\pspolygon[fillstyle=solid,fillcolor=black](1,2)(1,1)(3,1)(3,2)}}00
  \piece{\Ir[red]}13
\end{block}
\\\\
\begin{block}85
  \background{6}{5}{lightgray}
  \column35
  \piece{\annotation{\pspolygon[fillstyle=solid,fillcolor=black](1,2)(1,1)(3,1)(3,2)}}00
  \piece{\Ir[green]}12
  \piece{\annotation{\pscircle[fillstyle=solid,fillcolor=black](0,0){0.2}}}{3.5}{2.5}
\end{block}
& ~~~~ & 
\begin{block}85
  \background{6}{5}{lightgray}
  \column35
  \piece{\Iu[red]}30
  \piece{\annotation{\pspolygon[fillstyle=solid,fillcolor=black](1,2)(1,1)(3,1)(3,2)}}00
\end{block}
&
\begin{block}85
  \background{6}{5}{lightgray}
  \column35
  \piece{\Iu[red]}31
  \piece{\annotation{\pspolygon[fillstyle=solid,fillcolor=black](1,2)(1,1)(3,1)(3,2)}}00
\end{block}
&
\begin{block}85
  \background{6}{5}{lightgray}
  \column35
  \piece{\annotation{\pspolygon[fillstyle=solid,fillcolor=black](1,2)(1,1)(3,1)(3,2)}}00
  \piece{\Ir[green]}02
\end{block}
&
\begin{block}85
  \background{6}{5}{lightgray}
  \column35
  \piece{\annotation{\pspolygon[fillstyle=solid,fillcolor=black](1,2)(1,1)(3,1)(3,2)}}00
  \piece{\Ir[red]}22
\end{block}
&
\begin{block}85
  \background{6}{5}{lightgray}
  \column35
  \piece{\annotation{\pspolygon[fillstyle=solid,fillcolor=black](1,2)(1,1)(3,1)(3,2)}}00
  \piece{\Ir[red]}13
\end{block}

\end{tabular}
\else
\includegraphics{figures/rotations/inst/floor.epsi.clean}
\fi


    \end{center}
    In each of these configurations but the third in each row, the
    previous position overlaps an occupied gridsquare---one of the
    assumed gridsquares in the first two configurations of the first
    row, and an initially-filled gridsquare in the notch columns in
    the remainder.
    
    Thus the only way for the $\tI$ to fully enter the notch is for it
    to have previously been partially in the notch.  Now we consider
    the previous position of the $\tI$ before it was partially in the
    notch.  The following configurations are arranged analogously to
    the above.
   \begin{center}
      \ifredraw
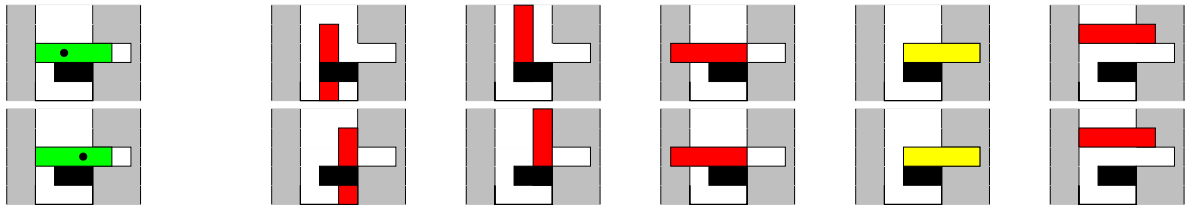
\begin{tabular}{lllllllll}
\begin{block}85
  \piece{\background{7}{5}{lightgray}}{-1}0
  \piece{\column35}00
  \piece{\annotation{\pspolygon[fillstyle=solid,fillcolor=black](1,2)(1,1)(3,1)(3,2)}}00
  \piece{\Ir[green]}02
  \piece{\annotation{\pscircle[fillstyle=solid,fillcolor=black](0,0){0.2}}}{1.5}{2.5}
\end{block}
& ~~~~ & 
\begin{block}85
  \piece{\background{7}{5}{lightgray}}{-1}0
  \piece{\column35}00
  \piece{\Iu[red]}10
  \piece{\annotation{\pspolygon[fillstyle=solid,fillcolor=black](1,2)(1,1)(3,1)(3,2)}}00
\end{block}
&
\begin{block}85
  \piece{\background{7}{5}{lightgray}}{-1}0
  \piece{\column35}00
  \piece{\Iu[red]}11
  \piece{\annotation{\pspolygon[fillstyle=solid,fillcolor=black](1,2)(1,1)(3,1)(3,2)}}00
\end{block}
&
\begin{block}85
  \piece{\background{7}{5}{lightgray}}{-1}0
  \piece{\column35}00
  \piece{\annotation{\pspolygon[fillstyle=solid,fillcolor=black](1,2)(1,1)(3,1)(3,2)}}00
  \piece{\Ir[red]}{-1}2
\end{block}
&
\begin{block}85
  \piece{\background{7}{5}{lightgray}}{-1}0
  \piece{\column35}00
  \piece{\annotation{\pspolygon[fillstyle=solid,fillcolor=black](1,2)(1,1)(3,1)(3,2)}}00
  \piece{\Ir[yellow]}12
\end{block}
&
\begin{block}85
  \piece{\background{7}{5}{lightgray}}{-1}0
  \piece{\column35}00
  \piece{\annotation{\pspolygon[fillstyle=solid,fillcolor=black](1,2)(1,1)(3,1)(3,2)}}00
  \piece{\Ir[red]}03
\end{block}
\\\\
\begin{block}85
  \piece{\background{7}{5}{lightgray}}{-1}0
  \piece{\column35}00
  \piece{\annotation{\pspolygon[fillstyle=solid,fillcolor=black](1,2)(1,1)(3,1)(3,2)}}00
  \piece{\Ir[green]}02
  \piece{\annotation{\pscircle[fillstyle=solid,fillcolor=black](0,0){0.2}}}{2.5}{2.5}
\end{block}
& ~~~~ & 
\begin{block}85
  \piece{\background{7}{5}{lightgray}}{-1}0
  \piece{\column35}00
  \piece{\Iu[red]}20
  \piece{\annotation{\pspolygon[fillstyle=solid,fillcolor=black](1,2)(1,1)(3,1)(3,2)}}00
\end{block}
&
\begin{block}85
  \piece{\background{7}{5}{lightgray}}{-1}0
  \piece{\column35}00
  \piece{\Iu[red]}21
  \piece{\annotation{\pspolygon[fillstyle=solid,fillcolor=black](1,2)(1,1)(3,1)(3,2)}}00
\end{block}
&
\begin{block}85
  \piece{\background{7}{5}{lightgray}}{-1}0
  \piece{\column35}00
  \piece{\annotation{\pspolygon[fillstyle=solid,fillcolor=black](1,2)(1,1)(3,1)(3,2)}}00
  \piece{\Ir[red]}{-1}2
\end{block}
&
\begin{block}85
  \piece{\background{7}{5}{lightgray}}{-1}0
  \piece{\column35}00
  \piece{\annotation{\pspolygon[fillstyle=solid,fillcolor=black](1,2)(1,1)(3,1)(3,2)}}00
  \piece{\Ir[yellow]}12
\end{block}
&
\begin{block}85
  \piece{\background{7}{5}{lightgray}}{-1}0
  \piece{\column35}00
  \piece{\annotation{\pspolygon[fillstyle=solid,fillcolor=black](1,2)(1,1)(3,1)(3,2)}}00
  \piece{\Ir[red]}03
\end{block}

\end{tabular}
\else
\includegraphics{figures/rotations/inst/partialfloor.epsi.clean}
\fi


    \end{center}
    But, in all but the fourth position of each row, the $\tI$
    overlaps a filled square, as before; in the fourth, the previous
    position is fully in the notch.  But, by our assumption that we
    are considering the \emph{first} entry into the notch, this cannot
    have been the previous configuration.
    
    Therefore the $\tI$ cannot fully enter the \OXX or \XXX.
  \end{enumerate}
  Thus the instantaneous model is \reasonable.
\end{proof}

\subsection{Continuous Rotation}

\begin{oldtheorem}[Lemma \ref{lemma:continuous-reasonable}]
  The continuous rotation model is \reasonable.
\end{oldtheorem}
\begin{proof}
  Any move which is allowed in the continuous rotation model is also
  allowed in the instanteous rotation model, so this follows
  immediately from Lemma \ref{lemma:instantaneous-reasonable}.
\end{proof}

\subsection{Tetris Rotations}

\begin{oldtheorem}[Lemma \ref{lemma:tetris-reasonable}]
  The Tetris rotation model is \reasonable.
\end{oldtheorem}
\begin{proof}  We show that the four requirements are met:
  \begin{enumerate}
  \item \textbf{No piece can fill a hole.}  Immediate from Lemma
    \ref{lemma:stationary-unfillable-holes} and inspection of the
    rotations.
    
  \item \textbf{An $\tLG$ cannot enter a \spurn.}  Suppose that the
    $\tLG$ somehow managed to enter the notch.  Then it must be in the
    first of the following configurations after it enters the notch,
    and must have previously been in either the remaining
    configurations immediately before (respectively, arising from a
    clockwise rotation, counterclockwise rotation, left slide, right
    slide, and drop).  Note that it is impossible for the $\tLG$ to
    have entered the notch via a fix, which does not change the
    piece's position or orientation.
    \begin{center}
      \ifredraw
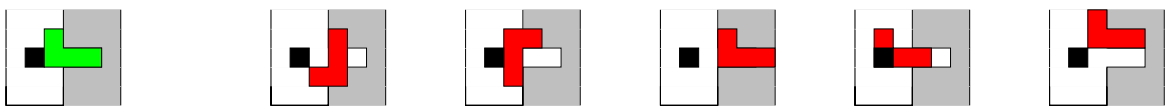
\begin{tabular}{lllllll}
\begin{block}85
  \background{6}{5}{lightgray}
  \column35
  \piece{\annotation{\pspolygon[fillstyle=solid,fillcolor=black](1,3)(1,2)(2,2)(2,3)}}00
  \piece{\LGr[green]}22
\end{block}
& ~~~~ & 
\begin{block}85
  \background{6}{5}{lightgray}
  \column35
  \piece{\annotation{\pspolygon[fillstyle=solid,fillcolor=black](1,3)(1,2)(2,2)(2,3)}}00
  \piece{\LGu[red]}21
\end{block}
&
\begin{block}85
  \background{6}{5}{lightgray}
  \column35
  \piece{\LGd[red]}21
  \piece{\annotation{\pspolygon[fillstyle=solid,fillcolor=black](1,3)(1,2)(2,2)(2,3)}}00
\end{block}
&
\begin{block}85
  \background{6}{5}{lightgray}
  \column35
  \piece{\annotation{\pspolygon[fillstyle=solid,fillcolor=black](1,3)(1,2)(2,2)(2,3)}}00
  \piece{\LGr[red]}32
\end{block}
&
\begin{block}85
  \background{6}{5}{lightgray}
  \column35
  \piece{\LGr[red]}12
  \piece{\annotation{\pspolygon[fillstyle=solid,fillcolor=black](1,3)(1,2)(2,2)(2,3)}}00
\end{block}
&
\begin{block}85
  \background{6}{5}{lightgray}
  \column35
  \piece{\annotation{\pspolygon[fillstyle=solid,fillcolor=black](1,3)(1,2)(2,2)(2,3)}}00
  \piece{\LGr[red]}23
\end{block}
\end{tabular}
\else
\includegraphics{figures/rotations/tetris/spurn.epsi.clean}
\fi


    \end{center}
    In each of these configurations, the $\tLG$ intersects a filled
    gridsquare---in the fourth, in the second column of the notch row;
    for the others, one or more of the initially-filled gridsquares in
    the notch columns of the bucket.
    
  \item \textbf{Only an $\tI$ can fully pass a \balcony.}  Immediate
    from Lemma \ref{lemma:heavy-line-unpassable-balcony} and
    inspection of the rotation rules.
    
  \item \textbf{An $\tI$ cannot enter a \OXX or \XXX.}  This is
    analagous to this condition for instantaneous rotation,
    but simpler.
    
    Suppose that the $\tI$ managed to enter the notch, and consider
    the first time that it did so.  It is in the first configuration
    of the following when this occurs.  In the move before the $\tI$'s
    first entry into the notch, it must be in one of the remaining
    configurations---respectively, arising from a rotation (clockwise
    and counterclockwise have the same result), right slide, left
    slide, and drop.  Note that it is impossible for the $\tI$ to
    have entered the notch via a fix, which does not change the
    piece's position or orientation.
   \begin{center}
      \ifredraw
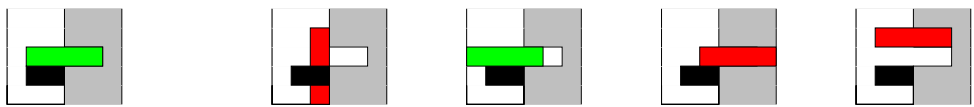
\begin{tabular}{lllllllll}
\begin{block}85
  \background{6}{5}{lightgray}
  \column35
  \piece{\annotation{\pspolygon[fillstyle=solid,fillcolor=black](1,2)(1,1)(3,1)(3,2)}}00
  \piece{\Ir[green]}12
\end{block}
& ~~~~ & 
\begin{block}85
  \background{6}{5}{lightgray}
  \column35
  \piece{\Iu[red]}20
  \piece{\annotation{\pspolygon[fillstyle=solid,fillcolor=black](1,2)(1,1)(3,1)(3,2)}}00
\end{block}
&
\begin{block}85
  \background{6}{5}{lightgray}
  \column35
  \piece{\annotation{\pspolygon[fillstyle=solid,fillcolor=black](1,2)(1,1)(3,1)(3,2)}}00
  \piece{\Ir[green]}02
\end{block}
&
\begin{block}85
  \background{6}{5}{lightgray}
  \column35
  \piece{\annotation{\pspolygon[fillstyle=solid,fillcolor=black](1,2)(1,1)(3,1)(3,2)}}00
  \piece{\Ir[red]}22
\end{block}
&
\begin{block}85
  \background{6}{5}{lightgray}
  \column35
  \piece{\annotation{\pspolygon[fillstyle=solid,fillcolor=black](1,2)(1,1)(3,1)(3,2)}}00
  \piece{\Ir[red]}13
\end{block}
\end{tabular}
\else
\includegraphics{figures/rotations/tetris/floor.epsi.clean}
\fi


    \end{center}
    In each of these configurations but that arising from a right
    slide, the previous position overlaps an occupied gridsquare.
    
    Thus the only way for the $\tI$ to fully enter the notch is for it
    to have previously been partially in the notch.  Now we consider
    the previous position of the $\tI$ before it was partially in the
    notch.  The following configurations are arranged analogously to
    the above.
    \begin{center}
      \ifredraw
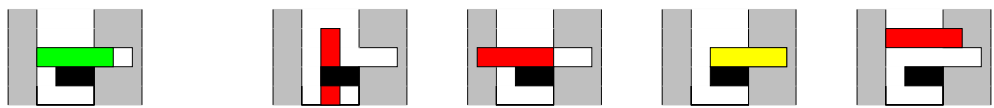
\begin{tabular}{lllllllll}
\begin{block}85
  \piece{\background{7}{5}{lightgray}}{-1}0
  \piece{\column35}00
  \piece{\annotation{\pspolygon[fillstyle=solid,fillcolor=black](1,2)(1,1)(3,1)(3,2)}}00
  \piece{\Ir[green]}02
\end{block}
& ~~~~ & 
\begin{block}85
  \piece{\background{7}{5}{lightgray}}{-1}0
  \piece{\column35}00
  \piece{\Iu[red]}10
  \piece{\annotation{\pspolygon[fillstyle=solid,fillcolor=black](1,2)(1,1)(3,1)(3,2)}}00
\end{block}
&
\begin{block}85
  \piece{\background{7}{5}{lightgray}}{-1}0
  \piece{\column35}00
  \piece{\annotation{\pspolygon[fillstyle=solid,fillcolor=black](1,2)(1,1)(3,1)(3,2)}}00
  \piece{\Ir[red]}{-1}2
\end{block}
&
\begin{block}85
  \piece{\background{7}{5}{lightgray}}{-1}0
  \piece{\column35}00
  \piece{\annotation{\pspolygon[fillstyle=solid,fillcolor=black](1,2)(1,1)(3,1)(3,2)}}00
  \piece{\Ir[yellow]}12
\end{block}
&
\begin{block}85
  \piece{\background{7}{5}{lightgray}}{-1}0
  \piece{\column35}00
  \piece{\annotation{\pspolygon[fillstyle=solid,fillcolor=black](1,2)(1,1)(3,1)(3,2)}}00
  \piece{\Ir[red]}03
\end{block}
\end{tabular}
\else
\includegraphics{figures/rotations/tetris/partialfloor.epsi.clean}
\fi


    \end{center}
    In all but the position arising from the left slide, the
    $\tI$ overlaps a filled square, as before.  By our assumption
    that we are considering the \emph{first} entry into the notch,
    this cannot have been the previous configuration.
    
    Therefore the $\tI$ cannot fully enter the \OXX or \XXX.
  \end{enumerate}
  Thus the Tetris model is \reasonable.
\end{proof}

\section{Deriving Unfillability from \Reasonable Rotation Models}
\label{app:unfillable}


Recall that a bucket is \emph{unfillable} if it cannot be filled
completely using arbitrarily many pieces from the set $\{\tLG, \tLS,
\tSq, \tI\}$.  Note that, for the purposes of proving unfillability,
we do not need to worry about the piece $\tRG$.

In this appendix, we show that a variety of configurations are
unfillable in any \reasonable rotation model.  Note that by clause (1)
of \reasonable, we can limit our attention to a single bucket when
proving unfillability.

\begin{lemma}  \mylabel{lemma:spurned}
  In any \reasonable rotation model, a bucket with (a) a hole, (b) a
  \spurn, (c) a \bOXX, or (d) a \bXXX is unfillable.
\end{lemma}
\begin{proof}
  Claim (a) is immediate from clause (1) of the definition.  Since an
  $\tI$ in the \spurn would have to occupy the (filled) gridsquare in
  the second column of that row, an $\tLG$ is the only piece that
  could fill the notch; thus claim (b) follows from clause (2) of the
  definition of \reasonable.  For claims (c) and (d), the unfilled
  notch must be filled with an $\tI$ by clause (3), and by clause (4)
  no $\tI$ can enter the notch.
\end{proof}

\begin{lemma}  \mylabel{lemma:bXXO}
  In any \reasonable rotation model, a bucket with a \bXXO is
  unfillable.
\end{lemma}
\begin{proof}
  Say the \bXXO is in row $n$.  By clause (3), only an $\tI$ can fill
  the notch in row $n$.  If the notch is filled by an $\tI$ before the
  empty gridsquare in the third column of row $n-1$ is filled, then
  the result is a hole.  If the empty gridsquare in the third column
  of row $n-1$ is filled before the notch in row $n$ is filled, then
  the result is a \bXXX, or a hole if filling the third gridsquare in
  row $n-1$ simultaneously fills the third gridsquare of row $n$.
  Either way, the result is unfillable.
\end{proof}

\begin{lemma}  \mylabel{lemma:balcony-notch}
  Consider a bucket with a \balcony in row $h$, and, somewhere below
  row $h$, an \bl{\alpha}.  In any \reasonable rotation model, the
  bucket is unfillable if either 
  \begin{CompactEnumerate}
  \item the \bl{\alpha} spans no notch rows, and $\alpha \not\equiv 0
    ~(\mod 4)$, or
  \item the \bl{\alpha} spans one notch row, and $\alpha \not\in
    \{0,1\} ~(\mod 4)$.
  \end{CompactEnumerate}
\end{lemma}
\begin{proof}
  By clause (3) of the definition of \reasonable, the \bl{\alpha} can
  only be filled by $\tI$'s.  (The \bl{\alpha} cannot be located in the
  exception of clause (3)---in the $i$th column in the $h$ and
  $(h-1)$st rows, where the \balcony leaves the $i$th column of row
  $h$ open---because otherwise the \bl{\alpha} would be above row
  $h$.)
  \begin{enumerate}
  \item An $\tI$ cannot fit horizontally into any non-notch row, so
    the only way to fill the \bl{\alpha} is with vertically placed
    $\tI$'s.  But since $\alpha \not\equiv 0 ~(\mod 4)$, the \bl\alpha
    cannot be completely filled by vertical $\tI$'s.
  \item Just as in case (1), since $\alpha \not\equiv 0 ~(\mod 4)$ we
    cannot fill the \bl{\alpha} using vertical $\tI$'s alone.  The only
    other possibility is to place an $\tI$ horizontally into the notch
    row, filling some gridsquare in the middle of the \bl\alpha.  This
    results in a \bl\beta and \bl\gamma, where the notch was in the
    $(\beta+1)$th-highest row of the \bl\alpha and $\beta + 1 + \gamma
    = \alpha$.  
    
    By the assumption that $\alpha - 1 \not\equiv 0 ~(\mod 4)$, we
    know that $\beta + \gamma \not\equiv 0 ~(\mod 4)$, which implies
    that at least one of \bl\beta and \bl\gamma falls into case (1).
  \end{enumerate}
  Thus such a bucket is unfillable.
\end{proof}

\begin{lemma}   \mylabel{lemma:b1-b2}
  In any \reasonable rotation model, a bucket with a \bver or \btwo is
  unfillable.
\end{lemma}
\begin{proof}
  Immediate from Lemma \ref{lemma:balcony-notch}, since at most one of
  the rows spanned by a \btwo is a notch row.
\end{proof}

\begin{lemma}  \mylabel{lemma:rect}
  In any \reasonable rotation model, a bucket with a \brect is
  unfillable.
\end{lemma}
\begin{proof}
  Suppose the notch in row $n$ has the rectangle beneath it unfilled.
  By clause (3) of the definition of \reasonable, only $\tI$'s can be
  used to fill the notch in row $n$ and the unfilled gridsquares in
  the rectangle below the notch.  
  
  Consider the unfilled gridsquare(s) in the $i$th column in rows
  $n-1, \ldots, n-j$, $i\in \{2,3\}$ and $j\in\{1,2,3\}$, and the
  filled gridsquare in row $n-j-1$.  These unfilled gridsquares must
  be filled by a vertically-placed $\tI$, since they are in non-notch
  rows.
  
  If $i=3$ and the gridsquare in column $3$ and row $n-1$ is filled
  before the notch in row $n$ is filled, then the result is a hole:
  since $j \le 3$, the vertical $\tI$ extends into row $n$.  If $i=2$
  and the gridsquare in column $2$ and row $n-1$ is filled before the
  notch in row $n$ is filled, then the result is a \spurn: again, the
  vertical $\tI$ extends into row $n$.
  
  If, on the other hand, the notch is filled before the unfilled
  gridsquare in row $n-1$, then the result is a \bl{j}: in column $i$,
  rows $n$ and $n-j-1$ are filled and rows $n-1,\ldots, n-j$ are
  unfilled.  Furthermore, for $j\le 3$, none of the unfilled rows are
  notch rows (since $n$ is a notch row).  This is unfillable by Lemma
  \ref{lemma:balcony-notch}.
  
  Regardless of which we try to fill first, the resulting
  configuration is unfillable.
\end{proof}

\begin{lemma}  \mylabel{lemma:bmod}
  In any reasonable rotation model, a bucket with a \bmod is
  unfillable.
\end{lemma}
\begin{proof}
  Suppose that in row $f$ the gridsquares in both the second and third
  columns are filled, and in rows $f+1, \ldots, f+\alpha$ both are
  empty, and that the second gridsquare in row $f+\alpha+1$ is filled.
  Further suppose that the second column is full and the third column
  empty in rows $f+\alpha+1, \ldots, f+\alpha+\beta$, and is filled in
  row $f+\alpha+\beta+1$, where $\beta \not\equiv 0 ~(\mod 4)$.  (The
  case when columns two and three are swapped is analogous.)  By
  clause (3) of the definition of \reasonable, only $\tI$'s can be
  used to fill the unfilled gridsquares beneath row
  $f+\alpha+\beta+1$.
  
  If $f+\alpha+1$ is a notch row, then it is a \spurn---or a hole for
  the case in which the unfilled segment of column three is the
  shorter one---which is unfillable by Lemma \ref{lemma:spurned}.  If
  all of $f+1, \ldots, f+\alpha$ are non-notch rows, then we have an
  \bl{\alpha} and an \bl{(\alpha+1)} spanning no notch rows.  By Lemma
  \ref{lemma:balcony-notch}, this is unfillable.  Otherwise, there is
  at least one unfilled notch row in rows $f+1, \ldots, f+\alpha$;
  consider the highest such notch row $r$.
  
  If an $\tI$ is placed horizontally to fill the notch in row $r$
  before any gridsquares in the second or third columns at or above
  row $r$ are filled, then the resulting configuration has a \bmod
  spanning only non-notch rows.  By the previous case, this is
  unfillable.
  
  Thus some $\tI$ must be placed vertically to fill gridsquares in the
  second or third columns at or above row $r$ before the notch is
  filled.  Such a placement of an $\tI$ must fill either the second or
  third gridsquare of row $r$ as well (a vertical $\tI$ fills four
  unfilled gridsquares above a filled gridsquare in the same column).
  This creates a \spurn or a hole, respectively.

  In any case, then, the configuration is unfillable.
\end{proof}

\begin{lemma}  \mylabel{lemma:bgap}
  In any reasonable rotation model, a bucket with a \bgap is
  unfillable.
\end{lemma}
\begin{proof}
  By clause (3) of the definition of \reasonable, only $\tI$'s can be
  used to fill the spanned notch.  Once this notch is filled, the
  result is a \bmod.
\end{proof}

\begin{oldtheorem}[Lemma \ref{lemma:bad-configs}]
  In any \reasonable rotation model, any bucket containing any of the
  following is unfillable:
  \begin{CompactEnumerate}
  \item a hole.  
  \item a \spurn.
  \item a \bXXO, \OXX, or \XXX.
  \item a \brect.
  \item a \bver.
  \item a \btwo.
  \item a \bmod.
  \item a \bgap.
  \end{CompactEnumerate}
\end{oldtheorem}
\begin{proof}
  Immediate from Lemmas \ref{lemma:spurned}, \ref{lemma:bXXO},
  \ref{lemma:b1-b2}, \ref{lemma:rect}, \ref{lemma:bmod}, and
  \ref{lemma:bgap}.
\end{proof}

\section{Proof of Soundness Propositions}
\mylabel{sec:soundness-proofs}
\mylabel{app:soundness}

In this section, we show that, under any \reasonable rotation model,
there is no possible way to win the game $\mathcal{G(P)}$ without
playing ``right.''  In Figure \ref{fig:bad-configs}, we show the
configurations that will be relevant in our proofs.


\begin{figure}[tb!]
  \begin{center}
    \ifredraw
\psset{unit=0.06in}
\begin{tabular}{lllllllllllll}
  \begin{block}{7}{8}
    \column{5}{8}
    \piece{\floor[lightgray]044}00
  \end{block}
  &
  \begin{block}{7}{8}
    \column{5}{8}
    \piece{\floor[lightgray]440}00
  \end{block}
  &
  \begin{block}{7}{8}
    \column{5}{8}
    \piece{\floor[lightgray]666}00
    \piece{\LGr[lightgray]}24
  \end{block}
  &
  \begin{block}{7}{14}
    \column{5}{14}
    \piece{\floor[lightgray]998}00
    \piece{\LGr[lightgray]}24
  \end{block}
  &
  \begin{block}{7}{14}
    \column{5}{14}
    \piece{\floor[lightgray]099}00
    \piece{\LGr[lightgray]}24
  \end{block}
  &
  \begin{block}{5}{8}
  \column{5}{8}
  \piece{\floor[lightgray]444}00
  \end{block}
  &
  \begin{block}{5}{14}
    \column{5}{14}
    \piece{\floor[lightgray]840}00
  \end{block}
&
  \begin{block}{5}{14}
    \column{5}{14}
    \piece{\floor[lightgray]044}00
    \piece{\LGr[green]}2{10}
  \end{block}
  &
  \begin{block}{5}{14}
    \column{5}{14}
    \piece{\floor[lightgray]440}00
    \piece{\LGr[green]}2{10}
  \end{block}
\\\\  
  (a) & (b) & (c) & (d) & (e) & (f) & (g) & (h) & (i) 
\\\\
  \begin{block}{5}{14}
    \column{5}{14}
    \piece{\floor[lightgray]840}00
    \piece{\LGr[green]}2{10}
  \end{block}
  &
  \begin{block}{5}{26}
    \column{5}{26}
    \piece{\floor[lightgray]440}00
    \piece{\LGr[green]}2{10}
    \piece{\LGr[green]}2{22}
  \end{block}
&
  \begin{block}{5}{26}
    \column{5}{26}
    \piece{\floor[lightgray]044}00
    \piece{\LGr[green]}2{10}
    \piece{\LGr[green]}2{22}
  \end{block}
&
\begin{block}{5}{20}
  \column{5}{20}
  \piece{\floor[lightgray]077}00
  \piece{\LGr[lightgray]}24
\end{block}
&
\begin{block}{5}{20}
  \column{5}{20}
  \piece{\floor[lightgray]099}00
  \piece{\LGr[lightgray]}24
  \piece{\LGr[green]}2{16}
\end{block}
&
  \begin{block}{5}{14}
    \column{5}{14}
    \piece{\floor[lightgray]444}00
    \piece{\LGr[green]}2{10}
  \end{block}
&
\begin{block}{5}{14}
  \column{6}{14}
  \piece{\floor[lightgray]344}00
  \piece{\LGr[green]}2{11}
\end{block}
&
\begin{block}{5}{22}
  \column{6}{22}
  \piece{\floor[lightgray]344}00
  \piece{\LGr[green]}2{11}
  \piece{\LGr[green]}2{17}
\end{block}
&
      \begin{block}{5}{14}
        \column{5}{14}
        \piece{\floor[lightgray]666}00
        \piece{\LGr[lightgray]}24
        \piece{\LGu[green]}06
      \end{block}
\\\\
(j) & (k) & (l) & (m) & (n) &  (o) & (p) & (q) & (r) 
\\\\
      \begin{block}{5}{14}
        \column{5}{14}
        \piece{\floor[lightgray]666}00
        \piece{\LGr[lightgray]}24
        \piece{\LGu[green]}16
      \end{block}
&
      \begin{block}{5}{14}
        \column{5}{14}
        \piece{\floor[lightgray]666}00
        \piece{\LGr[lightgray]}24
        \piece{\LGr[green]}06
      \end{block}
&
      \begin{block}{5}{14}
        \column{5}{14}
        \piece{\floor[lightgray]666}00
        \piece{\LGr[lightgray]}24
        \piece{\LGr[green]}2{10}
      \end{block}
&
  \begin{block}{5}{20}
    \column{5}{20}
    \piece{\floor[lightgray]998}00
    \piece{\LGr[lightgray]}24
    \piece{\LGr[green]}2{16}
  \end{block}
&
      \begin{block}{5}{26}
        \column{5}{26}
        \piece{\floor[lightgray]998}00
        \piece{\LGr[lightgray]}24
        \piece{\LGr[green]}2{16}
        \piece{\LGr[green]}2{22}
      \end{block}
  &  
  \\\\
  (s) & (t) & (u) & (v) & (w)


\end{tabular}

\else
\includegraphics[height=5.2in]{figures/configs/dead.epsi.clean}
\fi


    \caption{Configurations that can arise during play:
      (a,b) unprepped,
      (c)   \overflat,
      (d)   \thappy
      (e)   \tplat,
      (f)   \underflat, 
      (g)   \iprepped, 
      (h,i) \lgprepped{i},
      (j)   \ipreppedlg{i},
      (k,l) \lgpreppedlg{i}{j},
      (m)   \splat,
      (n)   \lgtplat{i},
      (o)   \lgunderflat{i},
      (p)   \lgunapproachable{i},
      (q)   \lglgunapproachable{i}{j},
      (r)   \lgoverflatA, 
      (s)   \lgoverflatB,
      (t)   \lgoverflatC,
      (u)   \lgoverflatD{i},
      (v)   \lgthappy{i},
      (w)   \lglgthappy{i}{j}.
      In each annotated configuration, the $i$th-highest (and
      $j$th-highest) notches have $\tLG$s filling them.}
    \mylabel{fig:bad-configs}
  \end{center}
\end{figure}

\subsection{Pieces and Configurations}

\subsubsection{\Unprepped Buckets}

\begin{proposition}  \mylabel{prop:i-unprepped}
  If $\tI$ is dropped validly into an \unprepped bucket, then it
  produces either an \underflat bucket or an \iprepped bucket.
\end{proposition}
\begin{proof}
  For an \unprepped bucket in which the first column is the unfilled
  one, the possible placements of the $\tI$ are as follows:
  \begin{center}
    \ifredraw
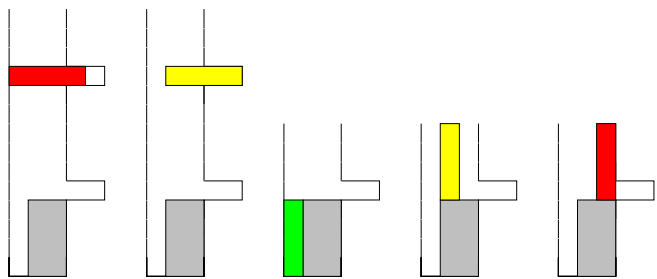
  \begin{center}
    \begin{tabular}{llllcllll}
      \begin{block}{5}{14}
        \column{5}{14}
        \piece{\floor[lightgray]044}00
        \piece{\Ir[red]}0{10}
      \end{block}
      &
      \begin{block}{5}{14}
        \column{5}{14}
        \piece{\floor[lightgray]044}00
        \piece{\Ir[yellow]}1{10}
      \end{block}
      &
      \begin{block}{5}{8}
        \column{5}{8}
        \piece{\floor[lightgray]044}00
        \piece{\Iu[green]}00
      \end{block}
      &
      \begin{block}{5}{8}
        \column{5}{8}
        \piece{\floor[lightgray]044}00
        \piece{\Iu[yellow]}14
      \end{block}
      &
      \begin{block}{5}{8}
        \column{5}{8}
        \piece{\floor[lightgray]044}00
        \piece{\Iu[red]}24
      \end{block}
      \\\\
    \end{tabular}
  \end{center}
\else
\includegraphics{figures/unprepped/ileft.epsi.clean}
\fi


  \end{center}
  (The first and second configurations have the $\tI$ horizontally in
  the $i$th-highest notch.  Both of these configurations are blocked
  for $i=1$, since this configuration has a \OXX and thus no $\tI$ can
  enter the bottom notch.)

  The first has a hole for any $i$.  In the second, for any $i\not=1$,
  the lowest notch is a \bOXX; for $i=1$, this is blocked.  The third
  configuration is underflat, the fourth configuration has a \spurn,
  and the last has a hole.

  For the other kind of \unprepped bucket, with the third column
  unfilled, the possible placements of the $\tI$ are as follows:
  \begin{center}
    \ifredraw
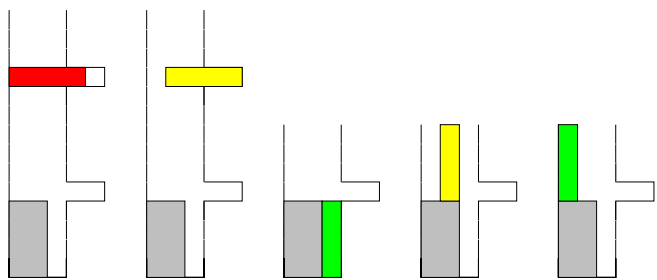
  \begin{center}
    \begin{tabular}{llllll}
      \begin{block}{5}{14}
        \column{5}{14}
        \piece{\floor[lightgray]440}00
        \piece{\Ir[red]}0{10}
      \end{block}
&
      \begin{block}{5}{14}
        \column{5}{14}
        \piece{\floor[lightgray]440}00
        \piece{\Ir[yellow]}1{10}
      \end{block}
      &
      \begin{block}{5}{8}
        \column{5}{8}
        \piece{\floor[lightgray]440}00
        \piece{\Iu[green]}20
      \end{block}
      &
      \begin{block}{5}{8}
        \column{5}{8}
        \piece{\floor[lightgray]440}00
        \piece{\Iu[yellow]}14
      \end{block}
      &
      \begin{block}{5}{8}
        \column{5}{8}
        \piece{\floor[lightgray]440}00
        \piece{\Iu[green]}04
      \end{block}\\\\
    \end{tabular}
  \end{center}
\else
\includegraphics{figures/unprepped/iright.epsi.clean}
\fi


  \end{center}
  (The first and second configurations have the $\tI$ horizontally in
  the $i$th-highest notch.)
  
  The first has a hole in the notch for any $i$.  The second
  configuration has a hole for $i=1$ and is a \bXXO for $i>1$.  The
  third is \underflat, and the fourth has a \spurn.  The fifth is
  \iprepped.
\end{proof}

\begin{proposition}  \mylabel{prop:lg-unprepped}
  If $\tLG$ is dropped validly into an \unprepped bucket, it produces
  \lgprepped{i} for some $i$.
\end{proposition}
\begin{proof}
  If the unprepped bucket has the first column empty and the bottom
  four rows of the second and third columns full, then the possible
  configurations are the following:
  \begin{center}
    \ifredraw 
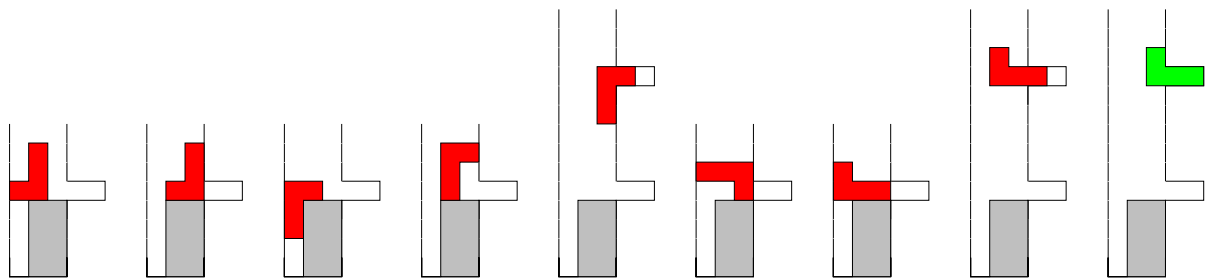
\begin{tabular}{ccccccccccc}
  \begin{block}{5}{8}
    \column{5}{8}
    \piece{\floor[lightgray]044}00
    \piece{\LGu[red]}04
  \end{block}
  &
  \begin{block}{5}{8}
    \column{5}{8}
    \piece{\floor[lightgray]044}00
    \piece{\LGu[red]}14
  \end{block}
  &
  \begin{block}{5}{8}
    \column{5}{8}
    \piece{\floor[lightgray]044}00
    \piece{\LGd[red]}02
  \end{block}
  &
  \begin{block}{5}{8}
    \column{5}{8}
    \piece{\floor[lightgray]044}00
    \piece{\LGd[red]}14
  \end{block}
  &
  \begin{block}{5}{14}
    \column{5}{14}
    \piece{\floor[lightgray]044}00
    \piece{\LGd[red]}2{8}
  \end{block}
  &
  \begin{block}{5}{10}
    \column{5}{8}
    \piece{\floor[lightgray]044}00
    \piece{\LGl[red]}04
  \end{block}
  &
  \begin{block}{5}{8}
    \column{5}{8}
    \piece{\floor[lightgray]044}00
    \piece{\LGr[red]}04
  \end{block}
  &
  \begin{block}{5}{14}
    \column{5}{14}
    \piece{\floor[lightgray]044}00
    \piece{\LGr[red]}1{10}
  \end{block}
  &
  \begin{block}{5}{14}
    \column{5}{14}
    \piece{\floor[lightgray]044}00
    \piece{\LGr[green]}2{10}
  \end{block}
\end{tabular}
\else
\includegraphics{figures/unprepped/lgleft.epsi.clean}
\fi


  \end{center}
  (In the fifth configuration, the $i$th-highest notch is partially
  filled, for any $i>1$; the eighth and ninth apply for any $i\geq 1$.)
  
  Each but the last of these configurations creates a hole (the fifth
  for any $i>1$ and the eighth for any $i\geq 1$); the last is \lgprepped{i}.

  For the other kind of unprepped bucket, with an empty third column,
  the possible configurations are the following:
  \begin{center}
    \ifredraw
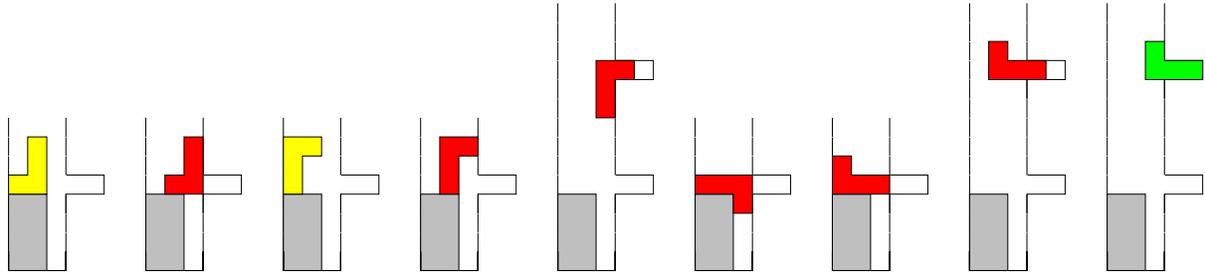
\begin{tabular}{ccccccccccc}
  \begin{block}{5}{8}
    \column{5}{8}
    \piece{\floor[lightgray]440}00
    \piece{\LGu[yellow]}04
  \end{block}
  &
  \begin{block}{5}{8}
    \column{5}{8}
    \piece{\floor[lightgray]440}00
    \piece{\LGu[red]}14
  \end{block}
  &
  \begin{block}{5}{8}
    \column{5}{8}
    \piece{\floor[lightgray]440}00
    \piece{\LGd[yellow]}04
  \end{block}
  &
  \begin{block}{5}{8}
    \column{5}{8}
    \piece{\floor[lightgray]440}00
    \piece{\LGd[red]}14
  \end{block}
  &
  \begin{block}{5}{14}
    \column{5}{14}
    \piece{\floor[lightgray]440}00
    \piece{\LGd[red]}2{8}
  \end{block}
  &
  \begin{block}{5}{10}
    \column{5}{8}
    \piece{\floor[lightgray]440}00
    \piece{\LGl[red]}03
  \end{block}
  &
  \begin{block}{5}{8}
    \column{5}{8}
    \piece{\floor[lightgray]440}00
    \piece{\LGr[red]}04
  \end{block}
  &
  \begin{block}{5}{14}
    \column{5}{14}
    \piece{\floor[lightgray]440}00
    \piece{\LGr[red]}1{10}
  \end{block}
  &
  \begin{block}{5}{14}
    \column{5}{14}
    \piece{\floor[lightgray]440}00
    \piece{\LGr[green]}2{10}
  \end{block}
\end{tabular}
\else
\includegraphics{figures/unprepped/lgright.epsi.clean}
\fi


  \end{center}
  (The fifth, eighth, and ninth have the $\tLG$ in the $i$th-highest
  notch; these apply for any $i$.)
  
  The first configuration has a \spurn, the third has a \bXXO, and the
  last is \lgprepped{i}.  The remainder of these configurations all
  have holes (the fifth and eighth for all $i\geq 1$).
\end{proof}

\begin{proposition} \mylabel{prop:sq-unprepped} 
  There is no valid move for $\tSq$ in an unprepped bucket.
\end{proposition}
\begin{proof}
  The possible placements for a $\tSq$ are as follows:
  \begin{center}
    \ifredraw
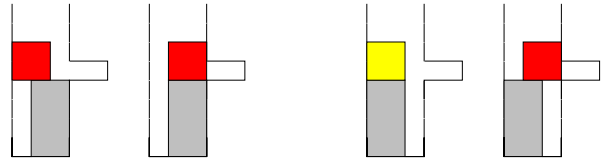
\begin{tabular}{lllll}
  \begin{block}{5}{8}
    \column{5}{8}
    \piece{\floor[lightgray]044}00
    \piece{\Sq[red]}04
  \end{block}
  &
  \begin{block}{5}{8}
    \column{5}{8}
    \piece{\floor[lightgray]044}00
    \piece{\Sq[red]}14
  \end{block}
  &
  ~~~~~
  &
  \begin{block}{5}{8}
    \column{5}{8}
    \piece{\floor[lightgray]440}00
    \piece{\Sq[yellow]}04
  \end{block}
  &
  \begin{block}{5}{8}
    \column{5}{8}
    \piece{\floor[lightgray]440}00
    \piece{\Sq[red]}14
  \end{block}\\\\
\end{tabular}
\else
\includegraphics{figures/unprepped/sq.epsi.clean}
\fi


  \end{center}
  The first, second, and fourth of these create holes; the third has a
  \spurn.
\end{proof}

\begin{proposition} \mylabel{prop:ls-unprepped}
  There is no valid move for $\tLS$ in an unprepped bucket.
\end{proposition}
\begin{proof}
  The possible configurations are the following:
  \begin{center}
    \ifredraw
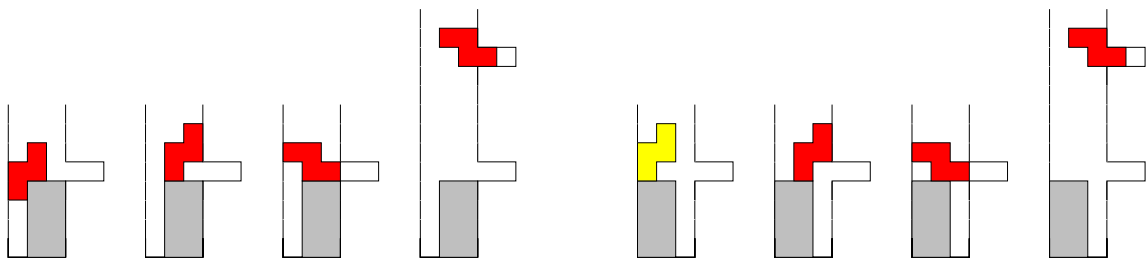
\begin{tabular}{cccccccccc}
  \begin{block}{5}{8}
    \column{5}{8}
    \piece{\floor[lightgray]044}00
    \piece{\LSu[red]}03
  \end{block}
  &
  \begin{block}{5}{8}
    \column{5}{8}
    \piece{\floor[lightgray]044}00
    \piece{\LSu[red]}14
  \end{block}
  &
  \begin{block}{5}{8}
    \column{5}{8}
    \piece{\floor[lightgray]044}00
    \piece{\LSl[red]}04
  \end{block}
  &
  \begin{block}{5}{13}
    \column{5}{13}
    \piece{\floor[lightgray]044}00
    \piece{\LSl[red]}1{10}
  \end{block}
  & ~~~~~ &
  \begin{block}{5}{8}
    \column{5}{8}
    \piece{\floor[lightgray]440}00
    \piece{\LSu[yellow]}04
  \end{block}
  &
  \begin{block}{5}{8}
    \column{5}{8}
    \piece{\floor[lightgray]440}00
    \piece{\LSu[red]}14
  \end{block}
  &
  \begin{block}{5}{8}
    \column{5}{8}
    \piece{\floor[lightgray]440}00
    \piece{\LSl[red]}04
  \end{block}
  &
  \begin{block}{5}{13}
    \column{5}{13}
    \piece{\floor[lightgray]440}00
    \piece{\LSl[red]}1{10}
  \end{block}\\\\
\end{tabular}
\else
\includegraphics{figures/unprepped/ls.epsi.clean}
\fi


  \end{center}
  (In the fourth and eighth of these configuration, the $\tLS$ is in
  the $i$th-highest notch, for any $i\geq 1$.)
  
  All of these configurations except the fifth have holes; the fifth
  has a \bXXO.
\end{proof}


\subsubsection{\iprepped Buckets}

\begin{proposition}  \mylabel{prop:lg-iprepped}
  If $\tLG$ is dropped validly into an \iprepped bucket, it produces
  \ipreppedlg{i} for some $i \ge 2$.
\end{proposition}
\begin{proof}
  The possible configurations are the following:
  \begin{center}
    \ifredraw
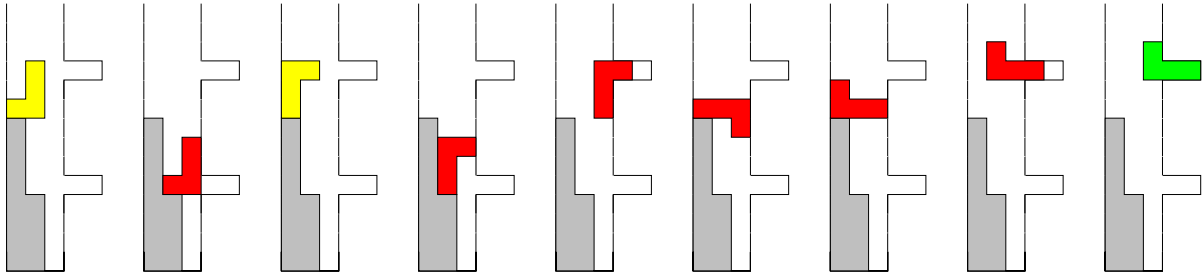
\begin{tabular}{ccccccccccc}
  \begin{block}{5}{14}
    \column{5}{14}
    \piece{\floor[lightgray]840}00
    \piece{\LGu[yellow]}08
  \end{block}
  &
  \begin{block}{5}{14}
    \column{5}{14}
    \piece{\floor[lightgray]840}00
    \piece{\LGu[red]}14
  \end{block}
  &
  \begin{block}{5}{14}
    \column{5}{14}
    \piece{\floor[lightgray]840}00
    \piece{\LGd[yellow]}08
  \end{block}
  &
  \begin{block}{5}{14}
    \column{5}{14}
    \piece{\floor[lightgray]840}00
    \piece{\LGd[red]}14
  \end{block}
  &
  \begin{block}{5}{14}
    \column{5}{14}
    \piece{\floor[lightgray]840}00
    \piece{\LGd[red]}2{8}
  \end{block}
  &
  \begin{block}{5}{10}
    \column{5}{14}
    \piece{\floor[lightgray]840}00
    \piece{\LGl[red]}07
  \end{block}
  &
  \begin{block}{5}{14}
    \column{5}{14}
    \piece{\floor[lightgray]840}00
    \piece{\LGr[red]}08
  \end{block}
  &
  \begin{block}{5}{14}
    \column{5}{14}
    \piece{\floor[lightgray]840}00
    \piece{\LGr[red]}1{10}
  \end{block}
  &
  \begin{block}{5}{14}
    \column{5}{14}
    \piece{\floor[lightgray]840}00
    \piece{\LGr[green]}2{10}
  \end{block}
\end{tabular}
\else
\includegraphics{figures/iprepped/lg.epsi.clean}
\fi


  \end{center}
  (In the fifth, eighth, and ninth configurations, the $i$th-highest
  notch is (partially) filled, for any $i\ge1$.)
  
  The first has a \bXXO, as does the third.  All other configurations
  except the last have holes (including the fifth and eighth for any
  $i$).  The last is \ipreppedlg{i}, though there is a hole if $i=1$.
\end{proof}

\begin{proposition}  \mylabel{prop:sq-iprepped}
  There is no valid placement of $\tSq$ in an \iprepped bucket.
\end{proposition}
\begin{proof}
  The possible configurations are the following:
  \begin{center}
    \ifredraw
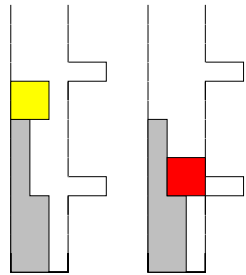
\begin{tabular}{lllll}
  \begin{block}{5}{14}
    \column{5}{14}
    \piece{\floor[lightgray]840}00
    \piece{\Sq[yellow]}08
  \end{block}
  &
  \begin{block}{5}{14}
    \column{5}{14}
    \piece{\floor[lightgray]840}00
    \piece{\Sq[red]}14
  \end{block}\\\\
\end{tabular}
\else
\includegraphics{figures/iprepped/sq.epsi.clean}
\fi


  \end{center}
  
  The first has a \bXXO, and the second has a hole.
\end{proof}

\begin{proposition}  \mylabel{prop:sq-iprepped-notched}
  There is no valid placement of $\tSq$ in an \ipreppedlg{i} bucket.
\end{proposition}
\begin{proof}
  The possible configurations are the following:
  \begin{center}
    \ifredraw
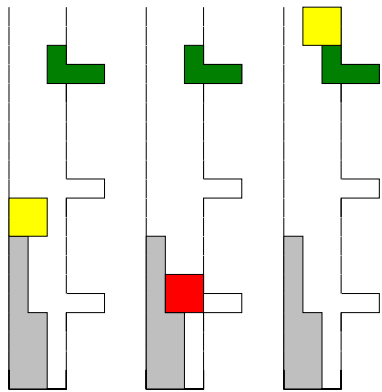
\begin{tabular}{lllll}
  \begin{block}{5}{20}
    \column{5}{20}
    \piece{\floor[lightgray]840}00
    \piece{\LGr[darkgreen]}2{16}
    \piece{\Sq[yellow]}08
  \end{block}
  &
  \begin{block}{5}{20}
    \column{5}{20}
    \piece{\floor[lightgray]840}00
    \piece{\LGr[darkgreen]}2{16}
    \piece{\Sq[red]}14
  \end{block}
  &
  \begin{block}{5}{20}
    \column{5}{20}
    \piece{\floor[lightgray]840}00
    \piece{\LGr[darkgreen]}2{16}
    \piece{\Sq[yellow]}1{18}
  \end{block}
\\\\
\end{tabular}
\else
\includegraphics{figures/iprepped/sqlg.epsi.clean}
\fi


  \end{center}
  
  (In each configuration, the LG is initially in the $i$th-highest
  notch for any $i>1$; if $i=1$, the configuration already has a
  hole.)
  
  The first and third configurations have a \bXXO; the second has a
  hole.
\end{proof}


\subsubsection{$\tLG$-prepped Buckets}

\begin{proposition} \mylabel{prop:sq-notched} 
  There is no valid move for $\tSq$ in \lgprepped{i}.
\end{proposition}
\begin{proof} 
  The possible configurations are as follows:
  \begin{center}
    \ifredraw
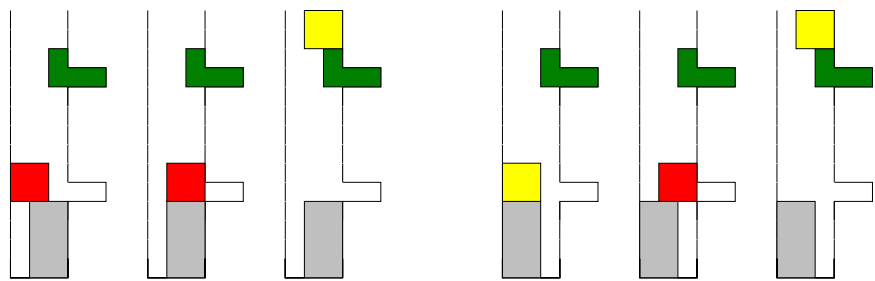
\begin{tabular}{lllllll}
  \begin{block}{5}{14}
    \column{5}{14}
    \piece{\floor[lightgray]044}00
    \piece{\LGr[darkgreen]}2{10}
    \piece{\Sq[red]}04
  \end{block}
  &
  \begin{block}{5}{14}
    \column{5}{14}
    \piece{\floor[lightgray]044}00
    \piece{\LGr[darkgreen]}2{10}
    \piece{\Sq[red]}14
  \end{block}
  &
  \begin{block}{5}{14}
    \column{5}{14}
    \piece{\floor[lightgray]044}00
    \piece{\LGr[darkgreen]}2{10}
    \piece{\Sq[yellow]}1{12}
  \end{block}
  &
  ~~~~~
  &
  \begin{block}{5}{14}
    \column{5}{14}
    \piece{\floor[lightgray]440}00
    \piece{\LGr[darkgreen]}2{10}
    \piece{\Sq[yellow]}04
  \end{block}
  &
  \begin{block}{5}{14}
    \column{5}{14}
    \piece{\floor[lightgray]440}00
    \piece{\LGr[darkgreen]}2{10}
    \piece{\Sq[red]}14
  \end{block}
  &
  \begin{block}{5}{14}
    \column{5}{14}
    \piece{\floor[lightgray]440}00
    \piece{\LGr[darkgreen]}2{10}
    \piece{\Sq[yellow]}1{12}
  \end{block}
\\\\
\end{tabular}
\else
\includegraphics{figures/notched/sq.epsi.clean}
\fi


  \end{center}
  (In each configuration, the $\tLG$ is initially in the
  $i$th-highest notch for any $i\geq 1$.  The second and fifth
  configurations are blocked if $i=1$.)
  
  The first, second (for $i>1$), and fifth (for $i>1$) configurations
  have holes.  The third has a \bOXX if $i>1$, or a \btwo if $i=1$.
  The fourth has a \spurn if $i>1$, and a hole if $i=1$.  The sixth
  has a \bXXO if $i>1$, or a hole if $i=1$.
\end{proof}

\begin{proposition}  \mylabel{prop:ls-notched}
  There is no valid move for $\tLS$ in \lgprepped{i}.
\end{proposition}
\begin{proof}
  With \lgprepped{i} so that the first column is unfilled in the
  bottom four rows, the possible configurations are the following:
  \begin{center}
    \ifredraw
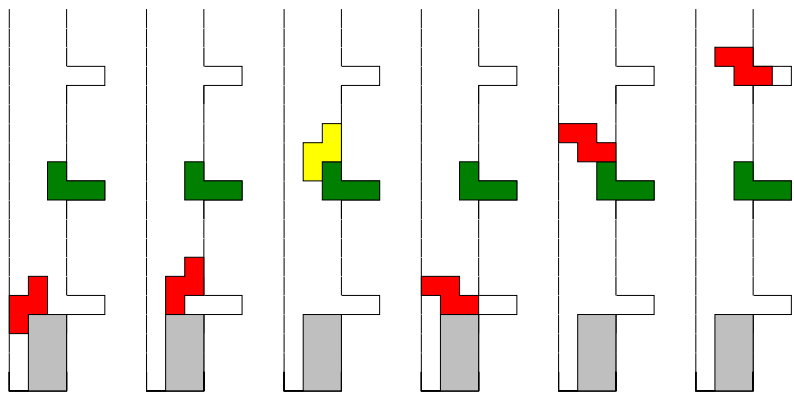
      \begin{tabular}{ccccccccc}
      \begin{block}{5}{20}
        \column{5}{20}
        \piece{\floor[lightgray]044}00
        \piece{\LGr[darkgreen]}2{10}
        \piece{\LSu[red]}03
      \end{block}
&
      \begin{block}{5}{20}
        \column{5}{20}
        \piece{\floor[lightgray]044}00
        \piece{\LGr[darkgreen]}2{10}
        \piece{\LSu[red]}14
      \end{block}
&
      \begin{block}{5}{20}
        \column{5}{20}
        \piece{\floor[lightgray]044}00
        \piece{\LGr[darkgreen]}2{10}
        \piece{\LSu[yellow]}1{11}
      \end{block}
&
      \begin{block}{5}{20}
        \column{5}{20}
        \piece{\floor[lightgray]044}00
        \piece{\LGr[darkgreen]}2{10}
        \piece{\LSl[red]}04
      \end{block}
&
      \begin{block}{5}{20}
        \column{5}{20}
        \piece{\floor[lightgray]044}00
        \piece{\LGr[darkgreen]}2{10}
        \piece{\LSl[red]}0{12}
      \end{block}
&
      \begin{block}{5}{20}
        \column{5}{20}
        \piece{\floor[lightgray]044}00
        \piece{\LGr[darkgreen]}2{10}
        \piece{\LSl[red]}1{16}
      \end{block}
\\\\
    \end{tabular}
\else
\includegraphics{figures/notched/lsright.epsi.clean}
\fi


  \end{center}
  (The $\tLG$ is in the $i$th-highest column initially.  In the last
  of these configurations, the $\tLS$ is in the $j$th-highest notch,
  for any $j\geq 1$ and $j \not= i$.  The second and fourth
  configurations are possible only if $i\not=1$.)
  
  Of these, all configurations but the third create holes (the last
  for any $j$), and the third has a \bOXX, or a \bone for $i=1$.
  
  With \lgprepped{i} so that the first four rows of the third column
  are unfilled, the possible configurations are the following:
  \begin{center}
    \ifredraw
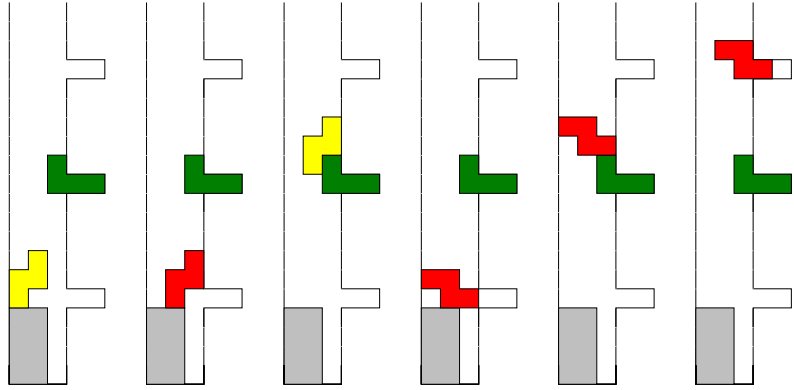
      \begin{tabular}{ccccccccc}
      \begin{block}{5}{20}
        \column{5}{20}
        \piece{\floor[lightgray]440}00
        \piece{\LGr[darkgreen]}2{10}
        \piece{\LSu[yellow]}04
      \end{block}
&
      \begin{block}{5}{20}
        \column{5}{20}
        \piece{\floor[lightgray]440}00
        \piece{\LGr[darkgreen]}2{10}
        \piece{\LSu[red]}14
      \end{block}
&
      \begin{block}{5}{20}
        \column{5}{20}
        \piece{\floor[lightgray]440}00
        \piece{\LGr[darkgreen]}2{10}
        \piece{\LSu[yellow]}1{11}
      \end{block}
&
      \begin{block}{5}{20}
        \column{5}{20}
        \piece{\floor[lightgray]440}00
        \piece{\LGr[darkgreen]}2{10}
        \piece{\LSl[red]}04
      \end{block}
&
      \begin{block}{5}{20}
        \column{5}{20}
        \piece{\floor[lightgray]440}00
        \piece{\LGr[darkgreen]}2{10}
        \piece{\LSl[red]}0{12}
      \end{block}
&
      \begin{block}{5}{20}
        \column{5}{20}
        \piece{\floor[lightgray]440}00
        \piece{\LGr[darkgreen]}2{10}
        \piece{\LSl[red]}1{16}
      \end{block}
\\\\
    \end{tabular}
\else
\includegraphics{figures/notched/lsleft.epsi.clean}
\fi


  \end{center}
  (Again, the $\tLG$ is in the $i$th-highest notch initially and the
  $\tLS$ in the last configuration is in the $j$th-highest notch, for
  any $j\geq 1$ and $j \not= i$.  The second and fourth configurations
  are possible only if $i\not=1$.)
  
  The first and third both have a \bXXO, or a hole if $i=1$.  The
  second (for $i>1$), fourth, fifth, and sixth (for any $j$) have a
  hole.
\end{proof}

\begin{proposition}  \mylabel{prop:lg-notched}
  If $\tLG$ is validly placed in \lgprepped{i}, then the result is a \splat or a \lgpreppedlg{i}{j} for some $j$.
\end{proposition}
\begin{proof}
  When the given ($\tLG$-UP-$i$) bucket has the empty rows in the
  first column, the possible configurations are as follows.  First,
  the configurations in which the $\tLG$ is placed vertically:
  \begin{center}
    \ifredraw
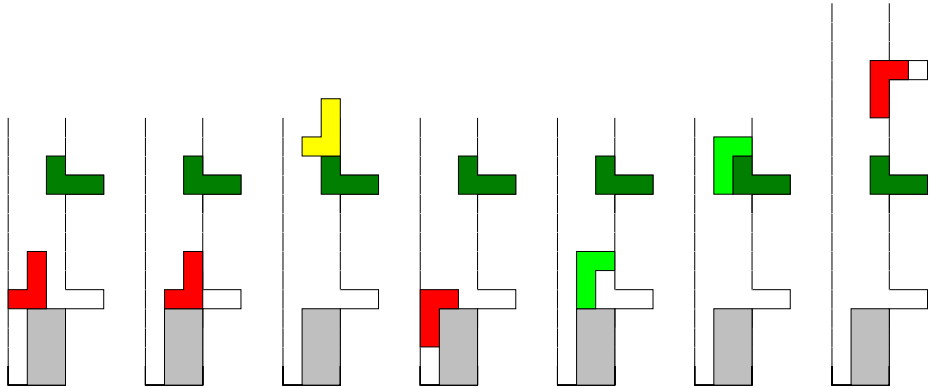
\begin{tabular}{llllllllllllll}
  \begin{block}{5}{14}
    \column{5}{14}
    \piece{\floor[lightgray]044}00
    \piece{\LGr[darkgreen]}2{10}
    \piece{\LGu[red]}04
  \end{block}
  &
  \begin{block}{5}{14}
    \column{5}{14}
    \piece{\floor[lightgray]044}00
    \piece{\LGr[darkgreen]}2{10}
    \piece{\LGu[red]}14
  \end{block}
  &
  \begin{block}{5}{14}
    \column{5}{14}
    \piece{\floor[lightgray]044}00
    \piece{\LGr[darkgreen]}2{10}
    \piece{\LGu[yellow]}1{12}
  \end{block}
  &
  \begin{block}{5}{14}
    \column{5}{14}
    \piece{\floor[lightgray]044}00
    \piece{\LGr[darkgreen]}2{10}
    \piece{\LGd[red]}02
  \end{block}
  &
  \begin{block}{5}{14}
    \column{5}{14}
    \piece{\floor[lightgray]044}00
    \piece{\LGr[darkgreen]}2{10}
    \piece{\LGd[green]}14
  \end{block}
  &
  \begin{block}{5}{14}
    \column{5}{14}
    \piece{\floor[lightgray]044}00
    \piece{\LGr[darkgreen]}2{10}
    \piece{\LGd[green]}1{10}
  \end{block}
  &
  \begin{block}{5}{20}
    \column{5}{20}
    \piece{\floor[lightgray]044}00
    \piece{\LGr[darkgreen]}2{10}
    \piece{\LGd[red]}2{14}
  \end{block}\\\\
\end{tabular}
\else
\includegraphics{figures/notched/lgleftV.epsi.clean}
\fi

  \end{center}
  (The notch initially filled with the $\tLG$ is the $i$th-highest; in
  the last configuration, the $j$th-highest notch is also filled for
  any $j\not=i$ and $j\geq 1$.  If $i=1$, the second configuration is
  impossible, and the fifth and sixth configurations are identical.)
  
  The first, second, fourth, fifth (for $i>1$), and seventh (for any
  $j \neq i$) configurations have holes.  The third has a \btwo for
  $i=1$ and a \bOXX for $i > 1$.  The sixth also has a \bOXX for $i >
  1$.  For $i=1$, the fifth and sixth configurations are identical,
  and form a \splat.

  And, now, those in which the $\tLG$ is placed horizontally:
  \begin{center}
    \ifredraw
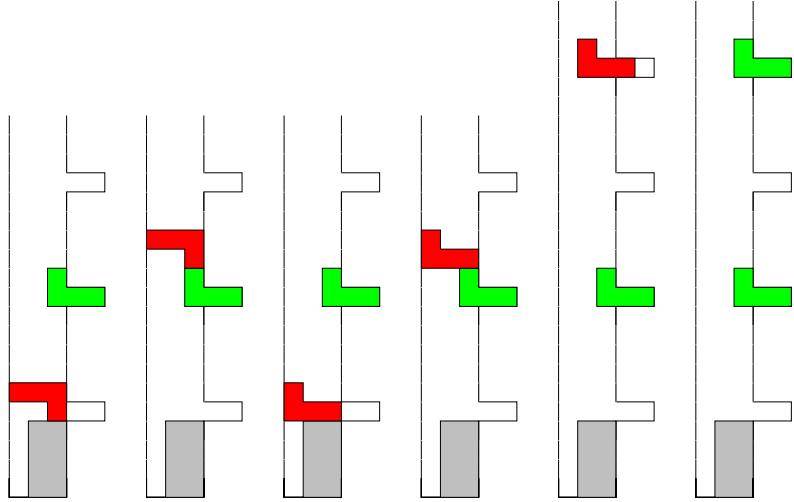
\begin{tabular}{llllllllllllll}
  \begin{block}{5}{10}
    \column{5}{20}
    \piece{\floor[lightgray]044}00
    \piece{\LGr[green]}2{10}
    \piece{\LGl[red]}04
  \end{block}
  &
  \begin{block}{5}{10}
    \column{5}{20}
    \piece{\floor[lightgray]044}00
    \piece{\LGr[green]}2{10}
    \piece{\LGl[red]}0{12}
  \end{block}
  &
  \begin{block}{5}{20}
    \column{5}{20}
    \piece{\floor[lightgray]044}00
    \piece{\LGr[green]}2{10}
    \piece{\LGr[red]}04
  \end{block}
  &
  \begin{block}{5}{20}
    \column{5}{20}
    \piece{\floor[lightgray]044}00
    \piece{\LGr[green]}2{10}
    \piece{\LGr[red]}0{12}
  \end{block}
  &
  \begin{block}{5}{26}
    \column{5}{26}
    \piece{\floor[lightgray]044}00
    \piece{\LGr[green]}2{10}
    \piece{\LGr[red]}1{22}
  \end{block}
  &
  \begin{block}{5}{26}
    \column{5}{26}
    \piece{\floor[lightgray]044}00
    \piece{\LGr[green]}2{10}
    \piece{\LGr[green]}2{22}
  \end{block}
\end{tabular}
\else
\includegraphics{figures/notched/lgleftH.epsi.clean}
\fi

  \end{center}
  (The $\tLG$ is initially in the $i$th-highest notch; in the last two
  configurations, the $j$th-highest notch is also (partially) filled
  by a new $\tLG$, for any $j\not=i$ and $j\ge 1$.  If $i=1$, the
  first and third configurations are impossible.)
  
  All but the last of these has a hole (and the fifth has a hole
  regardless of the value of $j$). The last configuration is
  \lgpreppedlg{i}{j}.

  \bigskip
  
  When the given \lgprepped{i} bucket has the empty rows in the third
  column, the possible configurations are as follows.  When the $\tLG$
  is placed vertically:
  \begin{center}
    \ifredraw
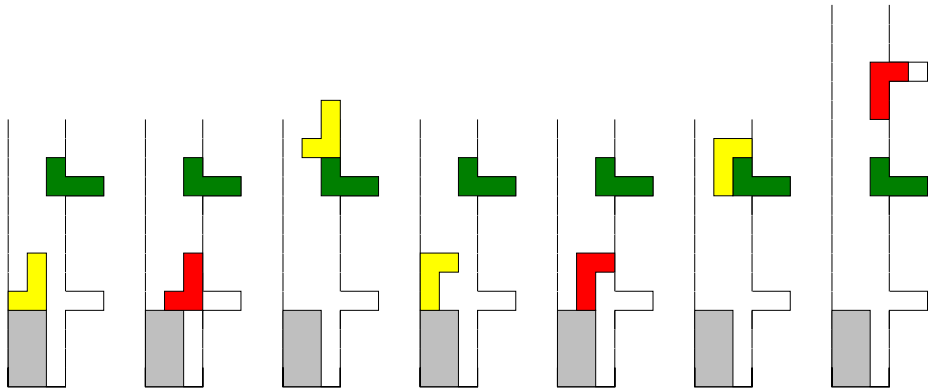
\begin{tabular}{llllllllllllll}
  \begin{block}{5}{14}
    \column{5}{14}
    \piece{\floor[lightgray]440}00
    \piece{\LGr[darkgreen]}2{10}
    \piece{\LGu[yellow]}04
  \end{block}
  &
  \begin{block}{5}{14}
    \column{5}{14}
    \piece{\floor[lightgray]440}00
    \piece{\LGr[darkgreen]}2{10}
    \piece{\LGu[red]}14
  \end{block}
  &
  \begin{block}{5}{14}
    \column{5}{14}
    \piece{\floor[lightgray]440}00
    \piece{\LGr[darkgreen]}2{10}
    \piece{\LGu[yellow]}1{12}
  \end{block}
  &
  \begin{block}{5}{14}
    \column{5}{14}
    \piece{\floor[lightgray]440}00
    \piece{\LGr[darkgreen]}2{10}
    \piece{\LGd[yellow]}04
  \end{block}
  &
  \begin{block}{5}{14}
    \column{5}{14}
    \piece{\floor[lightgray]440}00
    \piece{\LGr[darkgreen]}2{10}
    \piece{\LGd[red]}14
  \end{block}
  &
  \begin{block}{5}{14}
    \column{5}{14}
    \piece{\floor[lightgray]440}00
    \piece{\LGr[darkgreen]}2{10}
    \piece{\LGd[yellow]}1{10}
  \end{block}
  &
  \begin{block}{5}{20}
    \column{5}{20}
    \piece{\floor[lightgray]440}00
    \piece{\LGr[darkgreen]}2{10}
    \piece{\LGd[red]}2{14}
  \end{block}\\\\
\end{tabular}
\else
\includegraphics{figures/notched/lgrightV.epsi.clean}
\fi

  \end{center}
  (The notch initially filled with the $\tLG$ is the $i$th-highest for
  any $i\ge 1$; in the last configuration, the $j$th-highest notch is
  also filled for any $j\not=i$ and $j\geq 1$.  If $i=1$, then there
  is a hole initially, and all configurations are invalid.)
  
  The first has a \spurn.  The second, fifth, and seventh (for any
  $j$) have holes.  The third, fourth, and sixth all have a \bXXO.

  And, now, those in which the $\tLG$ is placed horizontally:
  \begin{center}
    \ifredraw
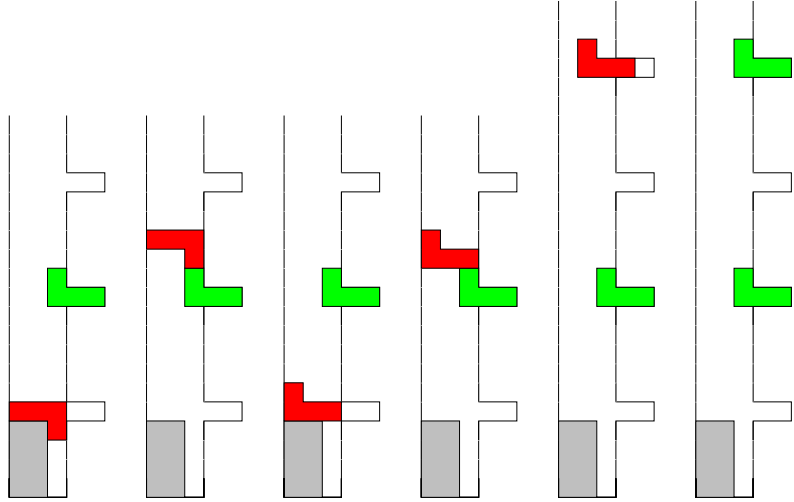
\begin{tabular}{llllllllllllll}
  \begin{block}{5}{10}
    \column{5}{20}
    \piece{\floor[lightgray]440}00
    \piece{\LGr[green]}2{10}
    \piece{\LGl[red]}03
  \end{block}
  &
  \begin{block}{5}{10}
    \column{5}{20}
    \piece{\floor[lightgray]440}00
    \piece{\LGr[green]}2{10}
    \piece{\LGl[red]}0{12}
  \end{block}
  &
  \begin{block}{5}{20}
    \column{5}{20}
    \piece{\floor[lightgray]440}00
    \piece{\LGr[green]}2{10}
    \piece{\LGr[red]}04
  \end{block}
  &
  \begin{block}{5}{20}
    \column{5}{20}
    \piece{\floor[lightgray]440}00
    \piece{\LGr[green]}2{10}
    \piece{\LGr[red]}0{12}
  \end{block}
  &
  \begin{block}{5}{26}
    \column{5}{26}
    \piece{\floor[lightgray]440}00
    \piece{\LGr[green]}2{10}
    \piece{\LGr[red]}1{22}
  \end{block}
  &
  \begin{block}{5}{26}
    \column{5}{26}
    \piece{\floor[lightgray]440}00
    \piece{\LGr[green]}2{10}
    \piece{\LGr[green]}2{22}
  \end{block}
\end{tabular}
\else
\includegraphics{figures/notched/lgrightH.epsi.clean}
\fi

  \end{center}
  (The $\tLG$ is initially in the $i$th-highest notch for any $i\ge
  1$; in the last two configurations, the $j$th-highest notch is
  (partially) filled by a new $\tLG$, for any $j\not=i$.  If $i =1$,
  there is a hole initially, and all configurations are invalid.)
  
  All but the last of these has a hole (and the next-to-last has a
  hole regardless of the value of $j$).  The last configuration is
  \lgpreppedlg{i}{j}.
\end{proof}

\begin{proposition} \mylabel{prop:sq-notched-two}
  There is no valid move for $\tSq$ in \lgpreppedlg{i}{j}.
\end{proposition}
\begin{proof}
  The possible configurations are:
  \begin{center}
    \ifredraw
\begin{tabular}{lllllllll}
\begin{block}{5}{26}
    \column{5}{26}
    \piece{\floor[lightgray]044}00
    \piece{\LGr[lightgray]}2{10}
    \piece{\LGr[lightgray]}2{22}
    \piece{\Sq[red]}04
  \end{block}
&
\begin{block}{5}{26}
    \column{5}{26}
    \piece{\floor[lightgray]044}00
    \piece{\LGr[lightgray]}2{10}
    \piece{\LGr[lightgray]}2{22}
    \piece{\Sq[red]}14
  \end{block}
&
\begin{block}{5}{26}
    \column{5}{26}
    \piece{\floor[lightgray]044}00
    \piece{\LGr[lightgray]}2{10}
    \piece{\LGr[lightgray]}2{22}
    \piece{\Sq[yellow]}1{12}
  \end{block}
&
  \begin{block}{5}{26}
    \column{5}{26}
    \piece{\floor[lightgray]044}00
    \piece{\LGr[lightgray]}2{10}
    \piece{\LGr[lightgray]}2{22}
    \piece{\Sq[yellow]}1{24}
  \end{block}
&
~~~~~~
&
\begin{block}{5}{26}
    \column{5}{26}
    \piece{\floor[lightgray]440}00
    \piece{\LGr[lightgray]}2{10}
    \piece{\LGr[lightgray]}2{22}
    \piece{\Sq[yellow]}04
  \end{block}
&
\begin{block}{5}{26}
    \column{5}{26}
    \piece{\floor[lightgray]440}00
    \piece{\LGr[lightgray]}2{10}
    \piece{\LGr[lightgray]}2{22}
    \piece{\Sq[red]}14
  \end{block}
&
\begin{block}{5}{26}
    \column{5}{26}
    \piece{\floor[lightgray]440}00
    \piece{\LGr[lightgray]}2{10}
    \piece{\LGr[lightgray]}2{22}
    \piece{\Sq[yellow]}1{12}
  \end{block}
&
  \begin{block}{5}{26}
    \column{5}{26}
    \piece{\floor[lightgray]440}00
    \piece{\LGr[lightgray]}2{10}
    \piece{\LGr[lightgray]}2{22}
    \piece{\Sq[yellow]}1{24}
  \end{block}
\\\\
\end{tabular}
\else
\includegraphics{figures/notched/sqlg.epsi.clean}
\fi

  \end{center}
  (The initially filled notches are the $i$th- and $j$th-highest, for
  $i, j \ge 1$ and $i\not= j$.  The second configuration is blocked
  for $\min(i,j) = 1$.  The fifth, sixth, seventh, and eighth
  configurations have holes and are thus invalid if $\min(i,j) = 1$.)
  
  In the first, second, and sixth configurations, there are holes.
  The third has a \btwo for $\min(i,j) =1$ and a \bOXX for $\min(i,j)
  >1$.  The fourth has a \bfive for $\max(i,j) = 2$, a \bgap for
  $\min(i,j) =1$ and $\max(i,j) >2$, and a \bOXX for $\min(i,j) >1$.
  In the fifth configuration, there is a \spurn.  The seventh and
  eighth each have a \bXXO for $\min(i,j) >1$.
\end{proof}


\subsubsection{\Splat Buckets}

\begin{proposition} \mylabel{prop:splat-tplat}
  If $\tSq$ is validly placed in a \splat bucket, then the
  result is a \tplat.
\end{proposition}
\begin{proof}
  The only possible configurations are
  \begin{center}
    \ifredraw
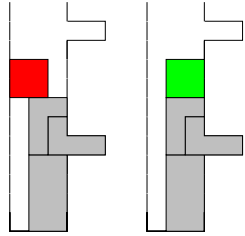
\begin{tabular}{llll}
\begin{block}{5}{12}
  \column{5}{12}
    \piece{\floor[lightgray]044}00
    \piece{\LGr[lightgray]}24
    \piece{\LGd[lightgray]}14
    \piece{\Sq[red]}07
  \end{block}
&
\begin{block}{5}{12}
  \column{5}{12}
    \piece{\floor[lightgray]044}00
    \piece{\LGr[lightgray]}24
    \piece{\LGd[lightgray]}14
    \piece{\Sq[green]}17
  \end{block}
\\\\
\end{tabular}
\else
\includegraphics{figures/rplat/create.epsi.clean}
\fi

  \end{center}
  
  The first has a hole; the second is a \tplat, as desired.
\end{proof}

\subsubsection{\Tplat Buckets}

\begin{proposition}
  \mylabel{prop:ls-rplat}  
  There is no valid move for $\tLS$ in a \tplat bucket.
\end{proposition}
\begin{proof}
  The possible configurations are the following:
  \begin{center}
    \ifredraw
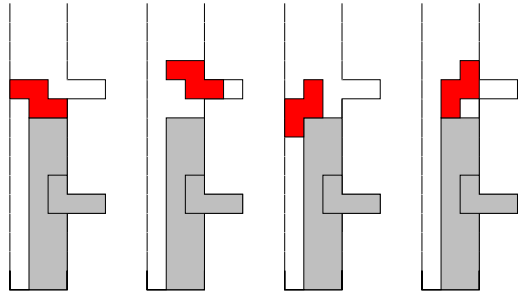
\begin{tabular}{llll}
\begin{block}{5}{15}
  \column{5}{15}
  \piece{\floor[lightgray]099}00
  \piece{\LGr[lightgray]}24
  \piece{\LSr[red]}09
\end{block}
&
\begin{block}{5}{15}
  \column{5}{15}
  \piece{\floor[lightgray]099}00
  \piece{\LGr[lightgray]}24
  \piece{\LSr[red]}1{10}
\end{block}
&
\begin{block}{5}{15}
  \column{5}{15}
  \piece{\floor[lightgray]099}00
  \piece{\LGr[lightgray]}24
  \piece{\LSu[red]}08
\end{block}
&
\begin{block}{5}{15}
  \column{5}{15}
  \piece{\floor[lightgray]099}00
  \piece{\LGr[lightgray]}24
  \piece{\LSu[red]}19
\end{block}
\\\\
\end{tabular}
\else
\includegraphics{figures/rplat/lsrplat.epsi.clean}
\fi

  \end{center}
  (The second configuration has the $\tLS$ in the $i$th-highest notch,
  for any $i \ge 1$.)
  
  All four have holes.
\end{proof}

\begin{proposition}
  \mylabel{prop:sq-rplat}  
  There is no valid move for $\tSq$ in a \tplat bucket.
\end{proposition}
\begin{proof}
  The possible configurations are the following:
  \begin{center}
    \ifredraw
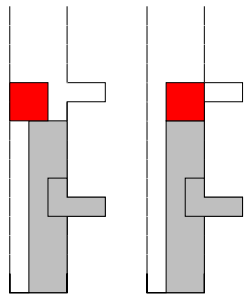
\begin{tabular}{ll}
\begin{block}{5}{15}
  \column{5}{15}
  \piece{\floor[lightgray]099}00
  \piece{\LGr[lightgray]}24
  \piece{\Sq[red]}09
\end{block}
&
\begin{block}{5}{15}
  \column{5}{15}
  \piece{\floor[lightgray]099}00
  \piece{\LGr[lightgray]}24
  \piece{\Sq[red]}19
\end{block}
\\\\
\end{tabular}
\else
\includegraphics{figures/rplat/sqrplat.epsi.clean}
\fi

  \end{center}
  Both have holes.
\end{proof}

\begin{proposition}
  \mylabel{prop:lg-rplat} If $\tLG$ is placed validly in a \tplat
  bucket, the result is a \lgtplat{i} for some $i$.
\end{proposition}
\begin{proof}
  The following configurations are possible:
  \begin{center}
    \ifredraw
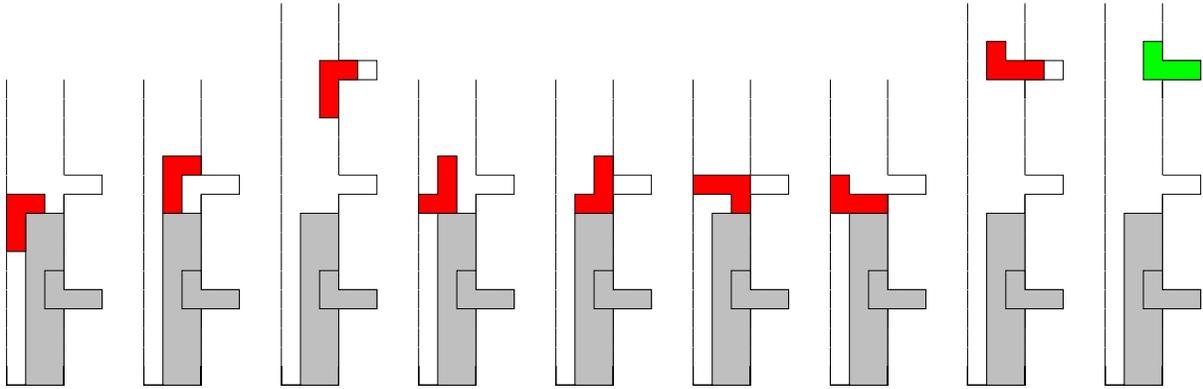
\begin{tabular}{lllllllll}
\begin{block}{5}{10}
  \column{5}{16}
  \piece{\floor[lightgray]099}00
  \piece{\LGr[lightgray]}24
  \piece{\LGd[red]}07
\end{block}
&
\begin{block}{5}{10}
  \column{5}{16}
  \piece{\floor[lightgray]099}00
  \piece{\LGr[lightgray]}24
  \piece{\LGd[red]}19
\end{block}
&
\begin{block}{5}{20}
  \column{5}{20}
  \piece{\floor[lightgray]099}00
  \piece{\LGr[lightgray]}24
  \piece{\LGd[red]}2{14}
\end{block}
&
\begin{block}{5}{10}
  \column{5}{16}
  \piece{\floor[lightgray]099}00
  \piece{\LGr[lightgray]}24
  \piece{\LGu[red]}09
\end{block}
&
\begin{block}{5}{10}
  \column{5}{16}
  \piece{\floor[lightgray]099}00
  \piece{\LGr[lightgray]}24
  \piece{\LGu[red]}19
\end{block}
&
\begin{block}{5}{10}
  \column{5}{16}
  \piece{\floor[lightgray]099}00
  \piece{\LGr[lightgray]}24
  \piece{\LGl[red]}09
\end{block}
&
\begin{block}{5}{10}
  \column{5}{16}
  \piece{\floor[lightgray]099}00
  \piece{\LGr[lightgray]}24
  \piece{\LGr[red]}09
\end{block}
&
\begin{block}{5}{16}
  \column{5}{20}
  \piece{\floor[lightgray]099}00
  \piece{\LGr[lightgray]}24
  \piece{\LGr[red]}1{16}
\end{block}
&
\begin{block}{5}{16}
  \column{5}{20}
  \piece{\floor[lightgray]099}00
  \piece{\LGr[lightgray]}24
  \piece{\LGr[green]}2{16}
\end{block}
\\\\
\end{tabular}
\else
\includegraphics{figures/rplat/lgrplat.epsi.clean}
\fi

  \end{center}
  (In the third configuration, the $\tLG$ is placed in the
  $i$th-highest bucket, for $i>1$, and in the eighth and ninth, for
  $i\geq 1$.)

  Of these, there is a hole in all but the last.  The last is a
  \lgtplat{i}.
\end{proof}

\begin{proposition}
  \mylabel{prop:ls-rplat-notched} There is no valid placement of
  $\tLS$ in \lgtplat{i} for any $i$.
\end{proposition}
\begin{proof}
  The following configurations are possible:
  \begin{center}
    \ifredraw
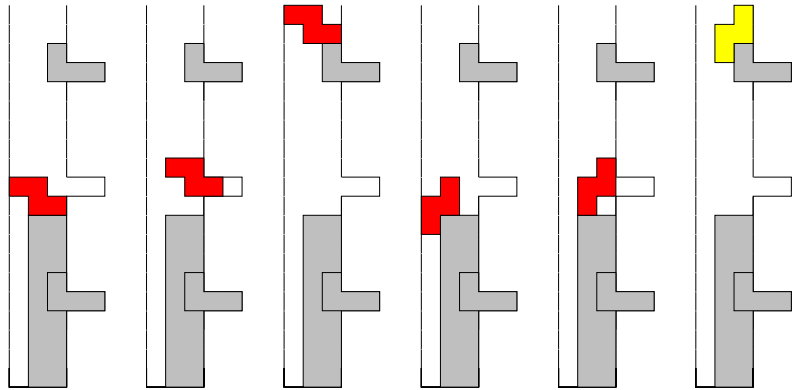
\begin{tabular}{lllllllll}
\begin{block}{5}{20}
  \column{5}{20}
  \piece{\floor[lightgray]099}00
  \piece{\LGr[lightgray]}24
  \piece{\LGr[lightgray]}2{16}
  \piece{\LSl[red]}09
\end{block}
&
\begin{block}{5}{20}
  \column{5}{20}
  \piece{\floor[lightgray]099}00
  \piece{\LGr[lightgray]}24
  \piece{\LGr[lightgray]}2{16}
  \piece{\LSl[red]}1{10}
\end{block}
&
\begin{block}{5}{20}
  \column{5}{20}
  \piece{\floor[lightgray]099}00
  \piece{\LGr[lightgray]}24
  \piece{\LGr[lightgray]}2{16}
  \piece{\LSl[red]}0{18}
\end{block}
&
\begin{block}{5}{20}
  \column{5}{20}
  \piece{\floor[lightgray]099}00
  \piece{\LGr[lightgray]}24
  \piece{\LGr[lightgray]}2{16}
  \piece{\LSu[red]}08
\end{block}
&
\begin{block}{5}{20}
  \column{5}{20}
  \piece{\floor[lightgray]099}00
  \piece{\LGr[lightgray]}24
  \piece{\LGr[lightgray]}2{16}
  \piece{\LSu[red]}19
\end{block}
&
\begin{block}{5}{20}
  \column{5}{20}
  \piece{\floor[lightgray]099}00
  \piece{\LGr[lightgray]}24
  \piece{\LGr[lightgray]}2{16}
  \piece{\LSu[yellow]}1{17}
\end{block}
\\\\
\end{tabular}
\else
\includegraphics{figures/rplat/lsrplatnotch.epsi.clean}
\fi

  \end{center}
  (Initially the $i$th-highest notch is filled by the $\tLG$.  The
  second configuration denotes the $\tLS$ in the $j$th-highest notch,
  for any $j \not= i$.  The fifth configuration is blocked for $i=1$.)
  
  Of these, there is a hole in all but the last, which has a \bmod (or
  a \btwo if $i=1$).
\end{proof}


\subsubsection{\Underflat Buckets}

\begin{proposition}   \mylabel{prop:lg-underflat}
  If $\tLG$ is dropped into an underflat bucket, it produces
  \lgunderflat{i} for some $i$.
\end{proposition}
\begin{proof}
  The possible configurations are the following:
 \begin{center}
   \ifredraw
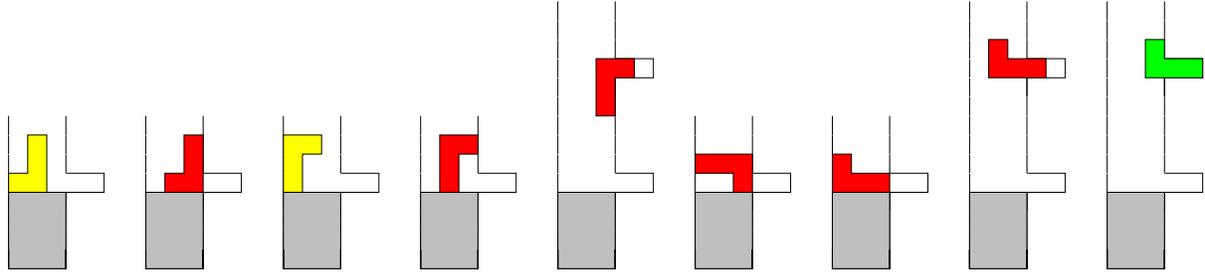
\begin{tabular}{cccccccccc}
  \begin{block}{5}{8}
    \column{5}{8}
    \piece{\floor[lightgray]444}00
    \piece{\LGu[yellow]}04
  \end{block}
  &
  \begin{block}{5}{8}
    \column{5}{8}
    \piece{\floor[lightgray]444}00
    \piece{\LGu[red]}14
  \end{block}
  &
  \begin{block}{5}{8}
    \column{5}{8}
    \piece{\floor[lightgray]444}00
    \piece{\LGd[yellow]}04
  \end{block}
  &
  \begin{block}{5}{8}
    \column{5}{8}
    \piece{\floor[lightgray]444}00
    \piece{\LGd[red]}14
  \end{block}
  &
  \begin{block}{5}{14}
    \column{5}{14}
    \piece{\floor[lightgray]444}00
    \piece{\LGd[red]}2{8}
  \end{block}
  &
  \begin{block}{5}{10}
    \column{5}{8}
    \piece{\floor[lightgray]444}00
    \piece{\LGl[red]}04
  \end{block}
  &
  \begin{block}{5}{8}
    \column{5}{8}
    \piece{\floor[lightgray]444}00
    \piece{\LGr[red]}04
  \end{block}
  &
  \begin{block}{5}{14}
    \column{5}{14}
    \piece{\floor[lightgray]444}00
    \piece{\LGr[red]}1{10}
  \end{block}
  &
  \begin{block}{5}{14}
    \column{5}{14}
    \piece{\floor[lightgray]444}00
    \piece{\LGr[green]}2{10}
  \end{block}\\
\end{tabular}
\else
\includegraphics{figures/underflat/lg.epsi.clean}
\fi


 \end{center}
 (The fifth of these applies to the $i$th-highest notch for any
 $i>1$, the eighth and ninth for any $i\geq 1$.)
 
 The first configuration creates a \spurn, and the third has a \bXXX.
 Of the others, all but the last---\lgunderflat{i}---create holes.
\end{proof}

\begin{proposition}  \mylabel{prop:sq-underflat}
  There is no valid move for $\tSq$ in an underflat bucket.
\end{proposition}
\begin{proof}
  The possible configurations are the following:
  \begin{center}
    \ifredraw
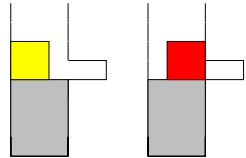
\begin{tabular}{cc}
\begin{block}{5}{8}
  \column{5}{8}
  \piece{\floor[lightgray]444}00
  \piece{\Sq[yellow]}04
\end{block}
&
\begin{block}{5}{8}
  \column{5}{8}
  \piece{\floor[lightgray]444}00
  \piece{\Sq[red]}14
\end{block}\\\\
\end{tabular}
\else
\includegraphics{figures/underflat/sq.epsi.clean}
\fi


  \end{center}
  The first of these configurations has a \spurn, and the second has a
  hole.
\end{proof}

\begin{proposition}  \mylabel{prop:ls-underflat}
  There is no valid move for $\tLS$ in an underflat bucket.
\end{proposition}
\begin{proof}
  The possible configurations are the following:
  \begin{center}
    \ifredraw
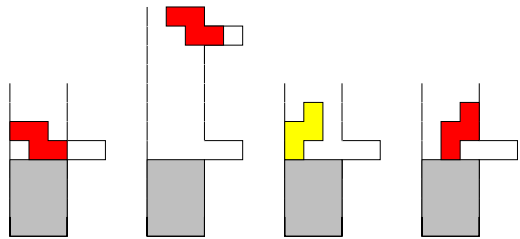
\begin{tabular}{cccc}
\begin{block}{5}{8}
  \column{5}{8}
  \piece{\floor[lightgray]444}00
  \piece{\LSl[red]}04
\end{block}
&
\begin{block}{5}{12}
  \column{5}{12}
  \piece{\floor[lightgray]444}00
  \piece{\LSl[red]}1{10}
\end{block}
&
\begin{block}{5}{8}
  \column{5}{8}
  \piece{\floor[lightgray]444}00
  \piece{\LSu[yellow]}04
\end{block}
&
\begin{block}{5}{8}
  \column{5}{8}
  \piece{\floor[lightgray]444}00
  \piece{\LSu[red]}14
\end{block}\\\\
\end{tabular}
\else
\includegraphics{figures/underflat/ls.epsi.clean}
\fi


  \end{center}
  (The second has the $\tLS$ in the $i$th-highest notch, for any $i
  \ge 1$.)
  
  The first, second, and fourth of these configurations have holes;
  the second has a \bXXX.
\end{proof}

\begin{proposition}  \mylabel{prop:sq-ufnotched}
  If $\tSq$ is dropped validly into \lgunderflat{i}, then (1) $i=1$,
  and (2) the result is an overflat bucket.
\end{proposition}
\begin{proof}  The possible configurations are as follows:
  \begin{center}
    \ifredraw
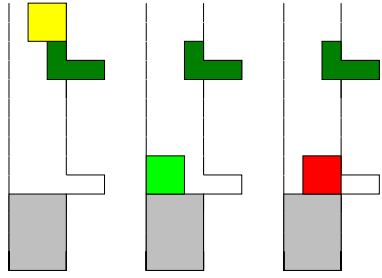
      \begin{tabular}{ccccc}
        \begin{block}{5}{14}
          \column{5}{14}
          \piece{\floor[lightgray]444}00
          \piece{\LGr[darkgreen]}2{10}
          \piece{\Sq[yellow]}1{12}
        \end{block}
        &
        \begin{block}{5}{14}
          \column{5}{14}
          \piece{\floor[lightgray]444}00
          \piece{\LGr[darkgreen]}2{10}
          \piece{\Sq[green]}04
        \end{block}
        &
        \begin{block}{5}{14}
          \column{5}{14}
          \piece{\floor[lightgray]444}00
          \piece{\LGr[darkgreen]}2{10}
          \piece{\Sq[red]}14
        \end{block}\\\\
      \end{tabular}
\else
\includegraphics{figures/underflat/sqlg.epsi.clean}
\fi


  \end{center}
  (In each of these configurations, the $i$th-highest notch is
  initially filled with the $\tLG$, for any $i\geq 1$.  The third
  configuration is impossible if $i=1$.)
  
  The first of these is a \bXXX for $i>1$ and a \btwo for $i=1$.  The
  second is a \spurn unless the lowest notch is filled, i.e., unless
  $i=1$; in this case, the reuslt is \overflat.  The third
  configuration has a hole.
\end{proof}


\subsubsection{\Overflat Buckets}

\begin{proposition}  \mylabel{prop:lg-overflat}
  If $\tLG$ is validly placed in an \overflat bucket, the result is
  \lgoverflatA, \lgoverflatB, \lgoverflatC, or \lgoverflatD{i} for
  some $i$.
\end{proposition}
\begin{proof}
  The possible configurations for placement of the $\tLG$ are as
  follows:

  \begin{center}
    \ifredraw
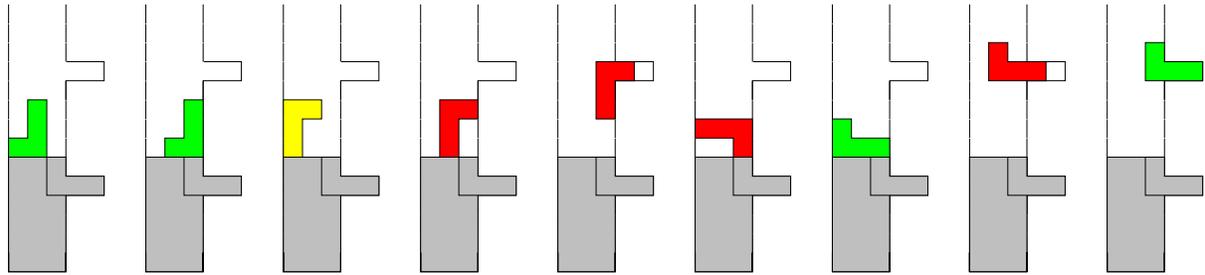
    \begin{tabular}{llllllllllllll}
      \begin{block}{5}{14}
        \column{5}{14}
        \piece{\floor[lightgray]666}00
        \piece{\LGr[lightgray]}24
        \piece{\LGu[green]}06
      \end{block}
      &
      \begin{block}{5}{14}
        \column{5}{14}
        \piece{\floor[lightgray]666}00
        \piece{\LGr[lightgray]}24
        \piece{\LGu[green]}16
      \end{block}
&
      \begin{block}{5}{14}
        \column{5}{14}
        \piece{\floor[lightgray]666}00
        \piece{\LGr[lightgray]}24
        \piece{\LGd[yellow]}06
      \end{block}
      &
      \begin{block}{5}{14}
        \column{5}{14}
        \piece{\floor[lightgray]666}00
        \piece{\LGr[lightgray]}24
        \piece{\LGd[red]}16
      \end{block}
      &
      \begin{block}{5}{14}
        \column{5}{14}
        \piece{\floor[lightgray]666}00
        \piece{\LGr[lightgray]}24
        \piece{\LGd[red]}2{8}
      \end{block}
&
      \begin{block}{5}{14}
        \column{5}{14}
        \piece{\floor[lightgray]666}00
        \piece{\LGr[lightgray]}24
        \piece{\LGl[red]}06
      \end{block}
&
      \begin{block}{5}{14}
        \column{5}{14}
        \piece{\floor[lightgray]666}00
        \piece{\LGr[lightgray]}24
        \piece{\LGr[green]}06
      \end{block}
      &
      \begin{block}{5}{14}
        \column{5}{14}
        \piece{\floor[lightgray]666}00
        \piece{\LGr[lightgray]}24
        \piece{\LGr[red]}1{10}
      \end{block}
      &
      \begin{block}{5}{14}
        \column{5}{14}
        \piece{\floor[lightgray]666}00
        \piece{\LGr[lightgray]}24
        \piece{\LGr[green]}2{10}
      \end{block}\\\\
    \end{tabular}
\else
\includegraphics{figures/overflat/lg.epsi.clean}
\fi


  \end{center}
  (The fifth, eight, and ninth denote the placement of the $\tLG$ in
  the $i$th-highest notch, for any $i\ge1$.)
  
  The first two configurations are \lgoverflatA and \lgoverflatB.  The
  third has a \btwo in the second column; the fourth, fifth (for any
  $i$), sixth and eighth (for any $i$) have holes.  The seventh is
  \lgoverflatC, and the last is \lgoverflatD{i}.
\end{proof}

\begin{proposition}  \mylabel{prop:ls-overflat}
  There is no valid move for $\tLS$ in an \overflat bucket.
\end{proposition}
\begin{proof}
  The possible configurations are:
  \begin{center}
    \ifredraw
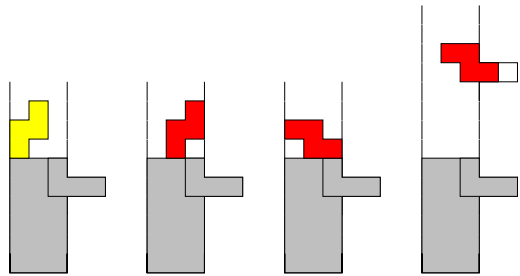
\begin{tabular}{ccccccccc}
  \begin{block}{5}{10}
    \column{5}{10}
    \piece{\floor[lightgray]666}00
    \piece{\LGr[lightgray]}24
    \piece{\LSu[yellow]}06
  \end{block}
  &
  \begin{block}{5}{10}
    \column{5}{10}
    \piece{\floor[lightgray]666}00
    \piece{\LGr[lightgray]}24
    \piece{\LSu[red]}16
  \end{block}
  &
  \begin{block}{5}{10}
    \column{5}{10}
    \piece{\floor[lightgray]666}00
    \piece{\LGr[lightgray]}24
    \piece{\LSl[red]}06
  \end{block}
  &
  \begin{block}{5}{14}
    \column{5}{14}
    \piece{\floor[lightgray]666}00
    \piece{\LGr[lightgray]}24
    \piece{\LSl[red]}1{10}
  \end{block}\\\\
\end{tabular}
\else
\includegraphics{figures/overflat/ls.epsi.clean}
\fi


  \end{center}
  (The fourth represents the placement of the $\tLS$ into the $i$th
  notch for any $i\ge1$.)
  
  The first of these configurations has a \bone; the remaining three
  have holes.
\end{proof}

\begin{proposition}  \mylabel{prop:sqsq-overflat}
  If $\tup{\tSq,\tSq}$ is validly placed in an \overflat bucket, then
  the result is an \unprepped bucket.
\end{proposition}
\begin{proof}
  The possible configurations after the placement of the first $\tSq$
  are as follows:
  \begin{center}
    \ifredraw
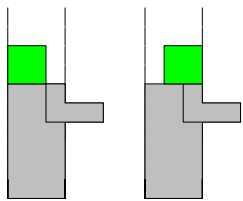
\begin{tabular}{ccccccccc}
  \begin{block}{5}{10}
    \column{5}{10}
    \piece{\floor[lightgray]666}00
    \piece{\LGr[lightgray]}24
    \piece{\Sq[green]}06
  \end{block}
  &
  \begin{block}{5}{10}
    \column{5}{10}
    \piece{\floor[lightgray]666}00
    \piece{\LGr[lightgray]}24
    \piece{\Sq[green]}16
  \end{block}
\\\\
\end{tabular}
\else
\includegraphics{figures/overflat/sq.epsi.clean}
\fi


  \end{center}
  Both of these configurations are valid.  Now, when we place the
  second $\tSq$, the possible results are the following:
  \begin{center}
    \ifredraw
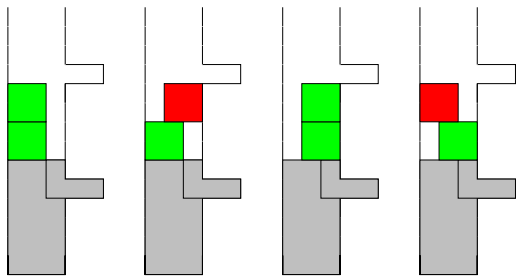
\begin{tabular}{ccccccccc}
  \begin{block}{5}{14}
    \column{5}{14}
    \piece{\floor[lightgray]666}00
    \piece{\LGr[lightgray]}24
    \piece{\Sq[green]}06
    \piece{\Sq[green]}08
  \end{block}
  &
  \begin{block}{5}{14}
    \column{5}{14}
    \piece{\floor[lightgray]666}00
    \piece{\LGr[lightgray]}24
    \piece{\Sq[green]}06
    \piece{\Sq[red]}18
  \end{block}
  &
  \begin{block}{5}{14}
    \column{5}{14}
    \piece{\floor[lightgray]666}00
    \piece{\LGr[lightgray]}24
    \piece{\Sq[green]}16
    \piece{\Sq[green]}18
  \end{block}
  &
  \begin{block}{5}{14}
    \column{5}{14}
    \piece{\floor[lightgray]666}00
    \piece{\LGr[lightgray]}24
    \piece{\Sq[green]}16
    \piece{\Sq[red]}08
  \end{block}
\\\\
\end{tabular}
\else
\includegraphics{figures/overflat/sqsq.epsi.clean}
\fi


  \end{center}
  The second and fourth configurations have holes; the first and third
  are both \unprepped.
\end{proof}

\begin{proposition}  \mylabel{prop:ls-over-1}
  There is no valid move for $\tLS$ in \lgoverflatA.
\end{proposition}
\begin{proof}
  The only possible configurations are as follows:
  \begin{center}
    \ifredraw
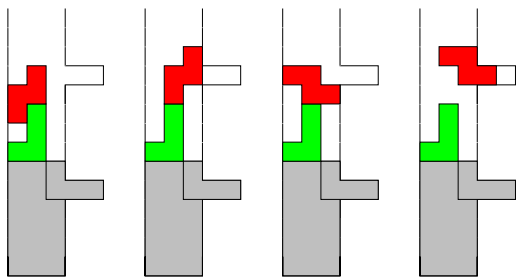
\begin{tabular}{cccccccccc}
  \begin{block}{5}{14}
    \column{5}{14}
    \piece{\floor[lightgray]666}00
    \piece{\LGr[lightgray]}24
    \piece{\LGu[green]}06
    \piece{\LSu[red]}08
  \end{block}
  &
  \begin{block}{5}{14}
    \column{5}{14}
    \piece{\floor[lightgray]666}00
    \piece{\LGr[lightgray]}24
    \piece{\LGu[green]}06
    \piece{\LSu[red]}19
  \end{block}
  &
  \begin{block}{5}{14}
    \column{5}{14}
    \piece{\floor[lightgray]666}00
    \piece{\LGr[lightgray]}24
    \piece{\LGu[green]}06
    \piece{\LSr[red]}09
  \end{block}
  &
  \begin{block}{5}{14}
    \column{5}{14}
    \piece{\floor[lightgray]666}00
    \piece{\LGr[lightgray]}24
    \piece{\LGu[green]}06
    \piece{\LSr[red]}1{10}
  \end{block}\\\\
\end{tabular}
\else
\includegraphics{figures/overflat/lslgA.epsi.clean}
\fi


  \end{center}
  (The fourth denotes the placement of the $\tLS$ into the
  $i$th-highest notch, for any $i\ge1$.)
  
  All of these configurations have holes.
\end{proof}

\begin{proposition}  \mylabel{prop:ls-over-2}
  There is no valid placement of $\tLS$ into \lgoverflatB.
\end{proposition}
\begin{proof}
  Then the possible configurations are the following:
  \begin{center}
    \ifredraw
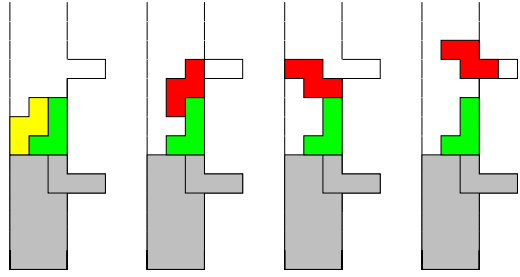
\begin{tabular}{cccccccccc}
  \begin{block}{5}{14}
    \column{5}{14}
    \piece{\floor[lightgray]666}00
    \piece{\LGr[lightgray]}24
    \piece{\LGu[green]}16
    \piece{\LSu[yellow]}06
  \end{block}
  &
  \begin{block}{5}{14}
    \column{5}{14}
    \piece{\floor[lightgray]666}00
    \piece{\LGr[lightgray]}24
    \piece{\LGu[green]}16
    \piece{\LSu[red]}18
  \end{block}
  &
  \begin{block}{5}{14}
    \column{5}{14}
    \piece{\floor[lightgray]666}00
    \piece{\LGr[lightgray]}24
    \piece{\LGu[green]}16
    \piece{\LSr[red]}09
  \end{block}
  &
  \begin{block}{5}{14}
    \column{5}{14}
    \piece{\floor[lightgray]666}00
    \piece{\LGr[lightgray]}24
    \piece{\LGu[green]}16
    \piece{\LSr[red]}1{10}
  \end{block}\\\\
\end{tabular}
\else
\includegraphics{figures/overflat/lslgB.epsi.clean}
\fi


  \end{center}
  (The fourth denotes the placement of the $\tLS$ into the
  $i$th-highest notch, for any $i\geq 1$.)
  
  The first configuration is \unapproachable, and the other three have
  holes.
\end{proof}

\begin{proposition}  \mylabel{prop:ls-over-3}
  If $\tLS$ is placed validly in \lgoverflatC, the 
  result is a \thappy bucket.
\end{proposition}
\begin{proof}
  The only possible configurations are the following:
  \begin{center}
    \ifredraw
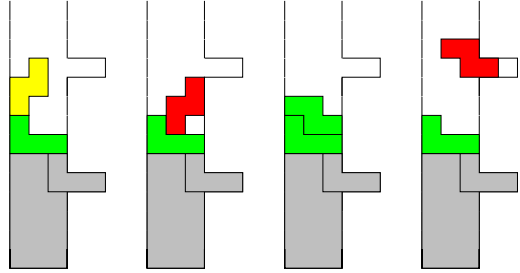
\begin{tabular}{cccccccccc}
  \begin{block}{5}{14}
    \column{5}{14}
    \piece{\floor[lightgray]666}00
    \piece{\LGr[lightgray]}24
    \piece{\LGr[green]}06
    \piece{\LSu[yellow]}08
  \end{block}
  &
  \begin{block}{5}{14}
    \column{5}{14}
    \piece{\floor[lightgray]666}00
    \piece{\LGr[lightgray]}24
    \piece{\LGr[green]}06
    \piece{\LSu[red]}17
  \end{block}
  &
  \begin{block}{5}{14}
    \column{5}{14}
    \piece{\floor[lightgray]666}00
    \piece{\LGr[lightgray]}24
    \piece{\LGr[green]}06
    \piece{\LSl[green]}07
  \end{block}
  &
  \begin{block}{5}{14}
    \column{5}{14}
    \piece{\floor[lightgray]666}00
    \piece{\LGr[lightgray]}24
    \piece{\LGr[green]}06
    \piece{\LSr[red]}1{10}
  \end{block}\\\\
\end{tabular}
\else
\includegraphics{figures/overflat/lslgC.epsi.clean}
\fi


  \end{center}
  (The fourth denotes the placement of the $\tLS$ into the
  $i$th-highest notch, for any $i\ge 1$.)
  
  The first has a \spurn, and the second and fourth have holes.  The
  third is \thappy, as desired.
\end{proof}

\begin{proposition}  \mylabel{prop:ls-over-4}
  There is no valid move for $\tLS$ in \lgoverflatD{i} for any $i$.
\end{proposition}
\begin{proof}
  Then the possible configurations are the following:
  \begin{center}
    \ifredraw
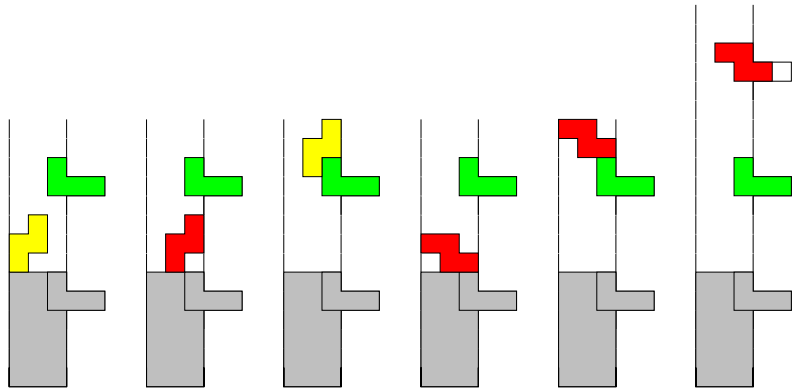
\begin{tabular}{cccccccccc}
  &
  \begin{block}{5}{14}
    \column{5}{14}
    \piece{\floor[lightgray]666}00
    \piece{\LGr[lightgray]}24
    \piece{\LGr[green]}2{10}
    \piece{\LSu[yellow]}06
  \end{block}
  &
  \begin{block}{5}{14}
    \column{5}{14}
    \piece{\floor[lightgray]666}00
    \piece{\LGr[lightgray]}24
    \piece{\LGr[green]}2{10}
    \piece{\LSu[red]}16
  \end{block}
  &
  \begin{block}{5}{14}
    \column{5}{14}
    \piece{\floor[lightgray]666}00
    \piece{\LGr[lightgray]}24
    \piece{\LGr[green]}2{10}
    \piece{\LSu[yellow]}1{11}
  \end{block}
  &
  \begin{block}{5}{14}
    \column{5}{14}
    \piece{\floor[lightgray]666}00
    \piece{\LGr[lightgray]}24
    \piece{\LGr[green]}2{10}
    \piece{\LSl[red]}06
  \end{block}
  &
  \begin{block}{5}{14}
    \column{5}{14}
    \piece{\floor[lightgray]666}00
    \piece{\LGr[lightgray]}24
    \piece{\LGr[green]}2{10}
    \piece{\LSl[red]}0{12}
  \end{block}
  &
  \begin{block}{5}{20}
    \column{5}{20}
    \piece{\floor[lightgray]666}00
    \piece{\LGr[lightgray]}24
    \piece{\LGr[green]}2{10}
    \piece{\LSr[red]}1{16}
  \end{block}\\\\
\end{tabular}
\else
\includegraphics{figures/overflat/lslgD.epsi.clean}
\fi


  \end{center}
  (Initially, in each configuration, there is an $\tLG$ in the
  $i$th-highest notch, for some $i\ge1$.  The sixth configuration
  denotes the placement of the $\tLS$ into the $j$th-highest notch,
  for any $j \not= i$ and $j\ge1$.)
  
  The first has a \bone.  The second, fourth, fifth, and sixth (for
  any $j$) have holes, and the third has a \bmod.
\end{proof}


\subsubsection{\Thappy Buckets}

\begin{proposition} \mylabel{prop:sq-thappy}
  There is no valid move for $\tSq$ in a \thappy bucket.
\end{proposition}
\begin{proof}
  The possible placements for a $\tSq$ in a \thappy bucket are as
  follows:
  \begin{center}
    \ifredraw
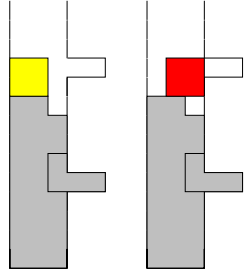
\begin{tabular}{llllllllllllll}
\begin{block}{5}{14}
  \column{5}{14}
  \piece{\floor[lightgray]998}00
  \piece{\LGr[lightgray]}24
  \piece{\Sq[yellow]}09
\end{block}
&
\begin{block}{5}{14}
  \column{5}{14}
  \piece{\floor[lightgray]998}00
  \piece{\LGr[lightgray]}24
  \piece{\Sq[red]}19
\end{block}\\\\
\end{tabular}
\else
\includegraphics{figures/thappy/sq.epsi.clean}
\fi


  \end{center}
  The first has a \spurn; the second has a hole.
\end{proof}

\begin{proposition}  \mylabel{prop:lg-thappy}
  If $\tLG$ is placed validly into a \thappy bucket, the resulting
  configuration is either \lgthappy{i} or \underflat.
\end{proposition}
\begin{proof}
  The possible configurations for the $\tLG$ in the \thappy bucket are
  as follows:
  \begin{center}
    \ifredraw
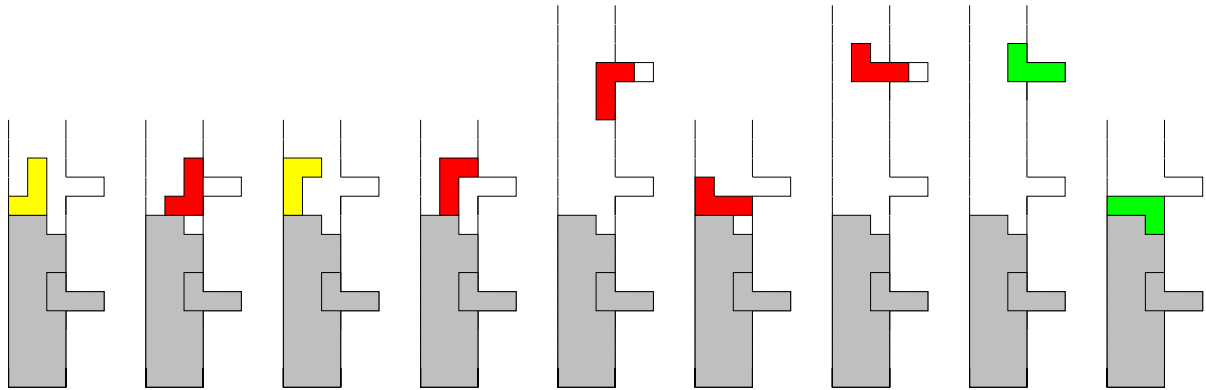
\begin{tabular}{llllllllllllll}
  \begin{block}{5}{14}
    \column{5}{14}
    \piece{\floor[lightgray]998}00
    \piece{\LGr[lightgray]}24
    \piece{\LGu[yellow]}09
  \end{block}
  &
  \begin{block}{5}{14}
    \column{5}{14}
    \piece{\floor[lightgray]998}00
    \piece{\LGr[lightgray]}24
    \piece{\LGu[red]}19
  \end{block}
  &
  \begin{block}{5}{14}
    \column{5}{14}
    \piece{\floor[lightgray]998}00
    \piece{\LGr[lightgray]}24
    \piece{\LGd[yellow]}09
  \end{block}
  &
  \begin{block}{5}{14}
    \column{5}{14}
    \piece{\floor[lightgray]998}00
    \piece{\LGr[lightgray]}24
    \piece{\LGd[red]}19
  \end{block}
  &
  \begin{block}{5}{20}
    \column{5}{20}
    \piece{\floor[lightgray]998}00
    \piece{\LGr[lightgray]}24
    \piece{\LGd[red]}2{14}
  \end{block}
  &
  \begin{block}{5}{14}
    \column{5}{14}
    \piece{\floor[lightgray]998}00
    \piece{\LGr[lightgray]}24
    \piece{\LGr[red]}09
  \end{block}
  &
  \begin{block}{5}{20}
    \column{5}{20}
    \piece{\floor[lightgray]998}00
    \piece{\LGr[lightgray]}24
    \piece{\LGr[red]}1{16}
  \end{block}
  &
  \begin{block}{5}{20}
    \column{5}{20}
    \piece{\floor[lightgray]998}00
    \piece{\LGr[lightgray]}24
    \piece{\LGr[green]}2{16}
  \end{block}
  &
  \begin{block}{5}{14}
    \column{5}{14}
    \piece{\floor[lightgray]998}00
    \piece{\LGr[lightgray]}24
    \piece{\LGl[green]}08
  \end{block}\\\\
\end{tabular}
\else
\includegraphics{figures/thappy/lg.epsi.clean}
\fi


  \end{center}
  (In the fifth, seventh, and eighth configurations, the $\tLG$ is in
  the $i$th-highest notch for any $i\ge 1$.)
  
  The first has a \spurn, and the third has a \btwo.  The eighth is
  \lgthappy{i}, and the ninth is \underflat.  The remainder have holes.
\end{proof}

\begin{proposition}  \mylabel{prop:ls-thappy}
  There is no valid move for $\tLS$ in a \thappy bucket.
\end{proposition}
\begin{proof}
  The possible configurations are as follows:
  \begin{center}
    \ifredraw
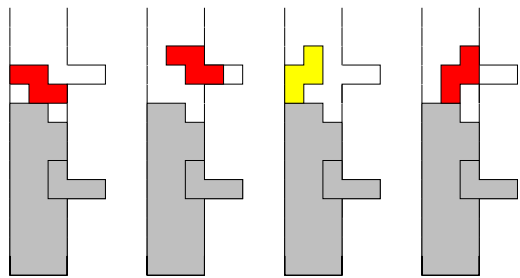
    \begin{tabular}{llllllllllllll}
      \begin{block}{5}{14}
        \column{5}{14}
        \piece{\floor[lightgray]998}00
        \piece{\LGr[lightgray]}24
        \piece{\LSl[red]}09
      \end{block}
&
      \begin{block}{5}{14}
        \column{5}{14}
        \piece{\floor[lightgray]998}00
        \piece{\LGr[lightgray]}24
        \piece{\LSl[red]}1{10}
      \end{block}
&
      \begin{block}{5}{14}
        \column{5}{14}
        \piece{\floor[lightgray]998}00
        \piece{\LGr[lightgray]}24
        \piece{\LSu[yellow]}09
      \end{block}
&
      \begin{block}{5}{14}
        \column{5}{14}
        \piece{\floor[lightgray]998}00
        \piece{\LGr[lightgray]}24
        \piece{\LSu[red]}19
      \end{block}
\\\\
    \end{tabular}
\else
\includegraphics{figures/thappy/ls.epsi.clean}
\fi


  \end{center}
  (The second configuration represents the placement of the $\tLS$
  into the $i$th-highest notch for any $i\geq 1$.)
  
  The first, second (for any $i$), and fourth configurations have
  holes; the third has a \spurn.
\end{proof}

\begin{proposition}  \mylabel{prop:sq-thappy-notched}
  There is no valid placement of $\tSq$ in \lgthappy{i}.
\end{proposition}
\begin{proof}
  The possible configurations are the following:
  \begin{center}
    \ifredraw
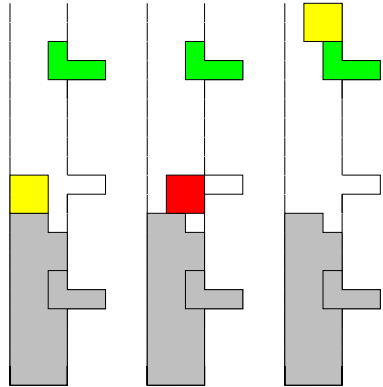
\begin{tabular}{ccc}
\begin{block}{5}{20}
  \column{5}{20}
  \piece{\floor[lightgray]998}00
  \piece{\LGr[lightgray]}24
  \piece{\LGr[green]}2{16}
  \piece{\Sq[yellow]}09
\end{block}
&
\begin{block}{5}{20}
  \column{5}{20}
  \piece{\floor[lightgray]998}00
  \piece{\LGr[lightgray]}24
  \piece{\LGr[green]}2{16}
  \piece{\Sq[red]}19
\end{block}
&
\begin{block}{5}{20}
  \column{5}{20}
  \piece{\floor[lightgray]998}00
  \piece{\LGr[lightgray]}24
  \piece{\LGr[green]}2{16}
  \piece{\Sq[yellow]}1{18}
\end{block}
\\\\
    \end{tabular}
\else
\includegraphics{figures/thappy/sqlg.epsi.clean}
\fi

  \end{center}
  (The $\tLG$ is initially in the $i$th-highest notch for $i\ge1$; the
  second configuration is blocked for $i=1$.)
  
  The first has a \spurn for $i>1$ and a hole for $i=1$.  The second
  has a hole.  The third has a \brect if $i>1$ and a \btwo if $i=1$.
\end{proof}

\begin{proposition}  \mylabel{prop:ls-thappy-notched}
  There is no valid placement of $\tLS$ in \lgthappy{i}.
\end{proposition}
\begin{proof}
  The possible configurations are the following:
  \begin{center}
    \ifredraw
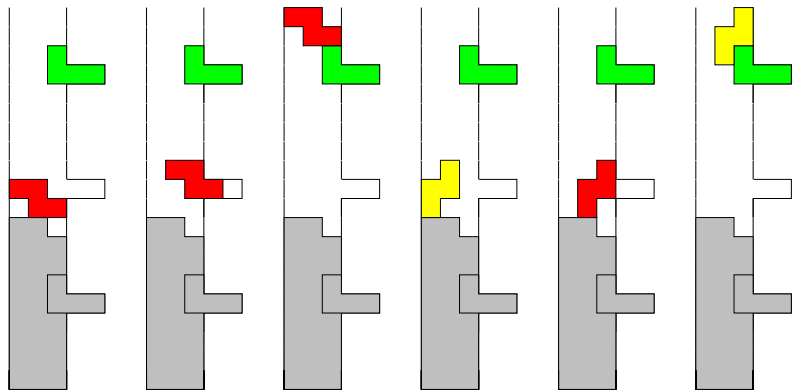
\begin{tabular}{cccccc}
\begin{block}{5}{20}
  \column{5}{20}
  \piece{\floor[lightgray]998}00
  \piece{\LGr[lightgray]}24
  \piece{\LGr[green]}2{16}
  \piece{\LSr[red]}09
\end{block}
&
\begin{block}{5}{20}
  \column{5}{20}
  \piece{\floor[lightgray]998}00
  \piece{\LGr[lightgray]}24
  \piece{\LGr[green]}2{16}
  \piece{\LSr[red]}1{10}
\end{block}
&
\begin{block}{5}{20}
  \column{5}{20}
  \piece{\floor[lightgray]998}00
  \piece{\LGr[lightgray]}24
  \piece{\LGr[green]}2{16}
  \piece{\LSr[red]}0{18}
\end{block}
&
\begin{block}{5}{20}
  \column{5}{20}
  \piece{\floor[lightgray]998}00
  \piece{\LGr[lightgray]}24
  \piece{\LGr[green]}2{16}
  \piece{\LSu[yellow]}09
\end{block}
&
\begin{block}{5}{20}
  \column{5}{20}
  \piece{\floor[lightgray]998}00
  \piece{\LGr[lightgray]}24
  \piece{\LGr[green]}2{16}
  \piece{\LSu[red]}19
\end{block}
&
\begin{block}{5}{20}
  \column{5}{20}
  \piece{\floor[lightgray]998}00
  \piece{\LGr[lightgray]}24
  \piece{\LGr[green]}2{16}
  \piece{\LSu[yellow]}1{17}
\end{block}
\\\\
    \end{tabular}
\else
\includegraphics{figures/thappy/lslg.epsi.clean}
\fi

  \end{center}
  (The $\tLG$ is initially in the $i$th-highest notch for $i\ge1$; the
  fifth configuration is blocked for $i=1$.  In the second
  configuration, the $\tLS$ is in the $j$th-highest notch for any
  $j\not=i$ and $j\ge1$.)
  
  The fourth configuration has a \spurn for $i>1$, and a hole for
  $i=1$.  The sixth has a \brect for $i>1$ and a \btwo for $i=1$.  All
  other configurations have holes.
\end{proof}

\begin{proposition}  \mylabel{lg-thappy-notched}
  If $\tLG$ is placed validly into \lgthappy{i} then the result is
  either \lgunderflat{i} or \lglgthappy{i}{j}.
\end{proposition}
\begin{proof}
  The possible configurations are as follows when the $\tLG$ is placed
  vertically:
  \begin{center}
    \ifredraw
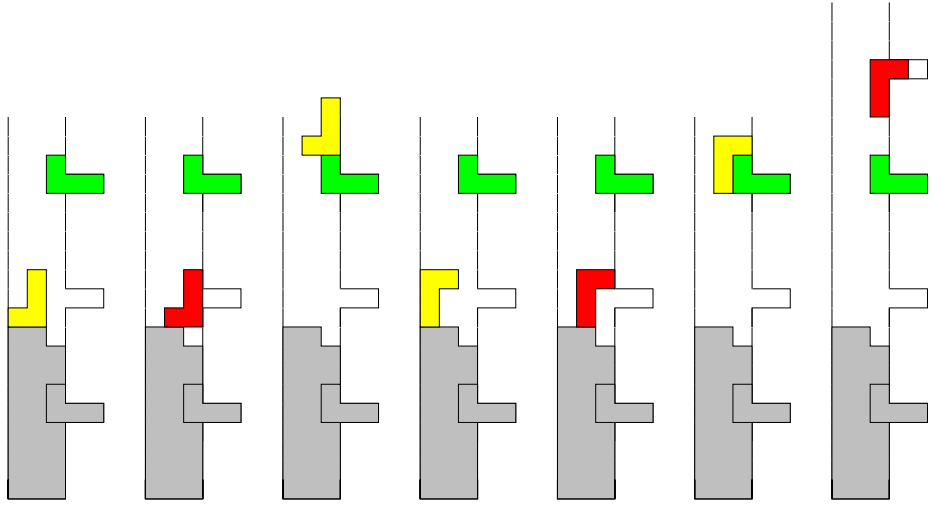
     \begin{tabular}{llllllllllllll}
      \begin{block}{5}{20}
        \column{5}{20}
        \piece{\floor[lightgray]998}00
        \piece{\LGr[lightgray]}24
        \piece{\LGr[green]}2{16}
        \piece{\LGu[yellow]}0{9}
      \end{block}
&
      \begin{block}{5}{20}
        \column{5}{20}
        \piece{\floor[lightgray]998}00
        \piece{\LGr[lightgray]}24
        \piece{\LGr[green]}2{16}
        \piece{\LGu[red]}1{9}
      \end{block}
&
      \begin{block}{5}{20}
        \column{5}{20}
        \piece{\floor[lightgray]998}00
        \piece{\LGr[lightgray]}24
        \piece{\LGr[green]}2{16}
        \piece{\LGu[yellow]}1{18}
      \end{block}
&
      \begin{block}{5}{20}
        \column{5}{20}
        \piece{\floor[lightgray]998}00
        \piece{\LGr[lightgray]}24
        \piece{\LGr[green]}2{16}
        \piece{\LGd[yellow]}0{9}
      \end{block}
&
      \begin{block}{5}{20}
        \column{5}{20}
        \piece{\floor[lightgray]998}00
        \piece{\LGr[lightgray]}24
        \piece{\LGr[green]}2{16}
        \piece{\LGd[red]}1{9}
      \end{block}
       &
       \begin{block}{5}{20}
         \column{5}{20}
         \piece{\floor[lightgray]998}00
         \piece{\LGr[lightgray]}24
         \piece{\LGr[green]}2{16}
         \piece{\LGd[yellow]}1{16}
      \end{block}
      &
       \begin{block}{5}{26}
         \column{5}{26}
         \piece{\floor[lightgray]998}00
         \piece{\LGr[lightgray]}24
         \piece{\LGr[green]}2{16}
         \piece{\LGd[red]}2{20}
      \end{block}

\\\\
    \end{tabular}
\else
\includegraphics{figures/thappy/lglgV.epsi.clean}
\fi

  \end{center}
  (The initially-placed $\tLG$ is in the $i$th-highest notch, for some
  $i\ge1$.  The second and fifth configurations are blocked for
  $i=1$.  In the last configuration, the $\tLG$ is in the
  $j$th-highest notch, for any $j\not=i$ and $j\ge1$.)
  
  The first configuration has a \spurn for $i>1$ and a hole for $i=1$.
  The second and fifth have a hole.  The third and sixth each have a
  \brect for $i>1$ and a \btwo for $i=1$.  The fourth has a \btwo for
  $i>1$ and a hole for $i=1$.  The last has a hole for any $j$.
  
  The possible configurations are as follows when the $\tLG$ is placed
  horizontally:
  \begin{center}
    \ifredraw
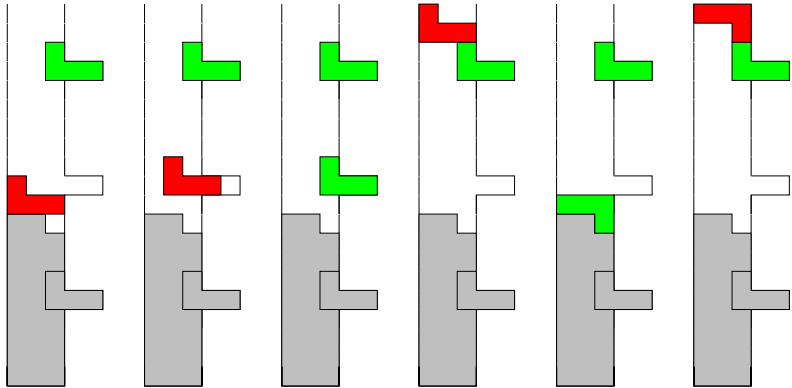
\begin{tabular}{llllllllllllll}
&
      \begin{block}{5}{20}
        \column{5}{20}
        \piece{\floor[lightgray]998}00
        \piece{\LGr[lightgray]}24
        \piece{\LGr[green]}2{16}
        \piece{\LGr[red]}0{9}
      \end{block}
&
      \begin{block}{5}{20}
        \column{5}{20}
        \piece{\floor[lightgray]998}00
        \piece{\LGr[lightgray]}24
        \piece{\LGr[green]}2{16}
        \piece{\LGr[red]}1{10}
      \end{block}
&
      \begin{block}{5}{20}
        \column{5}{20}
        \piece{\floor[lightgray]998}00
        \piece{\LGr[lightgray]}24
        \piece{\LGr[green]}2{16}
        \piece{\LGr[green]}2{10}
      \end{block}
&
      \begin{block}{5}{20}
        \column{5}{20}
        \piece{\floor[lightgray]998}00
        \piece{\LGr[lightgray]}24
        \piece{\LGr[green]}2{16}
        \piece{\LGr[red]}0{18}
      \end{block}
&
      \begin{block}{5}{20}
        \column{5}{20}
        \piece{\floor[lightgray]998}00
        \piece{\LGr[lightgray]}24
        \piece{\LGr[green]}2{16}
        \piece{\LGl[green]}0{8}
      \end{block}
&
      \begin{block}{5}{20}
        \column{5}{20}
        \piece{\floor[lightgray]998}00
        \piece{\LGr[lightgray]}24
        \piece{\LGr[green]}2{16}
        \piece{\LGl[red]}0{18}
      \end{block}
\\\\
    \end{tabular}
\else
\includegraphics{figures/thappy/lglgH.epsi.clean}
\fi

  \end{center}
  (The initially-placed $\tLG$ is in the $i$th-highest notch for some
  $i\ge1$; in the second and third configurations, the $\tLG$ is
  placed into the $j$th-highest notch, for $j \not=i$ and $j\ge1$.)
  
  The third configuration is \lglgthappy{i}{j}; the fifth is
  \lgunderflat{i}.  The remainder all have holes.
\end{proof}

\begin{proposition}  \mylabel{prop:sq-thappy-notched-two}
  There is no valid move for $\tSq$ in \lglgthappy{i}{j}.
\end{proposition}
\begin{proof}
  The possible configurations are the following:
  \begin{center}
    \ifredraw
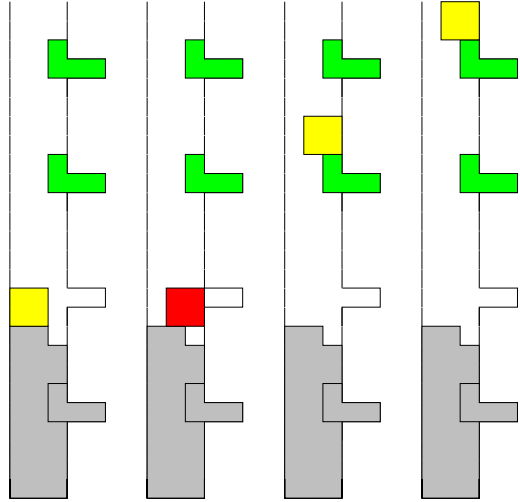
\begin{tabular}{cccc}
\begin{block}{5}{26}
  \column{5}{26}
  \piece{\floor[lightgray]998}00
  \piece{\LGr[lightgray]}24
  \piece{\LGr[green]}2{16}
  \piece{\LGr[green]}2{22}
  \piece{\Sq[yellow]}09
\end{block}
&
\begin{block}{5}{26}
  \column{5}{26}
  \piece{\floor[lightgray]998}00
  \piece{\LGr[lightgray]}24
  \piece{\LGr[green]}2{16}
  \piece{\LGr[green]}2{22}
  \piece{\Sq[red]}19
\end{block}
&
\begin{block}{5}{26}
  \column{5}{26}
  \piece{\floor[lightgray]998}00
  \piece{\LGr[lightgray]}24
  \piece{\LGr[green]}2{16}
  \piece{\LGr[green]}2{22}
  \piece{\Sq[yellow]}1{18}
\end{block}
&
\begin{block}{5}{26}
  \column{5}{26}
  \piece{\floor[lightgray]998}00
  \piece{\LGr[lightgray]}24
  \piece{\LGr[green]}2{16}
  \piece{\LGr[green]}2{22}
  \piece{\Sq[yellow]}1{24}
\end{block}
\\\\
\end{tabular}
\else
\includegraphics{figures/thappy/sqlglg.epsi.clean}
\fi


  \end{center}
  (The initial $\tLG$'s are in the $i$th- and $j$th-highest notches,
  for some $i,j \ge 1$.  The second configuration is blocked if
  $\min(i,j) = 1$.)
  
  The first has a \spurn (or a hole if $\min(i,j) = 1$), the second
  has a hole, and the third and fourth both have a \brect (or a \btwo
  if $\min(i,j) = 1$).
\end{proof}


\newpage

\subsection{Soundness Theorems}
\begin{oldtheorem}[Lemma \ref{lemma:initiator-correctness}]
  In an unprepped configuration, the only possibly valid strategy for
  $\tup{\tI,\tLG,\tSq}$ is to place all three pieces in some bucket to
  produce an overflat bucket, yielding a one-\overflat configuration.
\end{oldtheorem}
\begin{proof}
  Initially, all buckets are \unprepped.  By Proposition
  \ref{prop:i-unprepped}, the result of placing the $\tI$ in an
  unprepped bucket is either \underflat or \iprepped.

  \begin{itemize}
  \item \textbf{$\tI$ produces an \underflat bucket.}  Then the
    configuration is one-\underflat, and consists of unprepped buckets
    and an \underflat bucket.
    \begin{itemize}
    \item \textbf{The $\tLG$ goes into the \underflat bucket.}  By
      Proposition \ref{prop:lg-underflat}, the result is
      \lgunderflat{i} for some $i$.
      
      Then the current configuration consists of unprepped buckets and
      \lgunderflat{i}.  Now we consider where we can place the $\tSq$:
      \begin{itemize}
      \item \textbf{The $\tSq$ goes into the \lgunderflat{i} bucket.}
        By Proposition \ref{prop:sq-ufnotched}, this is valid iff
        $i=1$ and the result is \overflat.
      \item \textbf{The $\tSq$ goes into an unprepped bucket.}
        Invalid by Proposition \ref{prop:sq-unprepped}.
      \end{itemize}
      
    \item \textbf{The $\tLG$ goes into an unprepped bucket.} By
      Proposition \ref{prop:lg-unprepped} the result is \lgprepped{i}
      for some $i$.  Then the current configuration consists of
      unprepped buckets, an \underflat bucket, and \lgprepped{i}.  But
      now we must place the $\tSq$:
      \begin{itemize}
      \item \textbf{The $\tSq$ goes into the \lgprepped{i} bucket.}
        Invalid by Proposition \ref{prop:sq-notched}.
      \item \textbf{The $\tSq$ goes into the \underflat bucket.}
        Invalid by Proposition \ref{prop:sq-underflat}.
      \item \textbf{The $\tSq$ goes into an unprepped bucket.}
        Invalid by Proposition \ref{prop:sq-unprepped}.
      \end{itemize}
    \end{itemize}  
    Thus the only valid move sequence is to place the $\tLG$ in the
    same bucket to yield a \lgprepped{1}, and then the $\tSq$ in the
    same bucket to yield an \overflat bucket.

  \item \textbf{$\tI$ produces an \iprepped bucket.}  Then the current
    configuration consists of unprepped buckets and an \iprepped
    bucket.
    \begin{itemize}
    \item \textbf{The $\tLG$ goes into the \iprepped bucket.}  By
      Proposition \ref{prop:lg-iprepped}, the result is \ipreppedlg{i}
      for some $i$.
      
      Then the current configuration consists of unprepped buckets and
      \ipreppedlg{i}.  
      \begin{itemize}
      \item \textbf{The $\tSq$ goes into the \ipreppedlg{i} bucket.}
        Invalid by Proposition \ref{prop:sq-iprepped-notched}.
      \item \textbf{The $\tSq$ goes into an unprepped bucket.}
        Invalid by Proposition \ref{prop:sq-unprepped}.
      \end{itemize}
      
    \item \textbf{The $\tLG$ goes into an unprepped bucket.} By
      Proposition \ref{prop:lg-unprepped} the result is \lgprepped{i}
      for some $i$.  Then the current configuration consists of
      unprepped buckets, \lgprepped{i}, and \iprepped.  But now we
      must place the $\tSq$:
      \begin{itemize}
      \item \textbf{The $\tSq$ goes into the \lgprepped{i} bucket.}
        Invalid by Proposition \ref{prop:sq-notched}.
      \item \textbf{The $\tSq$ goes into the \iprepped bucket.}
        Invalid by Proposition \ref{prop:sq-iprepped}.
      \item \textbf{The $\tSq$ goes into an unprepped bucket.}
        Invalid by Proposition \ref{prop:sq-unprepped}.
      \end{itemize}
    \end{itemize}
  \end{itemize}
  Thus the only valid move sequence is to place the $\tI$ into a
  bucket to yield an \underflat configuration, then place the $\tLG$ in the
  same bucket to yield a \lgprepped{1}, and finally the $\tSq$ in the
  same bucket to yield an \overflat.
\end{proof}

\begin{lemma} \mylabel{prop:overflat-triggerhappy}
  In a one-\overflat configuration, the only possibly valid strategy
  for the sequence $\tup{\tLG,\tLS}$ is to place both pieces into the
  \overflat bucket, producing
  a \thappy bucket, yielding a
  one-\thappy configuration.
\end{lemma}
\begin{proof}
  The initial configuration consists of unprepped buckets and one
  overflat bucket.
  \begin{itemize}
  \item \textbf{The $\tLG$ goes into an unprepped bucket.}  By
    Proposition \ref{prop:lg-unprepped}, the result is \lgprepped{i}
    for some $i$.  Then the configuration now consists of unprepped
    buckets, one overflat bucket, and one \lgprepped{i} bucket.

    \begin{itemize}
    \item \textbf{The $\tLS$ goes into the overflat bucket.}  Invalid by
      Proposition \ref{prop:ls-overflat}.
    \item \textbf{The $\tLS$ goes into the \lgprepped{i} bucket.}
      Invalid by Proposition \ref{prop:ls-notched}.
    \item \textbf{The $\tLS$ goes into an unprepped bucket.}  Invalid by
      Proposition \ref{prop:ls-unprepped}.
    \end{itemize}
    
  \item \textbf{The $\tLG$ goes into the overflat bucket.}  By
    Proposition \ref{prop:lg-overflat}, the result is \lgoverflatA,
    \lgoverflatB, \lgoverflatC, or \lgoverflatD{i} for some $i$.  Then
    the current configuration consists of unprepped buckets and one
    \lgoverflatA, \lgoverflatB, \lgoverflatC, or \lgoverflatD{i}.
    \begin{itemize}
    \item \textbf{The $\tLS$ goes into an unprepped bucket.}  Invalid by
      Proposition \ref{prop:ls-unprepped}.
    \item \textbf{The $\tLS$ goes into the \lgoverflatA bucket.}
      Invalid by Proposition \ref{prop:ls-over-1}.
    \item \textbf{The $\tLS$ goes into the \lgoverflatB bucket.}  By
      Invalid by Proposition \ref{prop:ls-over-2}.
    \item \textbf{The $\tLS$ goes into the \lgoverflatC bucket.}  By
      Proposition \ref{prop:ls-over-3}, the result is \thappy.
    \item \textbf{The $\tLS$ goes into the \lgoverflatD{i} bucket.}
      Invalid by Proposition \ref{prop:ls-over-4}.
    \end{itemize}
  \end{itemize}
  Thus the only possibly valid move is the placement of the $\tLG$
  into the overflat bucket to yield either a \lgoverflatB or a
  \lgoverflatC, and the placement of the $\tLS$ into the same bucket
  to make it
  \thappy.  This results in a
  one-\thappy 
  configuration.
\end{proof}

\begin{lemma} \mylabel{prop:triggerhappy-overflat}
  In a one-\thappy configuration, the only possibly valid strategy for
  the sequence $\tup{\tLG,\tLG,\tSq}$ is (1) to place all three pieces
  in the \thappy bucket to yield a one-\overflat configuration, or (2)
  to place all three pieces in an unprepped bucket to yield a
  one-\tplat-one-\thappy configuration.
\end{lemma}
\begin{proof}
  Our initial configuration consists of unprepped buckets and one
  \thappy bucket.  First, we consider whether zero, one, or two of the
  $\tLG$'s go into the \thappy bucket:
  \begin{itemize}
  \item \textbf{Neither $\tLG$ goes into the \thappy bucket.}  Then
    the two $\tLG$'s go into unprepped buckets.
    \begin{itemize}
    \item \textbf{The two $\tLG$'s go into different unprepped
        buckets.}  Then by Proposition \ref{prop:lg-unprepped}, the
      result of dropping each $\tLG$ is \lgprepped{i} and
      \lgprepped{i'}, for some $i$ and $i'$.
      
      Thus our configuration consists of unprepped buckets,
      \lgprepped{i}, \lgprepped{i'}, and a \thappy bucket.

      \begin{itemize}
      \item \textbf{The $\tSq$ goes into an unprepped bucket.}
        Invalid by Proposition \ref{prop:sq-unprepped}.
        
      \item \textbf{The $\tSq$ goes into \lgprepped{i} or
          \lgprepped{i'}.}  Invalid by Proposition
        \ref{prop:sq-notched}.
        
      \item \textbf{The $\tSq$ goes into the \thappy bucket.}
        Invalid by Proposition \ref{prop:sq-thappy}.
      \end{itemize}

    \item \textbf{The two $\tLG$'s go into the same unprepped bucket.}
      Then by Proposition \ref{prop:lg-unprepped}, the result of
      dropping the first $\tLG$ into an unprepped bucket is
      \lgprepped{i} for some $i$.  By Proposition
      \ref{prop:lg-notched}, the result of the second is a \splat or
      \lgpreppedlg{i}{j}, for some $j$.
      
      Now our configuration consists of unprepped buckets, a \thappy
      bucket, and one \lgpreppedlg{i}{j} or \splat.
      \begin{itemize}
      \item \textbf{The $\tSq$ goes into an unprepped bucket.}
        Invalid by Proposition \ref{prop:sq-unprepped}.
        
      \item \textbf{The $\tSq$ goes into the \thappy bucket.}
        Invalid by Proposition \ref{prop:sq-thappy}.
        
      \item \textbf{The $\tSq$ goes into \lgpreppedlg{i}{j}.} 
        Invalid by Proposition \ref{prop:sq-notched-two}.
        
      \item \textbf{The $\tSq$ goes into the \splat bucket.}  By
        Proposition \ref{prop:splat-tplat}, the result is a \tplat.
      \end{itemize}
    \end{itemize}
    
  \item \textbf{Exactly one of the $\tLG$'s goes into the trigger-happy
      bucket.}  By Proposition \ref{prop:lg-thappy}, the result of
    placing the $\tLG$ into the \thappy bucket is either \lgthappy{i}
    or \underflat, and by Proposition \ref{prop:lg-unprepped}, the
    result of dropping the other $\tLG$ into an unprepped column is
    \lgprepped{i'}, for some $i$ and $i'$.
    
    The resulting configuration then consists of unprepped buckets,
    \lgprepped{i'} and either \lgthappy{i} or \underflat.
      \begin{itemize}
      \item \textbf{The $\tSq$ goes into an unprepped bucket.}
        Invalid by Proposition \ref{prop:sq-unprepped}.
        
      \item \textbf{The $\tSq$ goes into \lgprepped{i'}.}
        Invalid by Proposition \ref{prop:sq-notched}.
        
      \item \textbf{The $\tSq$ goes into \lgthappy{i}.}  Invalid by
        Proposition \ref{prop:sq-thappy-notched}.
        
      \item \textbf{The $\tSq$ goes into \underflat.}  Invalid by
        Proposition \ref{prop:sq-underflat}.
      \end{itemize}
      
    \item \textbf{Both of the $\tLG$'s go into the trigger-happy
        bucket.}  By Proposition \ref{prop:lg-thappy}, the result of
      placing the $\tLG$ into the \thappy bucket is either \underflat
      or \lgthappy{i} for some $i$.
    
      By Proposition \ref{lg-thappy-notched}, the result of placing
      the second $\tLG$ in \lgthappy{i} is either \lgunderflat{i} or
      \lglgthappy{i}{j} for some $j$.  By Proposition
      \ref{prop:lg-underflat}, the result of placing the $\tLG$ in the
      \underflat bucket is also \lgunderflat{i}.
      
      Then, regardless of whether the first $\tLG$ produced
      \lgthappy{i} or \underflat, the current configuration consists
      of unprepped buckets and either \lgunderflat{i} or
      \lglgthappy{i}{j}.
      \begin{itemize}
      \item \textbf{The $\tSq$ goes into an unprepped bucket.}
        Invalid by Proposition \ref{prop:sq-unprepped}.
        
      \item \textbf{The $\tSq$ goes into \lgunderflat{i}.}  By
        Proposition \ref{prop:sq-ufnotched}, this is valid iff $i=1$
        and the result is \overflat.
        
      \item \textbf{The $\tSq$ goes into \lglgthappy{i}{j}.}  Invalid by
        Proposition \ref{prop:sq-thappy-notched-two}.
      \end{itemize}
  \end{itemize}
  Thus the only possibly-valid moves are to place all three pieces
  into (1) the \thappy bucket to produce an \overflat bucket, yielding
  a one-\overflat configuration, or (2) an unprepped bucket to produce
  a \tplat bucket, yielding a one-\tplat-one-\thappy configuration.
\end{proof}

\begin{lemma} \mylabel{prop:filler-plateau-thappy-invalid}
  There is no valid move for the sequence $\tup{\tLG,\tLS}$ in a
  one-\tplat-one-\thappy configuration.
\end{lemma}
\begin{proof}
  Our initial configuration consists of unprepped buckets, one \tplat
  bucket, and one \thappy bucket.

  \begin{itemize}
    
  \item \textbf{The $\tLG$ goes into an unprepped bucket.}  Then by
    Proposition \ref{prop:lg-unprepped}, the result is \lgprepped{i} for
    some $i$.
    \begin{itemize}
      
    \item \textbf{The $\tLS$ goes into an unprepped bucket.}  Invalid
      by Proposition \ref{prop:ls-unprepped}.
      
    \item \textbf{The $\tLS$ goes into \lgprepped{i}.}  Invalid by
      Proposition \ref{prop:ls-notched}.
      
    \item \textbf{The $\tLS$ goes into the \tplat bucket.}  Invalid by
      Proposition \ref{prop:ls-rplat}.
      
    \item \textbf{The $\tLS$ goes into the \thappy bucket.}  Invalid
      by Proposition \ref{prop:ls-thappy}.

    \end{itemize}
    
  \item \textbf{The $\tLG$ goes into the \tplat bucket.}  Then by
    Proposition \ref{prop:lg-rplat}, the result is \lgtplat{i} for
    some $i$.

    \begin{itemize}
    \item \textbf{The $\tLS$ goes into an unprepped bucket.}  Invalid by
      Proposition \ref{prop:ls-unprepped}.
      
    \item \textbf{The $\tLS$ goes into \lgtplat{i}.}  Invalid by
      Proposition \ref{prop:ls-rplat-notched}.
    
    \item \textbf{The $\tLS$ goes into the \thappy bucket.}  Invalid by
      Proposition \ref{prop:ls-thappy}.
    \end{itemize}
    
  \item \textbf{The $\tLG$ goes into the \thappy bucket.}  Then by
    Proposition \ref{prop:lg-thappy}, the result is either underflat
    or \lgthappy{i} for some $i$.

    \begin{itemize}
    \item \textbf{The $\tLS$ goes into an unprepped bucket.}  Invalid by
      Proposition \ref{prop:ls-unprepped}.
      
    \item \textbf{The $\tLS$ goes into \lgthappy{i}.}  Invalid by
      Proposition \ref{prop:ls-thappy-notched}.
      
    \item \textbf{The $\tLS$ goes into the \underflat bucket.}
      Invalid by Proposition \ref{prop:ls-underflat}.
    
    \item \textbf{The $\tLS$ goes into the \tplat bucket.}  Invalid by
      Proposition \ref{prop:ls-rplat}.
    \end{itemize}

\end{itemize}

Thus there is no valid strategy for $\tup{\tLG,\tLS}$ in a
one-\tplat-one-\thappy configuration.
\end{proof}

\begin{oldtheorem}[Lemma \ref{lemma:filler-correctness}]
For the sequence $\tup{\tLG,\tLS,\tLG,\tLG,\tSq}$:
  \begin{enumerate}
  \item In a one-\overflat configuration, the only possibly valid
    strategy is either (1) to place all pieces in the \overflat bucket
    to produce an \overflat bucket, yielding a one-\overflat
    configuration, or (2) to place $\tup{\tLG,\tLS}$ into the
    \overflat bucket and $\tup{\tLG,\tLG,\tSq}$ in an unprepped
    bucket, yielding a one-\tplat-one-\thappy configuration.
    
  \item In a one-\tplat-one-\thappy configuration, there is no valid
    strategy.
\end{enumerate}
%
%
\end{oldtheorem}
\begin{proof}
  For the one-\overflat configuration, by Lemma
  \ref{prop:overflat-triggerhappy}, if the sequence $\tup{\tLG,\tLS}$
  is placed validly, both pieces are dropped into the \overflat
  bucket, and the result is a one-\thappy configuration.  By Lemma
  \ref{prop:triggerhappy-overflat}, the only valid placement for the
  sequence $\tup{\tLG,\tLG,\tSq}$ in a one-\thappy configuration
  yields a one-\overflat (placing all three pieces into the \thappy
  bucket) or one-\tplat-one-\thappy configuration (placing them into
  an unprepped bucket).
  
  In the one-\tplat-one-\thappy configuration, by Lemma
  \ref{prop:filler-plateau-thappy-invalid}, there is no valid
  trajectory sequence.
\end{proof}

\begin{oldtheorem}[Lemma \ref{lemma:terminator-correctness}]
  For the sequence $\tup{\tSq,\tSq}$,
  \begin{enumerate}
  \item In a one-\overflat configuration, the only possibly valid
    strategy is to place both pieces in the \overflat bucket to
    produce an unprepped bucket, yielding an unprepped configuration.
%
  \item In a one-\tplat-one-\thappy configuration, there is no valid
    strategy.
\end{enumerate}
\end{oldtheorem}
\begin{proof}
  By Propositions \ref{prop:sq-unprepped}, \ref{prop:sq-thappy}, and
  \ref{prop:sq-rplat}, no $\tSq$ can validly go into any unprepped,
  \thappy, or \tplat bucket.
  
  For a one-\overflat configuration, then, both $\tSq$'s must go into
  the \overflat bucket.  By Proposition \ref{prop:sqsq-overflat}, the
  result is an unprepped bucket.
  
  For a one-\tplat-one-\thappy configuration, there is no bucket into
  which the first $\tSq$ can be validly placed.
\end{proof}



\end{document}